\documentclass[aip,reprint]{revtex4-1}
\draft
\usepackage{amsmath}
\usepackage{graphicx}
\usepackage{xcolor}
\usepackage{caption}
\usepackage{subcaption}
\usepackage[colorinlistoftodos]{todonotes}

\usepackage{xr}
\usepackage[version=4]{mhchem}
\usepackage{mwe}
\usepackage[percent]{overpic}
\usepackage{floatrow}
\usepackage{multirow}

\usepackage{siunitx}
\DeclareSIUnit[number-unit-product = {\,}]
\cal{cal}
\DeclareSIUnit\kcal{\kilo\cal}
\DeclareSIUnit\kcal{\kilo\joule\per\mole}
\DeclareSIUnit\molar{\mole\per\cubic\deci\metre}
\DeclareSIUnit\Molar{\textsc{m}}
\DeclareSIUnit\nsec{\text{nanoseconds}}
    \DeclareSIUnit[number-unit-product = \ ]\nanosec{\text{nanoseconds}}
\usepackage{multirow}
\usepackage{xr}
\usepackage{cleveref}
\usepackage{chemformula}
\usepackage{chemfig}
\usepackage{floatrow}

\newcommand*\sfref[1]{%
    Supplementary Figure \ref{#1}}
 
\newcommand*\stref[1]{%
    Supplementary Table \ref{#1}}
 
\begin{document}

\title{Solvent quality and solvent polarity in polypeptides}

\author{Cedrix J. Dongmo Foumthuim}
\email{cedrix.dogmo@unive.it}
\affiliation{Dipartimento di Scienze Molecolari e Nanosistemi, 
Universit\`{a} Ca' Foscari di Venezia
Campus Scientifico, Edificio Alfa,
via Torino 155, 30170 Venezia Mestre, Italy}
\author{Achille Giacometti}
\email{achille.giacometti@unive.it}
\affiliation{Dipartimento di Scienze Molecolari e Nanosistemi, 
Universit\`{a} Ca' Foscari di Venezia
Campus Scientifico, Edificio Alfa,
via Torino 155, 30170 Venezia Mestre, Italy}
\affiliation{European Centre for Living Technology (ECLT)
Ca' Bottacin, 3911 Dorsoduro Calle Crosera, 
30123 Venice, Italy}

\date{\today}

\begin{abstract}
Using molecular dynamics and thermodynamic integration, we report on the solvation process in water and in cyclohexane of seven polypeptides (GLY, ALA, ILE, ASN, LYS, ARG, GLU).  The polypeptides are selected to cover the full hydrophobic scale while varying their chain length from tri- to undeca-homopeptides provide indications on possible non-additivity effects as well as the role of the peptide backbone in the overall stability of the polypeptides.  The use of different solvents and different polypeptides allows us to investigate the relation between \textit{solvent quality} -- the capacity of a given solvent to fold a given biopolymer often described on a scale ranging from "good" to "poor",  and \textit{solvent polarity} -- related to the specific interactions of any solvent with respect to a reference solvent. Undeca-glycine is found to be the only polypeptides to have a proper stable collapse in water (polar solvent), with the other hydrophobic polypetides displaying in water repeated folding and unfolding events and with polar polypeptides presenting a even more complex behavior. By contrast, all polypeptides but none are found to keep an extended conformation in cyclohexane, irrespective of their polarity. All considered polypeptides are also found to have a favorable solvation free energy independently of the solvent polarity and their intrinsic hydrophobicity, clearly highlighting the prominent stabilizing role of the peptide backbone, with the solvation process largely enthalpically dominated in polar polypeptides and partially entropically driven for hydrophobic polypeptides. Our study thus reveals the complexity of the solvation process of polypeptides defying the common view " like dissolves like", with the solute polarity playing the most prominent role.  The absence of a mirror symmetry upon the inversion of polarities of both the solvent and the polypeptides is confirmed.
\end{abstract}

\maketitle

\section{Introduction}
\label{sec:introduction}
\begin{figure}[h!]
\centering
\captionsetup{justification=raggedright,width=\linewidth}
\begin{subfigure}{6.5cm}
\includegraphics[width=\linewidth]{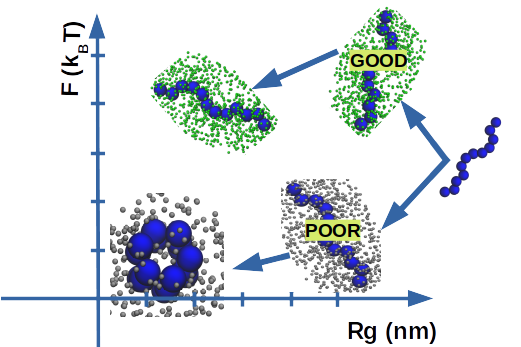}
\caption{}\label{fig:fig1a}
\end{subfigure}
\begin{subfigure}{6.5cm}
\includegraphics[width=\linewidth]{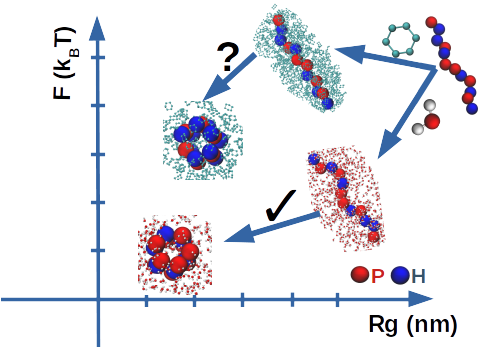}
\caption{}\label{fig:fig1b}
\end{subfigure}
\caption{Cartoon description of the solvophobic effects in different environments in the plane free energy $F$ (units of thermal energy $k_B T$) $F/k_B T$ with respect to gyration radius $R_g$ of the polymer. Panel (a) is for a synthetic homopolymer  which collapses into a globule in a "poor" solvent and remains extended in a "good" solvent. Panel (b) displays the question of whether a heteropolymer formed by hydrophobic (H) and polar (P) monomers assumed to be collapsing in water \ce{H2O} into a unique fold with preferential exposition of the polar residues P, does collapse in a non-polar solvent such as cyclohexane \ce{cC6H12} by reversing inside out its fold with H residues exposed to the solvent and P residues buried inside for the fold.} 
\label{fig:fig1}
\end{figure}

In polymer physics \cite{Flory69,Doi88,Khokhlov02,Rubinstein03} the term \textit{poor} solvent indicates that a synthetic polymer tends to collapse into a compact conformation because the effective intra-chain interactions occurring between different monomers composing the polymer overcome the monomer-solvent interactions. In the opposite limit of \textit{good} solvent, the polymer tends to remain into an extended conformation. This effect is pictorially represented in Fig. \ref{fig:fig1a} in a plot of the free energy $F/k_B T$, in units of the thermal energy $k_B T$, as a function of the mean radius of gyration $R_g$. In the case of poor solvent the polymer lowers its free energy by folding into a compact conformation, thus reducing $R_g$, whereas in the second case the free energy decreases but $R_g$ remains large because the polymer is solvophobic.
{\color{black} The distinction between good and bad solvent can be made more quantitative using familiar scaling arguments from polymer physics where $R_g \sim N^{\nu}$ where $\nu \approx 3/5$ for extended/swollen conformation and $\nu \approx 1/3$ for compact/globule conformation \cite{Flory69,deGennes79,Khokhlov02,Rubinstein03,Bhattacharjee13}. }
While this picture is very simple and handy, it clearly disregards the fact that it depends on the specific properties of the polymer as well as of the solvent. Hence \textit{solvent quality} is used to identify the relative character of one solvent with respect to a reference one in terms of the above picture. Thus one solvent can be a good solvent for one polymer and bad for another one, and this point becomes extremely important in the framework of biopolymers and biomolecules \cite{Bolen2008}.

The conformational freedom of biomolecules in general, and of proteins in particular, enables them to inter-convert between several states in solution, thereby adapting upon changing the solvent environments, for e.g. by changing from a polar to a non-polar solvent. The same flexibility allows them to perform various functions \textit{in vivo}. However, even though water  is undoubtedly the most-like biological milieu, the stability of these latter is not necessary compromised in non-polar solvents \cite{Wolynes1995, Meyer2013}. A protein can be regarded as a chain formed by a sequence of amino acids taken from a 20 alphabet letters, half of which have hydrophobic (H) character, so they tend to avoid contact with water, whereas the other half are polar (P) so they are happy to stay in contact with water. Proteins in water fold reproducibly and reliably to achieve their unique native states driven by several concurring interactions, including the tendency to avoid contact with water, denoted as the \textit{hydrophobic effect}, as indicated in Fig.\ref{fig:fig1b}. Note that \textit{solvent polarity} in fact refers to the polar character of a specific solvent as compared to water that is taken as a reference scale for an optimal polar solvent, and this is clearly different from the definition of \textit{solvent quality} defined earlier, albeit the two definitions are often interpreted as meaning the same thing.  However, the presence of the hydrophobic residues might suggest a similar folding event occurring also in non-aqueous mileu, such as for instance an organic solvent. In this case, it might happen that the " protein would turn inside out with its hydrophilic or polar residues inside and hydrophobic apolar residues outside", as suggested by Peter Wolynes sometime ago \cite{Wolynes1995}, and pictorially represented in Fig.\ref{fig:fig1b}.  To the best of our knowledge, no record of such event exists in literature. {\color{black} In a conventional surfactants framework  oil form droplets in water and water form droplets in oil. However, it has been recently shown \cite{Carrer20} that this " mirror symmetry" is not respected by using "unconventional" surfactants with hydrophobic head and polar tail that do not form micelles in apolar solvent in the same way as conventional surfactants do in polar solvents such as water. Hence there is no " mirror symmetry" in this more complex case, and the same appears to be true in proteins \cite{Hayashi17,Hayashi18}. Likely, this is because this argument overlooks the character of the peptide bond, a feature that might turn the delicate balance provided by the amino acid properties \cite{Bolen2008,Karandur2014}. In addition, the actual length and energy scales are different in the two cases: in water, the enthalpy gain in saturating hydrogen bonds as well as the entropy increase stemming from the additional free water molecules, have no counterpart in organic solvents where the van der Waals interactions are much weaker and the entropic gain significantly reduced \cite{Hayashi17,Hayashi18,Carrer20}.} A confirmation of this picture is the aim of the present study.

For a fully solvated analyte, the solvation free energy can be used as a good indicator of the overall stability of the studied system, in relation to the solvent considered, and we have already carried out a detailed analysis of the solvation free energy of each single amino acid side chain equivalent both in water \ce{H2O} and in cyclohexane \ce{cC6H12}, as paradigmatic representative of an organic, apolar solvent \cite{Dongmo2020}. It was found that the transfer free energy from water \ce{H2O} to cyclohexane \ce{cC6H12}, that is the work necessary to bring one single amino acid side chain from one solvent to the other, was respecting the expected hydrophobic scale of the amino acids. Hence, hydrophobic amino acid side chains have decreasing free energy transfer, whereas polar amino acid side chains have increasing free energy, in agreement with experimental findings \cite{Wolfenden15}. In this analysis, however, the backbone part of each amino acid was removed and replaced by a single hydrogen atom -- obtaining what is hereafter referred to as side chain amino acid equivalents, thus hindering the effect of the backbone part that it was already argued to play an important role \cite{Hajari15}. As experimentally the solubility of polypeptides in water \ce{H2O} decreases as the length increases \cite{Karandur2014}, this dependence should also been taken into account. Both aspects will be then considered in the present study.

Polyglycine peptides (GLY$_n$), formed by $n$ identical repeated residues, are a common model for the peptide units. Other polypeptides can be formed in the same way by using amino acids with different polarities as for instance those reported in Table \ref{tab:amino_acid}. The interest in understanding the $n$ dependence of the solvation free energy is twofold. On one hand, it constitutes one of the key ingredients of the forces stabilizing protein folding \cite{Bolen2008}. On the other hand, the solvation process is known to be significantly different above and below a critical size (of order of \SI{1}{nm}), at least in water \cite{Chandler2005}. For both these reasons, there were several studies in recent literature reporting several useful results. 

\begin{table}[h]
\caption{The correspondence between the seven amino acids with their tri- and uni-code nomenclature used in this work to build the homopeptides. See e.g. Ref. \cite{Voet2010}.}
  \label{tab:amino_acid}
  \begin{center}
    \begin{tabular}{l r c c }
      \hline
      \text{Character} & \text{Amino acid} & \text{Short name} & \text{Single letter} \\
      \hline
      \text{Hydrophobic} & \text{Glycine} & \text{GLY} & \text{G}\\
      \text{Hydrophobic} & \text{Alanine} & \text{ALA} & \text{A} \\
      \text{Hydrophobic} & \text{Isoleucine} & \text{ILE} & \text{I} \\
      \text{Polar} & \text{Asparagine} & \text{ASN} & \text{N} \\
      \text{Polar} & \text{Lysine} & \text{LYS} & \text{K} \\
      \text{Polar} & \text{Arginine} & \text{ARG} & \text{R} \\
      \text{Polar} & \text{Glutamic acid} & \text{GLU} & \text{E} \\
   \hline
   \end{tabular}
  \end{center}
\end{table}

\citeauthor{Tomar2013} \cite{Tomar2013} addressed the paradoxical difference between theory and experiments on the group-additivity of the solvation free energy in an osmolyte solution (water plus small organic cosolutes), and emphasized the importance of evaluating the transfer free energy from one solution to another. 


Using calorimetric measurements of the solvation enthalpies of some dipeptide analogs,  Avbelj and Baldwin \cite{Avbelj2009} have suggested that the principle of group additivity does not hold true for the interaction of the peptide group with water \ce{H2O}. According to their results, the main reason of this breakdown is the strong electrostatic interactions between neighbouring \ce{NHCO} units of peptides in water \ce{H2O}.

In 2013, Kokubo et \textit{al}. \cite{Kokubo2013} analysed the effect of flexibility on the solvation free energies of alanine peptides in water \ce{H2O}. They found a linear dependence with respect to the peptide length  $n$ for both electrostatics, van der Waals cavity-formation, and total solvation free energies. 

In an attempt to provide a general view on the additivity character of the solvation free energy, Staritzbichler and collaborators \cite{Staritzbichler2005} used multiconfiguration thermodynamic
integration, along with generalized-born surface area solvation model to compute the solvation free energy of different polypeptides in the form of rigid helices of various length $n$ in water \ce{H2O} and in chloroform \ce{CHCl3}. They preferentially considered uncharged amino acids while tuning their backbones to fit an ideal helix conformation. Their results suggest a nonlinearity in the solvation free energy in the case of short ($n \le 5$) peptide chains, turning to linear for longer chains. 

Hajari and van der Vegt \cite{Hajari15} performed a molecular simulation study on the temperature dependence of solvation free energy of both polar and hydrophobic tripeptides in water \ce{H2O}. They found a significant deviation from linearity in the case of hydrophobic polypeptides and a nearly linear dependence for polar polypeptides.  This latter result was ascribed to a nearly perfect enthalpy-entropy compensation, leading the overall solvation free energy nearly unaltered by the peptide backbone. Contrariwise, no such compensation was found for hydrophobic tripeptides.  

In their work, Konig et \textit{al}. \cite{Konig2013} addressed the extent to which the assumption of group additivity to the absolute solvation free energy can hold valid. In doing so, they made use of molecular dynamics–based free energy simulations to estimate the absolute solvation free energies for 15 N-acetyl-methylamide amino acids with neutral side chains. The authors have shown that values of solvation free energies of full amino acids based on group-additive approaches are systematically too negative while completely overestimating the hydrophobicity of glycine.     

A work from Montgomery group \cite{Char2010} explored the solvation free energy of polyglycines of different length $n$, in pure water \ce{H2O} and in the
osmolyte solutions, 2M urea and 2M trimethylamine N-oxide (TMAO). The solvation free energies were found linearly dependending on $n$ and they identified the dependence on the specific interactions (van der Waals, electrostatics, etc). 

While all these studies prove to be rather useful, a coherent picture of the solvation process is still lacking. Motivated by this, in the present work we first analyze the poor/good paradigm of water \ce{H2O} and cyclohexane \ce{cC6H12} on polypeptides of different length $n$ and different polarities (hydrophobic and polar), and then compute the corresponding solvation free energies, disentangling the enthalpic and entropic contributions. 

The remaining of the paper is organized as follows. In Section \ref{sec:theory} we describe the underlying theory and the simulation methods used in this study. Section \ref{sec:results} then includes all results and Section \ref{sec:conclusions} a summary of the results along with a discussion. Supplementary Information includes additional figures and tables relative to the results reported in the main text.
\section{Theory and Methods}
\label{sec:theory}
\subsection{Thermodynamic integration}
\label{subsec:thermodynamic}
The solvation free energy $\Delta G_{sol}$ can be defined as the difference between the free energy of a single analyte molecule in a specified solvent $G_{solvent}$ and in vacuum $G_{vacuum}$
\begin{eqnarray}
  \label{sec2:eq1}
     \Delta G_{sol} &=& G_{solvent} - G_{vacuum}
\end{eqnarray}
If $\Delta G_{sol}<0$ ($\Delta G_{sol}>0$) the solvent is stabilizing (destabilizing) the molecule with respect to vacuum. This concept can clearly be extended to the free energy transfer  $\Delta \Delta G (S_1 \to S_2)$ between two different solvents $S_1$ and $S_2$
\begin{eqnarray}
  \label{sec2:eq2}
 \Delta \Delta G \left(S_1 \to S_2\right) &=& \Delta G_{S_{2}} - \Delta G_{S_{1}}
\end{eqnarray}
where $\Delta G_{S_{1}}$ and $\Delta G_{S_{2}}$ are the solvation free energy for solvents $S_1$ and $S_2$, respectively.

From the numerical viewpoint, free energy differences can be conveniently computed by using thermodynamic integration \cite{Frenkel01}
\begin{eqnarray}
  \label{sec2:eq3}
  \Delta G_{AB} &=& \int_{\lambda_{A}}^{\lambda_{B}} d\lambda \left \langle \frac{\partial V\left(\mathbf{r};\lambda\right)}{\partial \lambda} \right \rangle_{\lambda}
\end{eqnarray}
where $V(\mathbf{r},\lambda)$ is the potential energy of the system as a function of the coordinate vector $\mathbf{r}$, and $\lambda$ is a switching-on parameter allowing to go from state A to state B by changing its value from $\lambda_{A}$ to $\lambda_{B}$. The average $\langle \ldots \rangle_{\lambda}$ in Eq.(\ref{sec2:eq3}) is the usual thermal average with potential $V(\mathbf{r},\lambda)$. The $\lambda$ interval $[\lambda_A,\lambda_B]$ is partitioned
into a grid of small intervals, molecular dynamics simulations are performed for each value of $\lambda$
belonging to each interval, and the results are then integrated over all values of $\lambda$ to obtain the final free energy
difference.

Assuming a constant heat capacity, the temperature dependence of the solvation free energy can be written as 
\begin{eqnarray}
  \label{sec2:eq4}
  \Delta G\left(T\right) &=& a + b T + cT \ln T
\end{eqnarray}
so that
\begin{eqnarray}
  \label{sec2:eq5}
  \Delta S\left(T\right) &=& - \left(\frac{\partial \Delta G\left(T\right)}{\partial T} \right)_{P} = -b -c\left[1+ \ln T\right]
\end{eqnarray}
with very little dependence on the choice of the specific functional form \cite{Hajari15}. The enthalpy change can then be obtained from
\begin{eqnarray}
  \label{sec2:eq6}
  \Delta H\left(T\right) &=& \Delta G\left(T\right) + T \Delta S\left(T\right)
\end{eqnarray}
A numerical fit of the parameters $a$, $b$, and $c$ appearing in Eq.(\ref{sec2:eq4}) based on the results of simulations at different temperatures, will provide the required expressions for the entropy (Eq.(\ref{sec2:eq5})) and for the enthalpy (Eq.(\ref{sec2:eq6})). Standard deviation can then be evaluated using error block analysis \cite{Hajari15}.

We remark here that this is neither the unique nor the most efficient way to compute $T\Delta S$ and $\Delta H$. Indeed, \citeauthor{Fogolari2016} \cite{Fogolari2016, Fogolari2018} and \citeauthor{Lai2012} \cite{Lai2012} looked for different ways to compute entropies and enthalpies directly thus avoiding the use of the phenomenological expression given in Eq.(\ref{sec2:eq4}). However, this analysis is much more computational demanding and it could not be afforded for the systematic investigation that we are presenting here.
We further note that Eq.(\ref{sec2:eq4}) is known to hold true only in water \ce{H2O} within the temperature range $270-330\SI{}{\kelvin}$ consider in the present study \cite{Hajari15}, and it also appears to work for single amino acid side chain equivalents in cyclohexane \ce{cC6H12} \cite{Dongmo2020}. 
\subsection{Numerical protocols}
\label{subsec:numerical}
The amino acid building blocks for the polypeptides selected in this work span the full hydrophobic scale ranging from polar uncharged (ASN) to hydrophobic ( GLY, ALA, ILE) through charged moieties (LYS, ARG, GLU). Moreover, most of these latter were recently shown to preferentially populate the $\alpha$-helical conformational space \cite{Skrbic2021}, one of the major secondary structural motif found in biopolymers. The initial structures for the polypeptides were prepared using the Avogadro tool (ver 1.2.0) \citep{Hanwell2012} in their extended configurations with the dihedral angles of ($\phi,\psi$)=($180^{\circ}, 180^{\circ}$) with the N- and C- termini capped with  the neutral acetate (ACE) and methylamine (NME), respectively. All the polypeptides were simulated in full atomistic details by employing the GROMOS96 (54a7) force field \cite{Schmid11} that appears to be an optimal compromise between precision and computational cost when computing hydration enthalpies as tested against experimental data \cite{Dongmo2020, Oostenbrink2004,Villa02}. A table summarizing the amino acids used to build the homopeptides, along with their common names, and both their simplified three letters codification with the corresponding uni-letter nomenclature is shown in table \ref{tab:amino_acid} above. 

It is worth stressing that in this work we have explicitly included charged residues, unlike previous works  that avoided this case because of the tremendous effort needed to model them \cite{Konig2013} as the charged moieties  require complex parameterization for the treatment of finite-range electrostatics interactions \cite{Reif2012, Shirts2003}. This endeavour then represents a significant step forward even at the computational implementation level with respect to previous studies.
  
The simulations were performed in water \ce{H2O} and cyclohexane \ce{cC6H12}, as paradigmatic representative of polar and hydrophobic solvents, and five polymers with length from tri- (n=3) to undeca-peptides (n=11) were considered. In all cases they were initially aligned along the $z$-axis as exemplified in \sfref{supp-fig:fig1}  in a rectangular box and subsequently solvated with the solvent. The box dimensions and the number of solvent molecules used are reported in table \ref{tab:sim_details} below. The simulations were performed with Gromacs simulation package (series 2018, 2020 and 2021) \cite{Abraham15} and all the solutes were modelled roughly at their physiological pH. Therefore, GLU was preferentially modelled in its conjugate base i.e. the singly-negative anion glutamate whilst the carboxylic acid of ARG was deprotonated and the amino and guanidino groups protonated leading to a singly-positive acid. Likewise, the carboxylic acid of LYS was deprotonated and both its $\alpha$-amino and side chain lysyl groups protonated resulting to a monocation. Accordingly, \ce{Na+} and \ce{Cl-} counterions were added to preserve the system's electroneutrality and achieve the physiological-like concentration of 0.15 M.
As detailed in Section \ref{sec:theory}, free energy differences as given by Eq.\ref{sec2:eq3} have been computed from the fully coupled ($\lambda=0$) to the fully uncoupled ($\lambda=1$) system, by gradually switching off all non-steric interactions. A grid of $\Delta \lambda =0.05$ has been used in all cases, resulting into 21 binning points.  Altogether, the data discussed throughout this study are the result of approximately 10290 individual runs running up to nearly 103 $\mu s$, and thus it represents a large scale extensive computational endeavour. 

\begin{table}[h]
  \begin{center}
  \caption{Simulation details including the unit box dimensions in $\SI{}{nm^3}$ and the number of solvent molecules used in the case of \ce{H2O} and \ce{cC6H12} for different polymer length. The table is meant to provide a general overview of the number of solvent molecules as subtle differences may arrive due to the size of the solute upon changing from GLY to ARG towards LYS and ILE.}
  \label{tab:sim_details}
    \begin{tabular}{l| r r r r r}
    \hline
      $n$                   & 3      &     5 & 7      & 9     & 11    \\
      box ($\SI{}{nm^3}$) & 3$\times$3$\times$3.5   &  3$\times$3$\times$4 &  3$\times$3$\times$4.5 &  3$\times$3$\times$5 &  3$\times$3$\times$5.5 \\
      \ce{H2O}            & 1007   & 1157  & 1251   & 1393  & 1517  \\
      \ce{cC6H12}         & 181    & 210   & 218    & 241   & 262  \\
    \hline
    \end{tabular}
  \end{center}
\end{table}

The simulations described herein follow our previous protocol \cite{Dongmo2020}. However unlike that case of single amino acid side chain equivalents,  here the full atomistic polypeptide structures of different length has been considered, and the fully fledged thermodynamics integration has been carried out. Throughout the thermodynamics integration calculations, the polymers were kept restrained in a stretched conformation by applying a force at the two CA end-points of the polymer, as illustrated in \sfref{supp-fig:fig1}. This maximizes the number of solute-solvent contacts and hence the solvation, thus allowing  a direct comparison between them. 

Following preliminary equilibration steps in the canonical \textit{NVT} and isobaric-isothermal \textit{NPT} ensembles, most of the thermodynamic integrations were performed with time step of $\SI{}{2\times10^{-15}\s}$, although in some cases stability tests suggested the use of time steps as low as $\SI{}{1\times10^{-15}\s}$.

In order to assess the enthalpic and entropic single contributions, a set of 7 different temperatures ranging from \SI{270}{\kelvin} to \SI{330}{\kelvin} were performed.
In the case of the undeca-polypeptides, an additional set of simulations of various time-scales were performed with the same conditions as above but in this case the polymers were unrestrained, closely following previous protocol \cite{Dongmo2018}. Those more conventional simulations were performed at room temperature \SI{300}{\kelvin} and the conformational freedom of the homopeptides enables them to explore the available phase space and thus adopting the most favourable conformation with respect to the solvent considered. 

Standard probes such as the radius of gyration $R_g$ \cite{Flory69} and the solvent accessible surface are (SASA) \cite{Eisenhaber1995} were used to provide a quantitative assessment of the peptides behaviours in the considered solvents. 
It is important to remark that while calculation of SASA in folded state is unambigously defined, corresponding values in the unfolded conformation is not \cite{Gong2008}. 

\section{Results}
\label{sec:results}

\subsection{Good and poor solvents}
\label{subsec:good}

As a preliminary step, we have performed molecular dynamics simulations of polypeptides formed by 11 identical residues ranging from hydrophobic (GLY, ALA, ILE), to polar (ASN) and charged (LYS, ARG, GLU). In the following, we will denote as ASN11, ALA11, etc. polypeptides formed by 11 identical ASN, ALA, etc.  {\color{black} Note that we are denoting them as "polypetides" even if it would not be strictly correct for a number of residues ranging from 3 to 11 as those considered here.} We included also GLY as glycine has essentially no side chain (its side chain reduces to an hydrogen atom), and hence {\color{black} it represents a very convenient benchmark to compare with.} It has been argued that water \ce{H2O} at room temperature is a poor solvent for GLY15 \cite{Tran2008} and more generally for a protein backbone \cite{Bolen2008}. We shall confirm this results here with GLY11. By contrast, we shall see that cyclohexane \ce{cC6H12} is a good solvent for the same chain, indicating the presence of preferential interactions between the backbone of GLY11 and cyclohexane molecules. Support to this interpretation stems from present calculation as well as from the linear decrease of the solvation free energy as a function of the number of repeated units, as it will be discussed further below. 

We performed molecular dynamics of GLY11 and ALA11 in both water \ce{H2O} and cyclohexane \ce{cC6H12} at room temperature ($T=\SI{300}{\kelvin}$). In all cases the initial condition was taken to be a random swollen conformation. {\color{black} Self-assembly of GLY and ALA oligopeptides in water were previously studied by Pettit and collaborators  \cite{Kokubo2013,Karandur2016} who observed a fast aggregation coherent with our results.} 
Results for the other considered polypeptides can be found in Supplementary Information.

\begin{figure}[h!]
\centering
\captionsetup{justification=raggedright,width=\linewidth}
\includegraphics[width=\linewidth, trim=10.2cm 1cm 3.8cm 5.7cm, clip=true, angle=0]{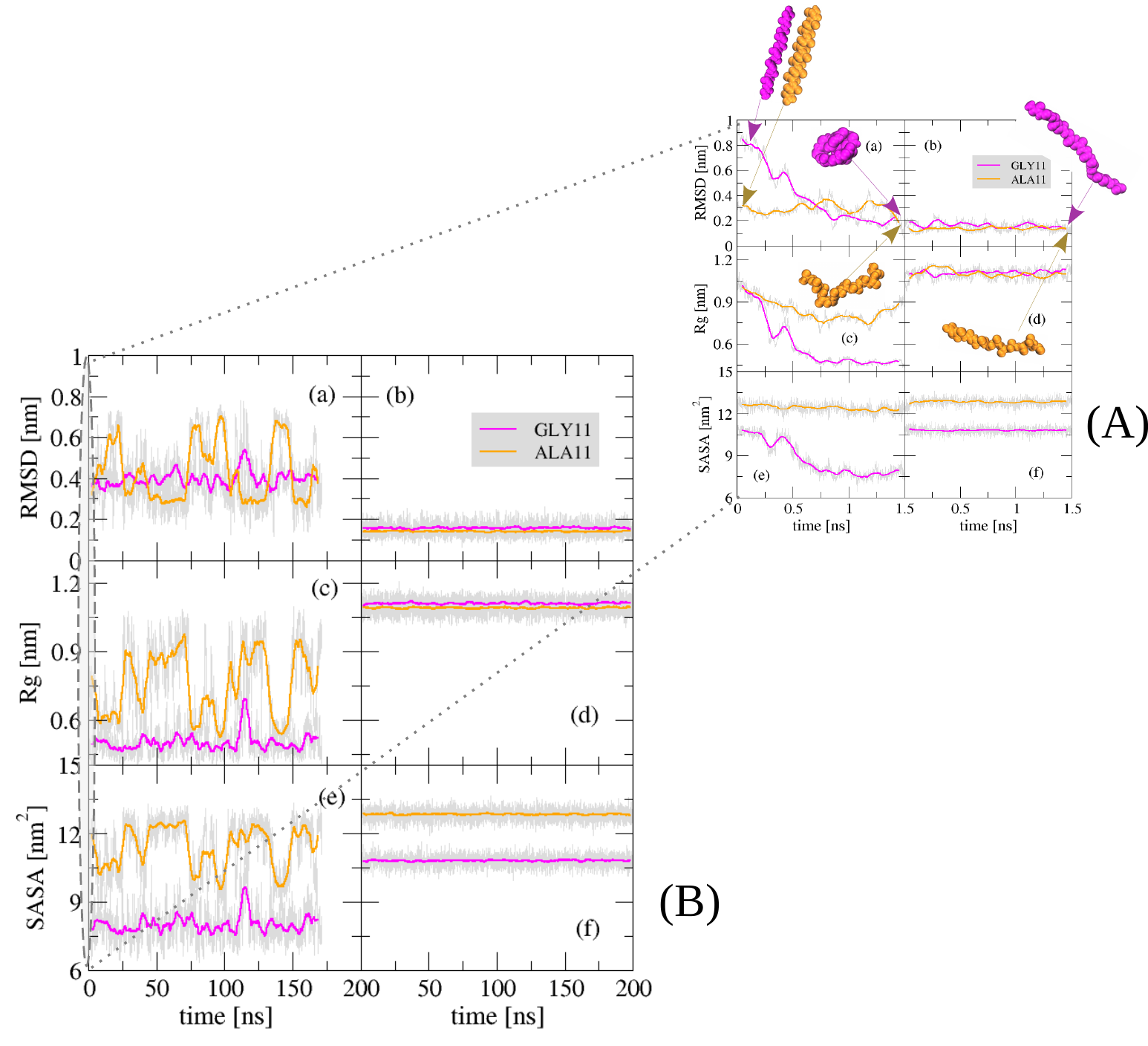}


 \caption{{\color{black} Initial (inset) and equilibrium probes of the conformational behaviour of GLY11 and ALA11 at the pre-production stages (\textit{NVT} and \textit{NPT} equilibration). Panels (a) and (b): root-mean-square-deviation (RMSD) from the initial state in water \ce{H2O} (a) and in cyclohexane \ce{cC6H12} (b). Panels (c) and (d): radius of gyration $R_g$  in water \ce{H2O} (c) and in cyclohexane \ce{cC6H12} (d). Panels (e) and (f): the solvent accessible surface area (SASA) in water \ce{H2O} (e) and in cyclohexane \ce{cC6H12} (f). In all cases the inset (A) report the few initial nanoseconds of the equilibration process. Results for GLY11 are displayed in magenta and for ALA11 in orange. 
 The insets also report representative snapshots of GLY11 (magenta) and ALA11 (orange) both at the initial and final stages. In all cases, the initial conformation is a random coil.}}
\label{fig:fig2}
\end{figure}

{\color{black} Figure \ref{fig:fig2} reports the behavior of the three selected probes to the conformational state: the root-mean-square-deviation from the initial state (RMSD) (top panels (a) for water \ce{H2O} and (b) for cyclohexane \ce{cC6H12}), the radius of gyration $R_g$ (middle panels (c) for water \ce{H2O} and (d) for cyclohexane \ce{cC6H12}), and the solvent accessible surface area (SASA) (bottom panels (e) for water \ce{H2O} and (f) for cyclohexane \ce{cC6H12}). The inset highlights the significant drop in all three probes in the case of GLY11 in water (magenta solid line in (A)-(a),(c),(e)) occurring  within the first \SI{1.5}{\nsec} from the initial extented conformation, followed by an equilibration around these values. A much more unstable trajectory is followed by ALA11 in water \ce{H2O} (orange line in panels (B)-(a),(c),(e)), with repeated folding and unfolding events occurring during the entire trajectory. By contrast, in cyclohexane (panels (A)\&(B)-(b),(d),(f)), both GLY11 and ALA11 display a fast settling into an extended conformation essentially equivalent to the initial conformation. Note that a more quantitative assessment on the difference between compact/globule and extended/swollen can be obtained by computing the $\nu$ exponent in $R_{g} \sim n^{\nu}$ with $\nu \approx 0.6$ in the extended (Flory) regime and $\nu \approx 0.33$ in the compact/globule regime \cite{Flory69,deGennes79,Khokhlov02,Rubinstein03,Bhattacharjee13}. However, it should be emphasized that the above scaling is strictly valid in the $n \gg 1$ limit (as it is the case in polymer physics), so its application to small polypeptides as those treated in this paper should be taken with great care. This is indeed shown in \sfref{supp-fig:rg_n} where we find $\nu$ unphysically small for all considered peptides irrespective of their polarity.

All in all, the results for GLY11 in water \ce{H2O} provide support to past evidence \cite{Karandur2014,Merlino2017} that water is a poor solvent for polyglycine, whereas the results for GLY11 in cyclohexane \ce{cC6H12} are consistent with the presence of a long-lived metastable state for globular proteins in cyclohexane \ce{cC6H12} \cite{Pace04}. The are also in line with the idea  \cite{Bolen2008} that water \ce{H2O} is a poor solvent for the protein backbone, and that this is one of the main driving force in the collapse of the chain to a globule-shaped structure, along with solvent entropy gain and the burial of the hydrophobic side chains  \cite{Merlino2017}. This is particularly effective in water \ce{H2O} because of its small size ($\approx 2.8$ \AA ~of diameter) and large number density (55.3 M under standard conditions).
Cyclohexane \ce{cC6H12} has size more than two times larger than water \ce{H2O} and significantly smaller number density, and the solvent entropic gain is reduced accordingly.

The behavior of ALA11 in water, which displays an erratic sequence of folding and unfolding events  for which no stable collapse is observed (see Fig. \ref{fig:fig2}), is more surprising. ALA is usually classified as a hydrophobic amino acid (see Table \ref{tab:amino_acid}), and hence a conformational folding akin to GLY11 may have been expected.  However, ALA has a larger side chain that provides a larger steric hindrance that may hamper the collapse of the small peptides such as those considered here. In addition the energetic interactions of the two polypeptides with water is different.  By contrast, the behaviour of the GLY11 and ALA11 is nearly identical in cyclohexane \ce{cC6H12} with both remaining extended throughout the full trajectory. This can be interpreted as cyclohexane \ce{cC6H12} being a good solvent for both, and it might provide one possible reason of the experimentally noted absence of a collapse of proteins in cyclohexane \ce{cC6H12}, and more generally in any non-polar solvent \cite{Pace04}. 
Table \ref{tab:tab1} summarizes all these results in a synoptic form where water \ce{H2O} is referred to as poor$+$ (i.e. with stable fold) solvent for GLY11 and as poor (no stable collapse) for ALA11. Likewise cyclohexane \ce{cC6H12} will be referred to as a good$+$ solvent for both. }

For the remaining 5 considered polypeptides, the results for RMSD (top panel), the radius of gyration $R_{g}$ (middle panel) and the SASA (bottom panel) for the full trajectory in water \ce{H2O} (left) and in cyclohexane \ce{cC6H12} (right) are reported in \sfref{supp-fig:structural_data}, and confirm a rather complex and diverse behaviour. In water \ce{H2O}, ILE11 (hydrophobic) displays an initial collapse followed by a fluctuating behaviour about a less compact conformation (black line left panel), whereas for ASN11 (polar, red line left panel) $R_g$ remains mostly stable throughout the full trajectory following an initial drop, but with a final large fluctuation. Interestingly, in cyclohexane \ce{cC6H12} ILE11 remains extended (black line right panel) whereas ASN11 collapses (red line right panel).  The other three polypeptides (LYS11, ARG11, and GLU11, polar because charged), display large fluctuations in water \ce{H2O} (left panel), and remain rather extended in cyclohexane \ce{cC6H12} (right panel). All these findings are summarized in Table \ref{tab:tab1}. 

The results of these last three polypeptides (LYS11, ARG11, and GLU11) show a complex behaviour that defies any simple description in terms of poor and good solvent. Of course, this was to be expected: each residue has its own characteristics that go beyond the operative description in term of good and bad solvents, and sometimes this matters for this kind of calculation. For instance, isoleucine ILE is known to be a strong hydrophobic amino acid and the corresponding marked collapse of ILE11 in water \ce{H2O} with a noticeable structural rearrangement in the course of the simulation as depicted in the RMSD (top), $R_g$ (middle) and SASA (bottom) plots in  \sfref{supp-fig:structural_data}, agrees with this picture. Also the corresponding absence of any collapse or structural rearrangement of ILE11 in cyclohexane \ce{cC6H12}, could be ascribed to the stabilizing effect of cyclohexane \ce{cC6H12} in line with its hydrophobic character. However, the negatively charged GLU11 polypeptide in water \ce{H2O}, adopts a U-like shape after a long equilibration, and subsequently collapses to a globule although with less compact shape. We surmise that the length and shape of the side chain arms are major factors prohibiting the proper collapse of GLU11 in water \ce{H2O}. In cyclohexane \ce{cC6H12}, after a short equilibration time a relatively steady and stable conformation is achieved, compatible with a favourable solute-solvent interactions over the solvent entropy promoting the collapse. 

Comparatively, ASN11 and ARG11 behave symmetrically, with water \ce{H2O} acting as a good solvent whereas cyclohexane \ce{cC6H12} as a poor one. Indeed, ASN11 in \ce{H2O} seems to remain marginally extended and undergo a number of noticeable conformational fluctuations as reported by the minor changes seen in its solvent accessible surface area plot and the root mean square deviation analysis, respectively. Transiently formed globular-like conformations are identified in the trajectory signalled by the significant decrease in the radius of gyration $R_g$ reported. 

In cyclohexane \ce{cC6H12}, after a short equilibration period corresponding to the coil-to-globule adaptation, all RMSD, $R_g$ and SASA level off and remain steady flat throughout the simulation timescale, implying undoubtedly a favourable and stable ASN11 - \ce{cC6H12} interactions. Furthermore, we monitored an increase in the number of intramolecular hydrogen bonds (see also further below) in ASN11 as shown in figure \sfref{supp-fig:hbond_rest}, a sign of an increased compactness of the globular shape obtained. ARG11, albeit simulated on a shorter time span, displays a behaviour in both water \ce{H2O} and cyclohexane \ce{cC6H12} mirroring that reported for ASN11.  Again, as already mentioned for GLU11, the long arms of ARG11 side chains are forming a cage-like network around the backbone, thus restraining the degrees of freedom of the latter thereby shielding its proper collapse to a globular state. Meanwhile, in cyclohexane \ce{cC6H12} a fast structural reorganization of ARG11 is seen wherein the polymer's side chains are preferentially folded back inside towards the core and the backbone rather exposed to the bulk. 

\begin{figure*}[hptb]
  \centering
  \begin{subfigure}{0.22\textwidth}
    \centering
    \includegraphics[width=0.50\textwidth,trim=6cm 6cm 6cm 6cm,clip]{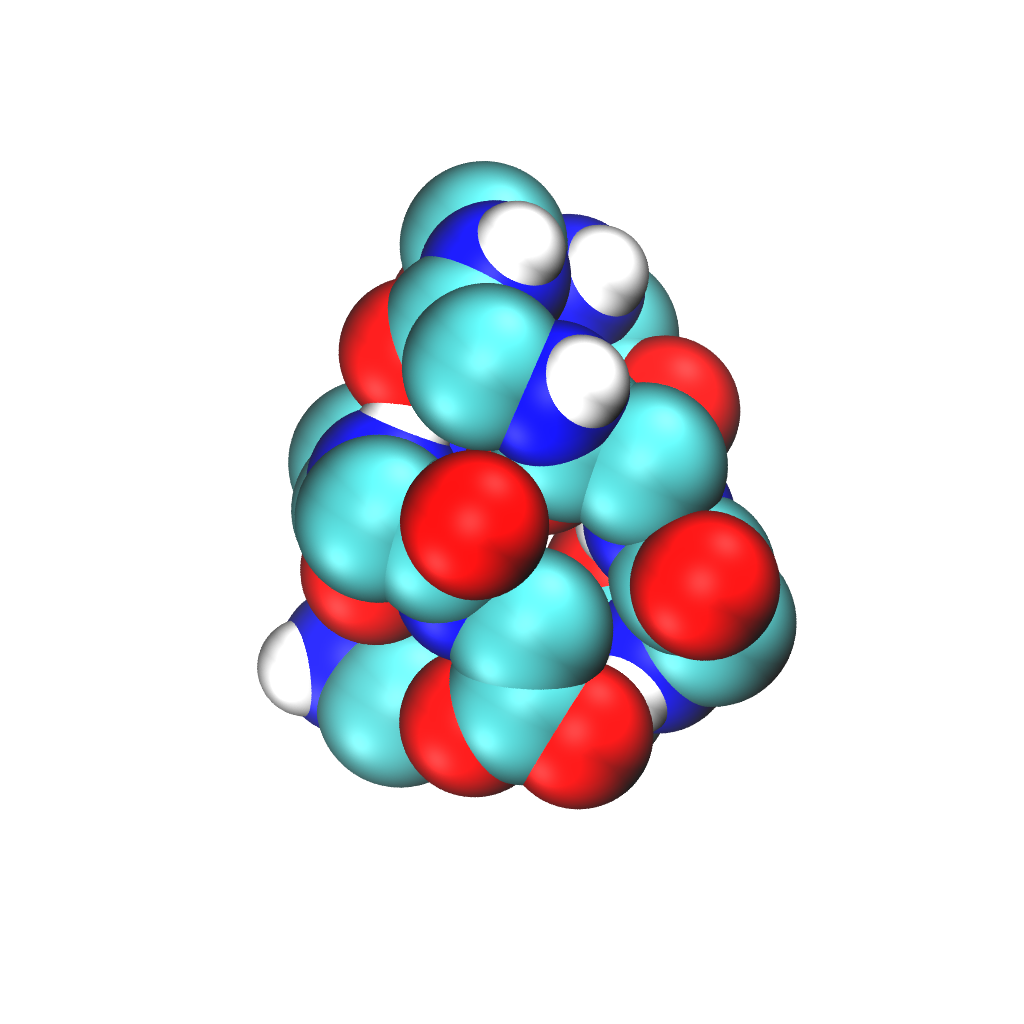}
     \caption{GLY11 in \ce{H2O}}
    \label{fig:fig3a} 
  \end{subfigure}
  \hfill
  \begin{subfigure}{0.22\textwidth}
    \centering
    \includegraphics[width=0.50\textwidth,trim=6cm 6cm 6cm 6cm,clip]{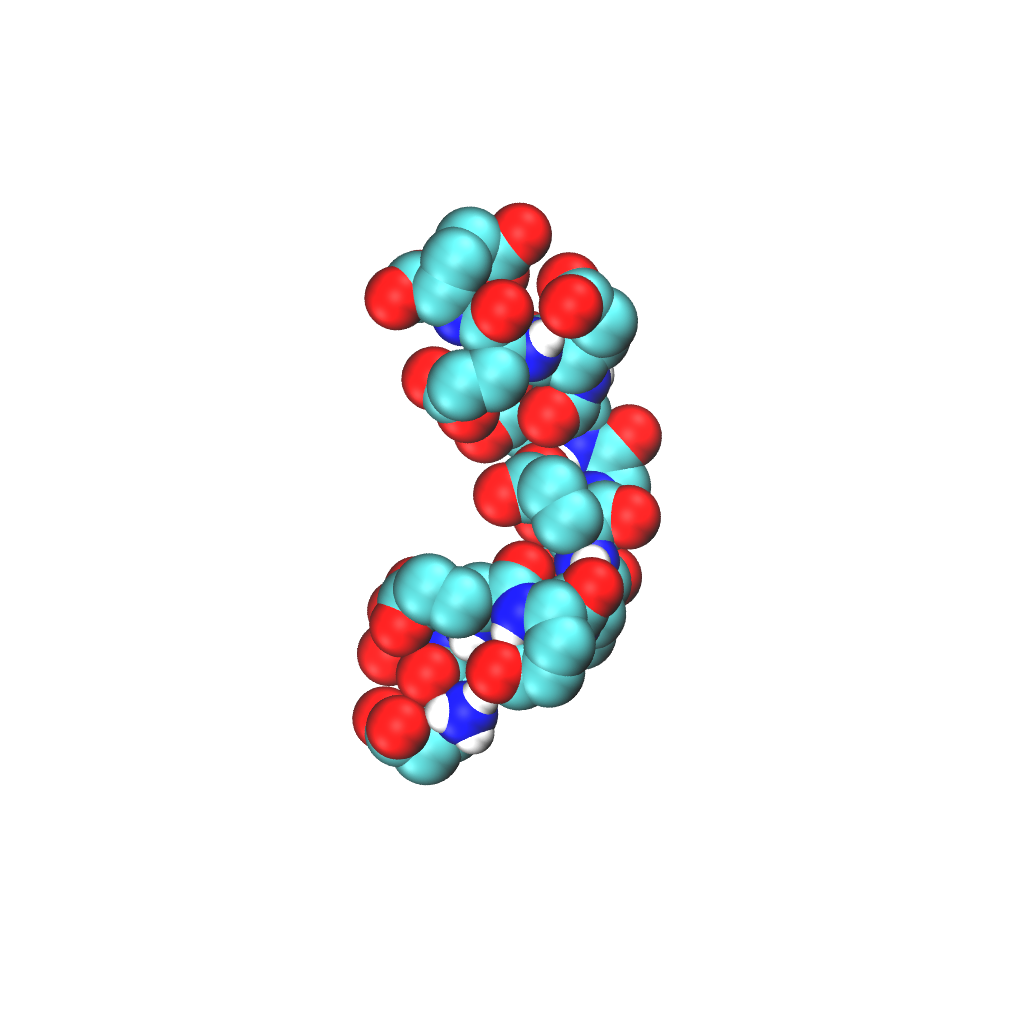}
    \caption{GLY11 in \ce{cC6H12}}
    \label{fig:fig3b} 
    \end{subfigure} 
    \hfill 
  \begin{subfigure}{0.22\textwidth}
    \centering
    \includegraphics[width=0.50\textwidth,trim=6cm 6cm 6cm 6cm,clip]{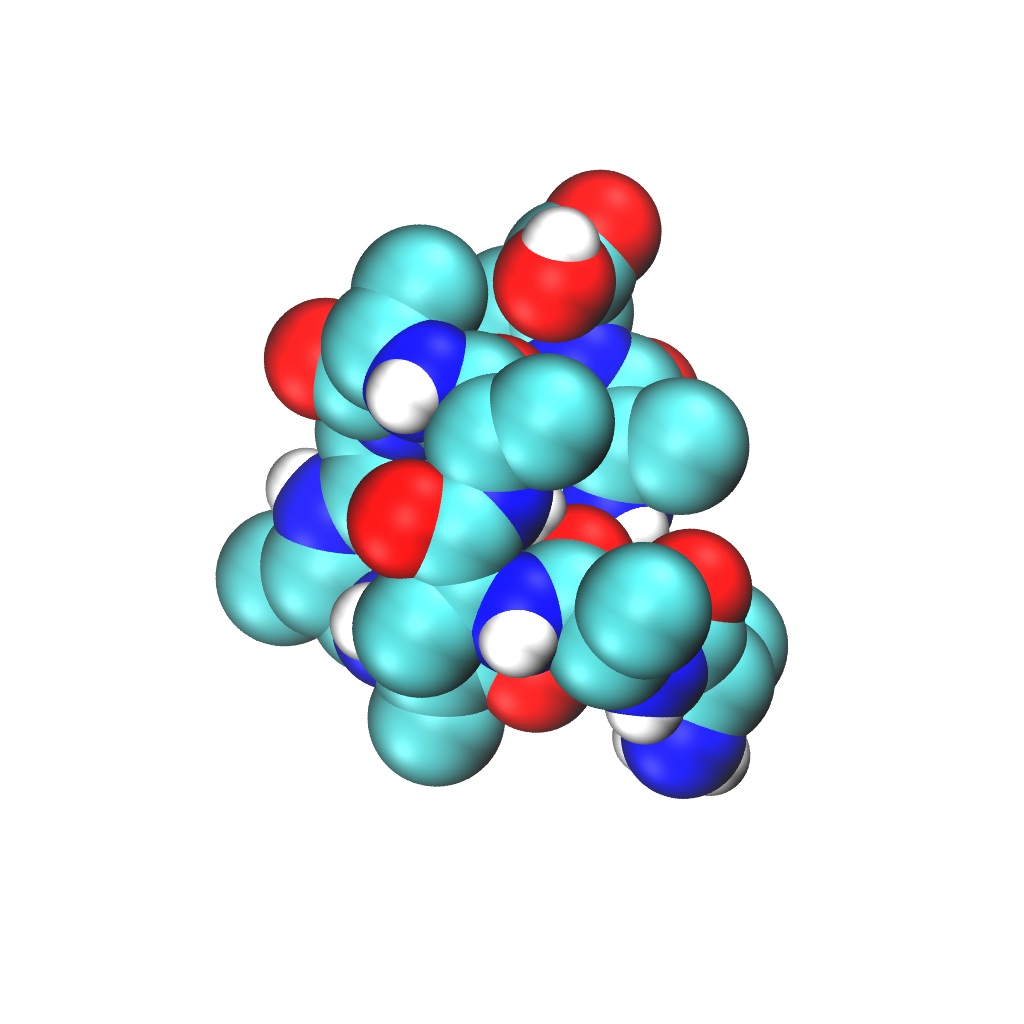}
    \caption{ALA11 in \ce{H2O}}
    \label{fig:fig3c}
  \end{subfigure}
  \hfill
  \begin{subfigure}{0.22\textwidth}
    \centering
    \includegraphics[width=0.50\textwidth,trim=15cm 15cm 15cm 15cm,clip]{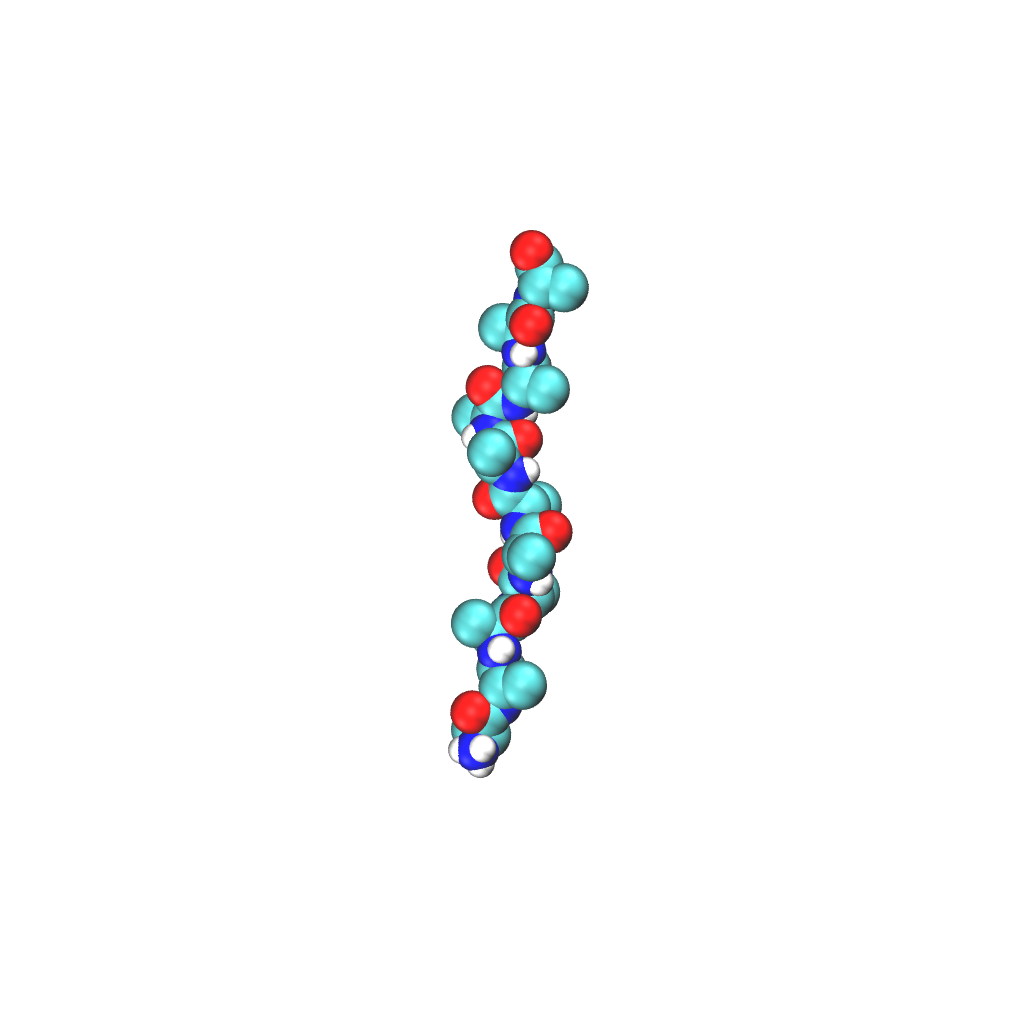}
    \caption{ALA11 in \ce{cC6H12}}
    \label{fig:fig3d}
 \end{subfigure} 
\hfill
\begin{subfigure}{0.22\textwidth}
    \centering
    \includegraphics[width=0.50\textwidth,trim=6cm 6cm 6cm 6cm,clip]{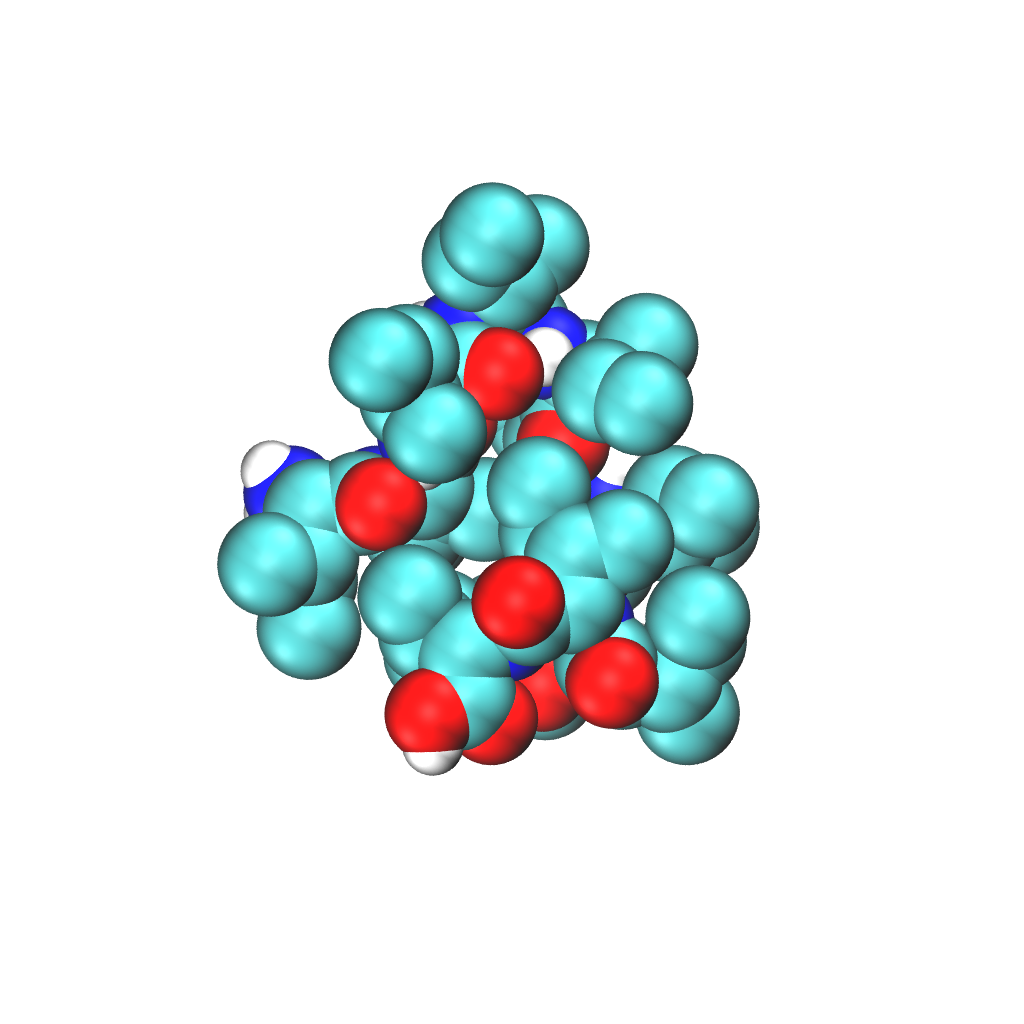}
    \caption{ILE11 in \ce{H2O}.}
    \label{fig:fig3e}
\end{subfigure}
\hfill
\begin{subfigure}{0.22\textwidth}
    \centering
    \includegraphics[width=0.50\textwidth,trim=6cm 6cm 6cm 6cm,clip]{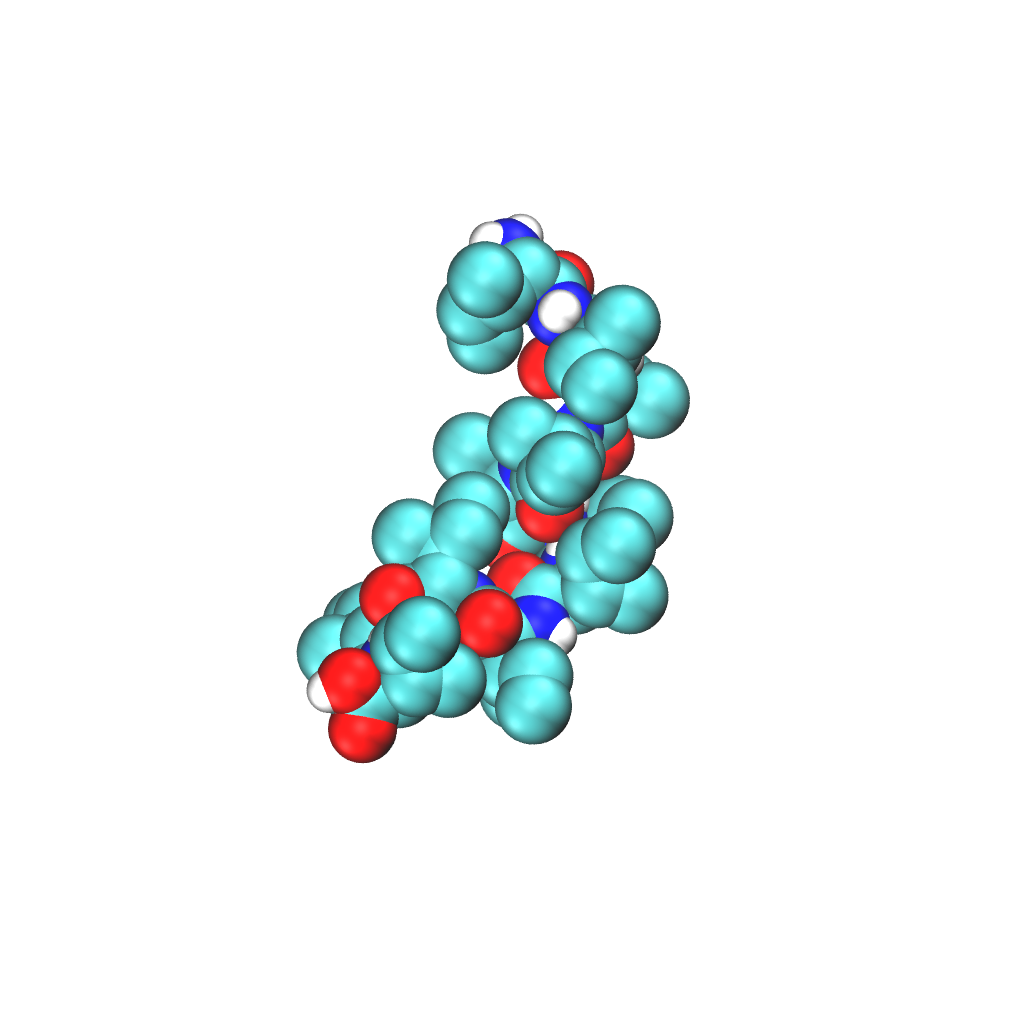}
    \caption{ILE11 in \ce{cC6H12}}
    \label{fig:fig3f}
\end{subfigure} 
\hfill
\begin{subfigure}{0.22\textwidth}
    \centering
    \includegraphics[width=0.50\textwidth,trim=6cm 6cm 6cm 6cm,clip]{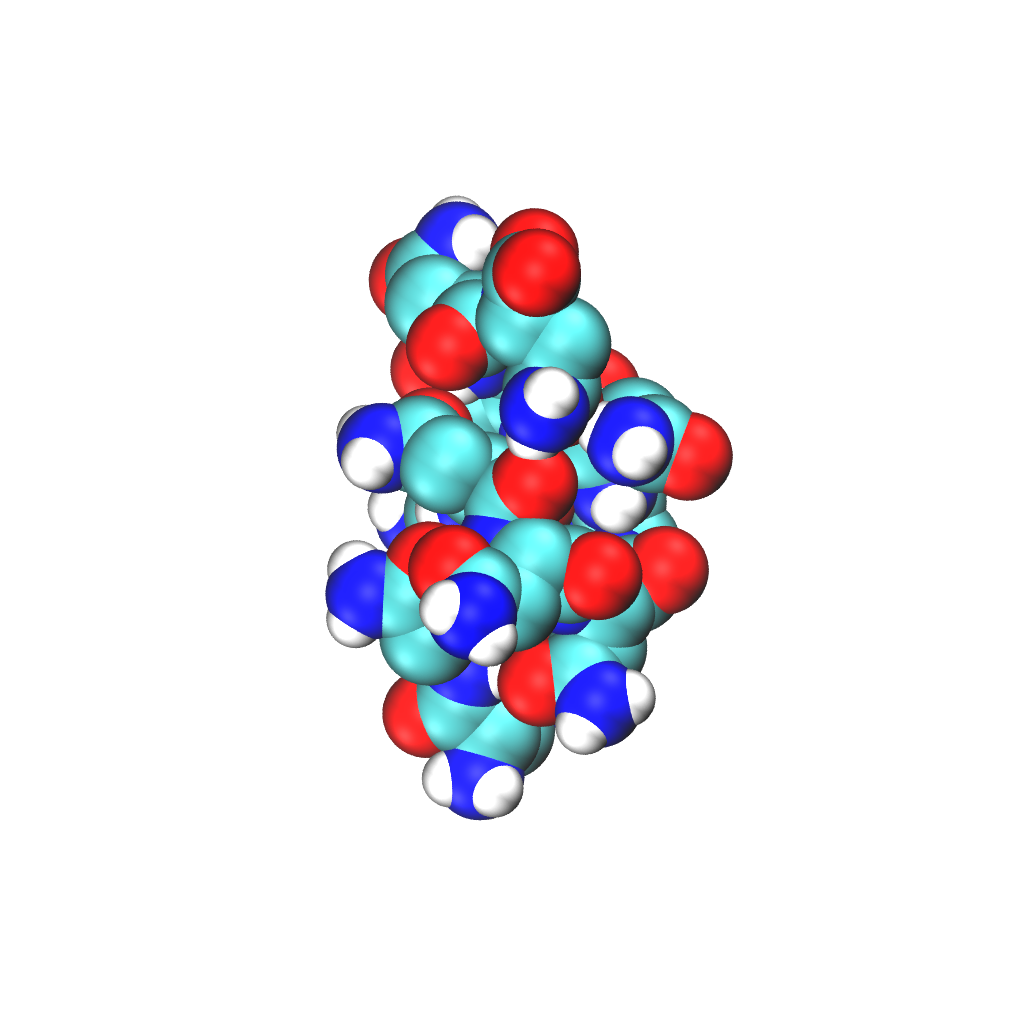}
    \caption{ASN11 in \ce{H2O}.}
    \label{fig:fig3g}
\end{subfigure}
\hfill
\begin{subfigure}{0.22\textwidth}
    \centering
    \includegraphics[width=0.50\textwidth,trim=6cm 6cm 6cm 6cm,clip]{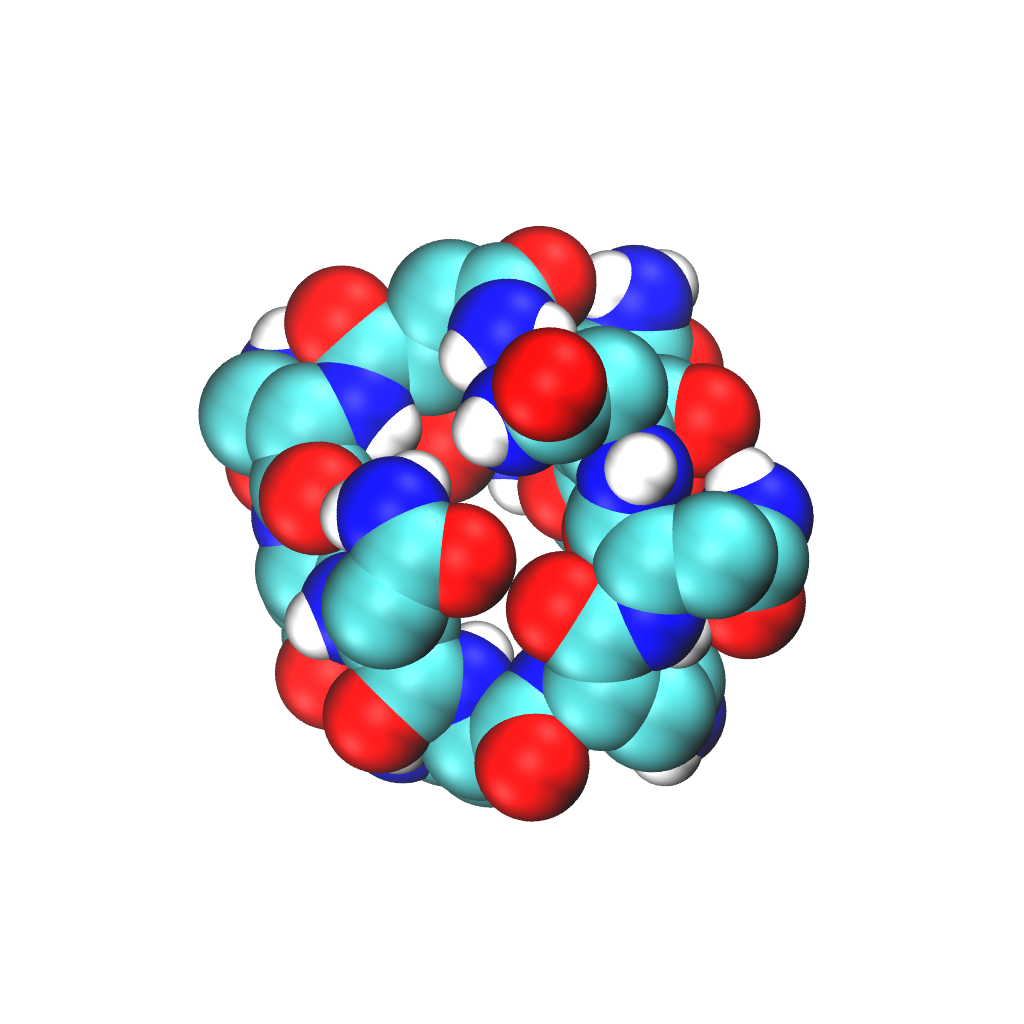}
    \caption{ASN11 in \ce{cC6H12}}
    \label{fig:fig3h}
\end{subfigure} 
\hfill
\begin{subfigure}{0.22\textwidth}
    \centering
    \includegraphics[width=0.50\textwidth,trim=6cm 6cm 6cm 6cm,clip]{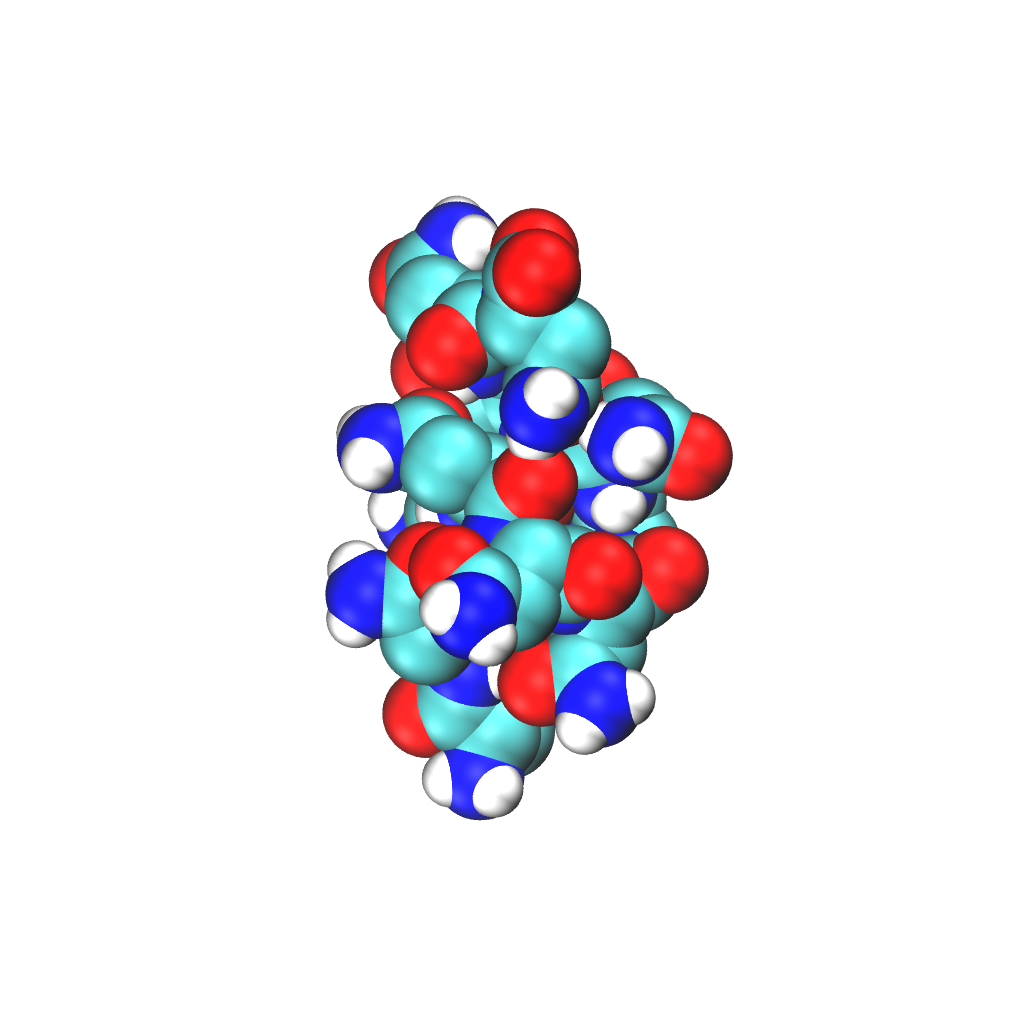}
    \caption{LYS11 in \ce{H2O}.}
    \label{fig:fig3i}
\end{subfigure}
\hfill
\begin{subfigure}{0.22\textwidth}
    \centering
    \includegraphics[width=0.50\textwidth,trim=6cm 6cm 6cm 6cm,clip]{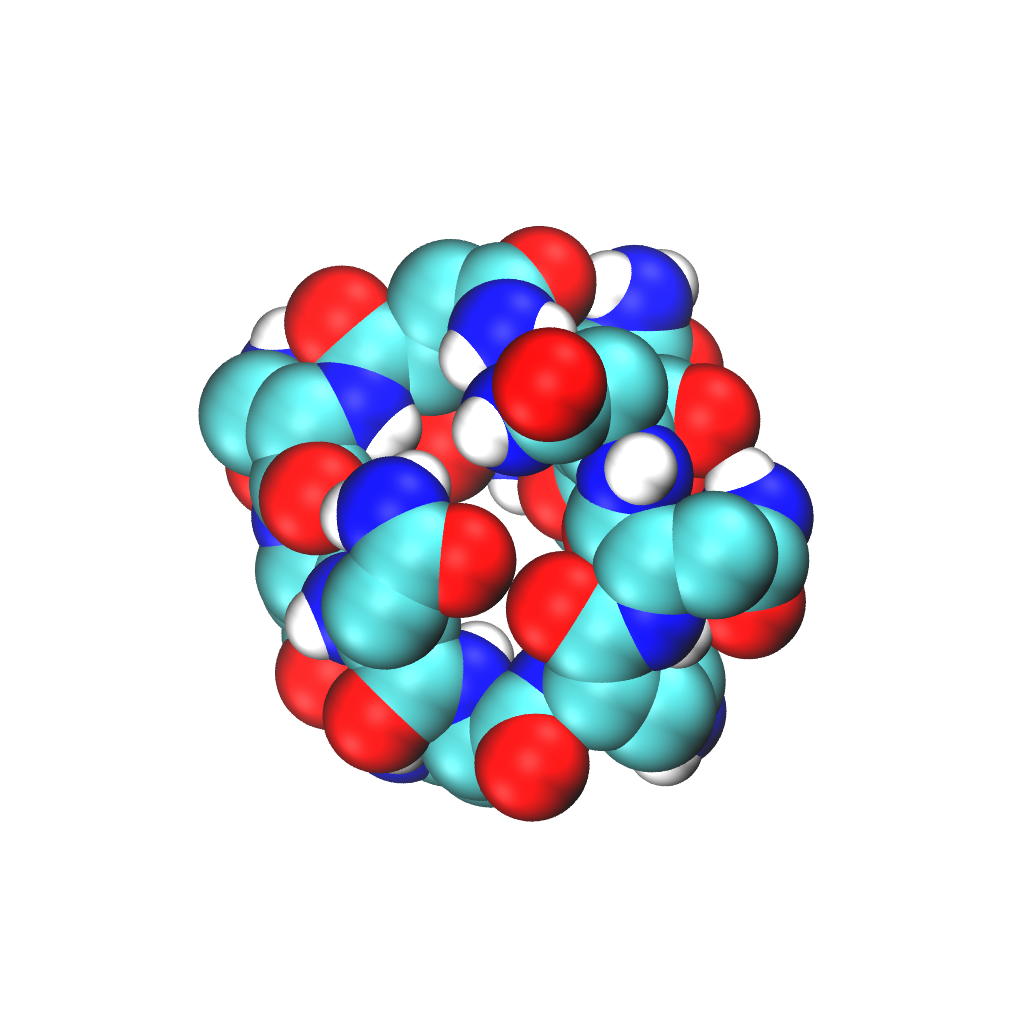}
    \caption{LYS11 in \ce{cC6H12}}
    \label{fig:fig3l}
\end{subfigure} 
\hfill
\begin{subfigure}{0.22\textwidth}
    \centering
    \includegraphics[width=0.50\textwidth,trim=6cm 6cm 6cm 6cm,clip]{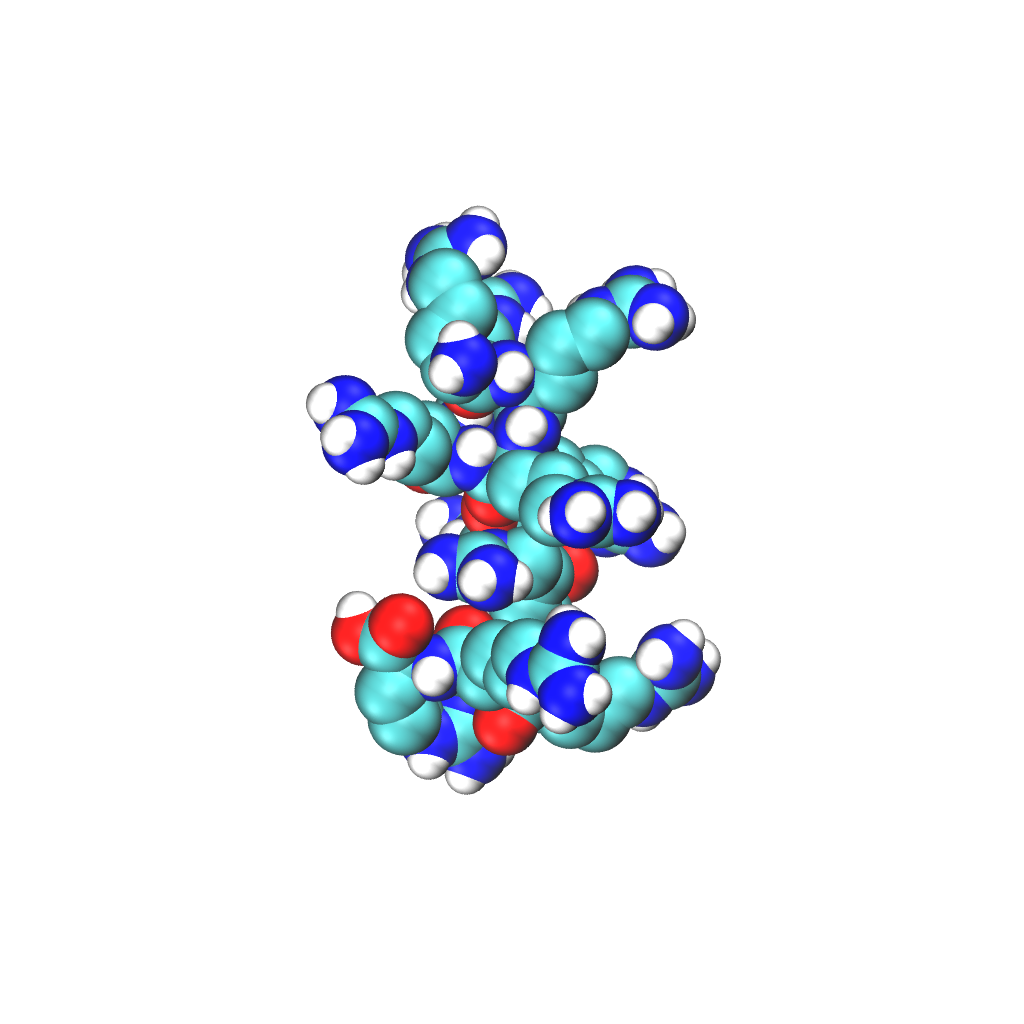}
    \caption{ARG11 in \ce{H2O}.}
    \label{fig:fig3m}
\end{subfigure}
\hfill
\begin{subfigure}{0.22\textwidth}
    \centering
    \includegraphics[width=0.50\textwidth,trim=6cm 6cm 6cm 6cm,clip]{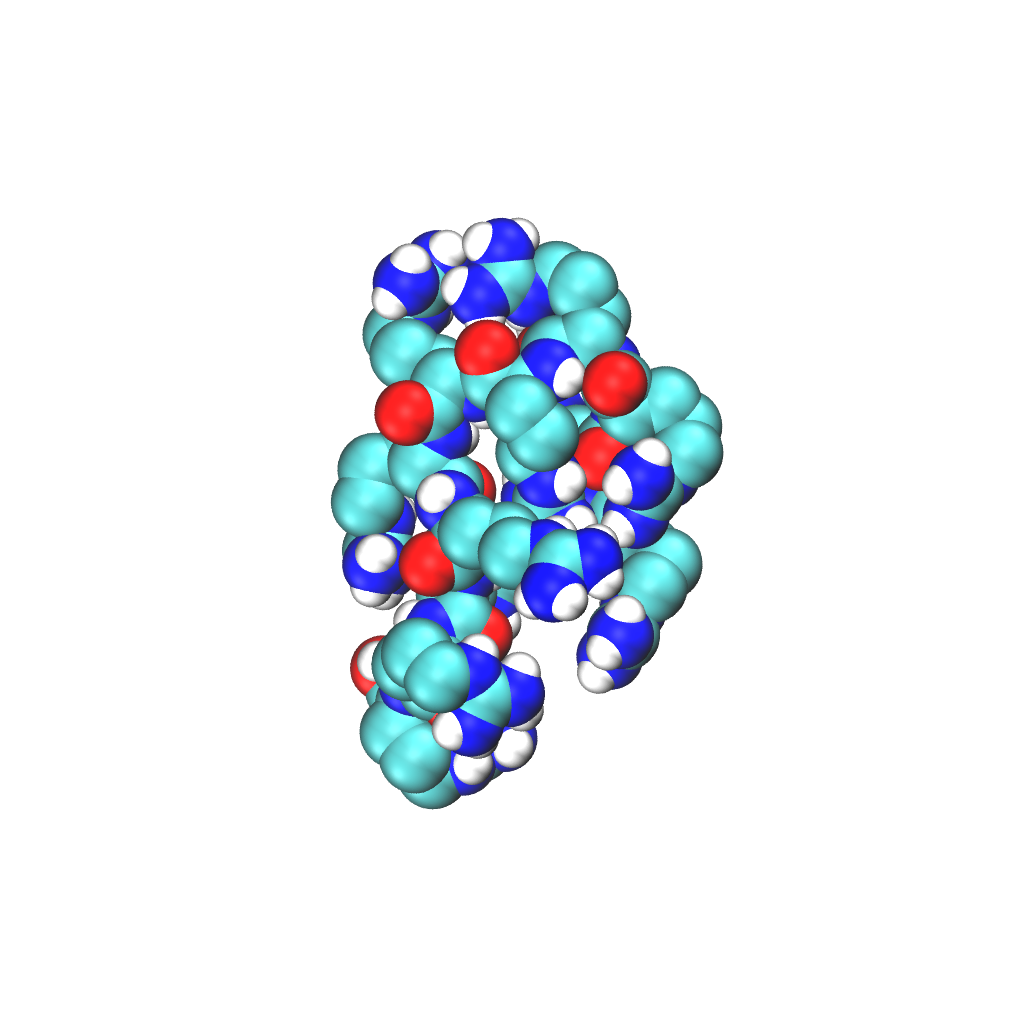}
    \caption{ARG11 in \ce{cC6H12}}
    \label{fig:fig3n}
\end{subfigure} 
\hfill
\begin{subfigure}{0.22\textwidth}
     \centering
    \includegraphics[width=0.50\textwidth,trim=6cm 6cm 6cm 6cm,clip]{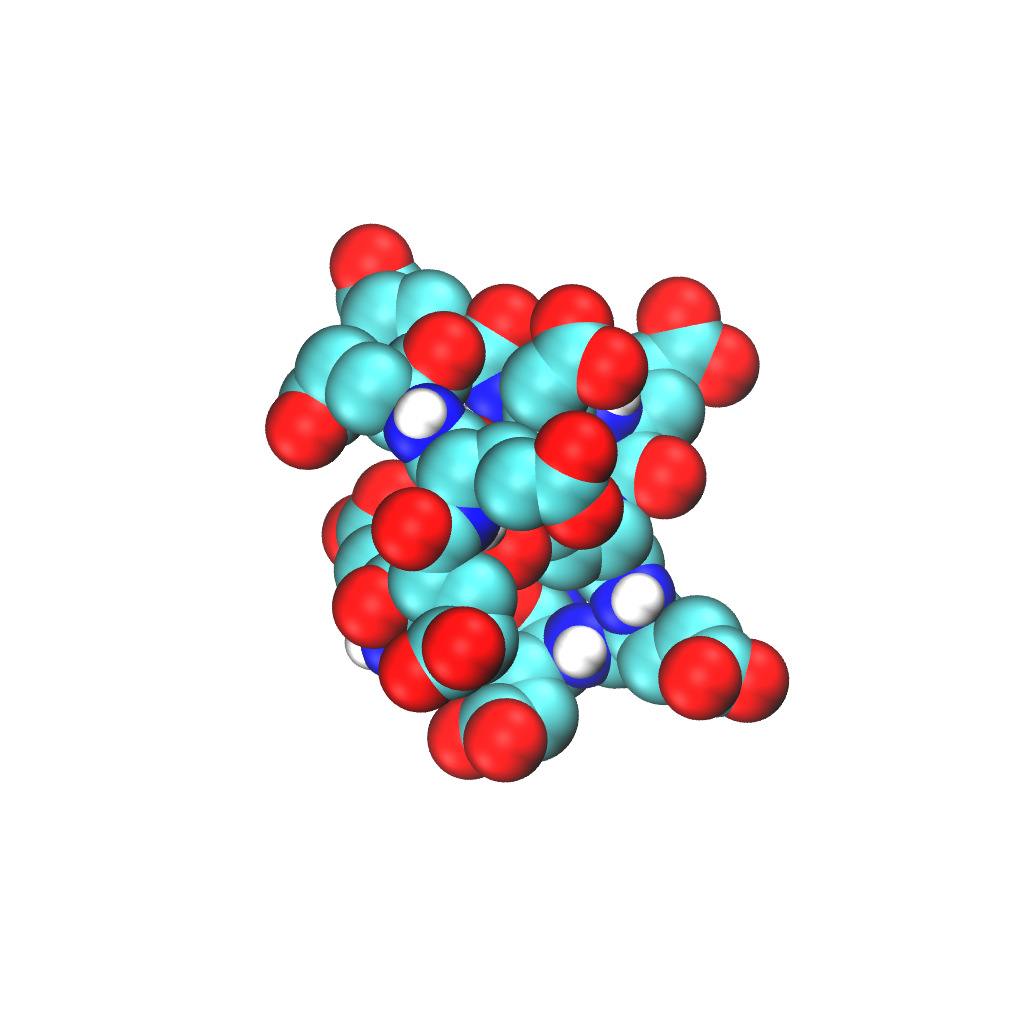}
    \caption{GLU11 in \ce{H2O}.}
    \label{fig:fig3o}
\end{subfigure}
\hfill
\begin{subfigure}{0.22\textwidth}
     \centering
    \includegraphics[width=0.50\textwidth,trim=6cm 6cm 6cm 6cm,clip]{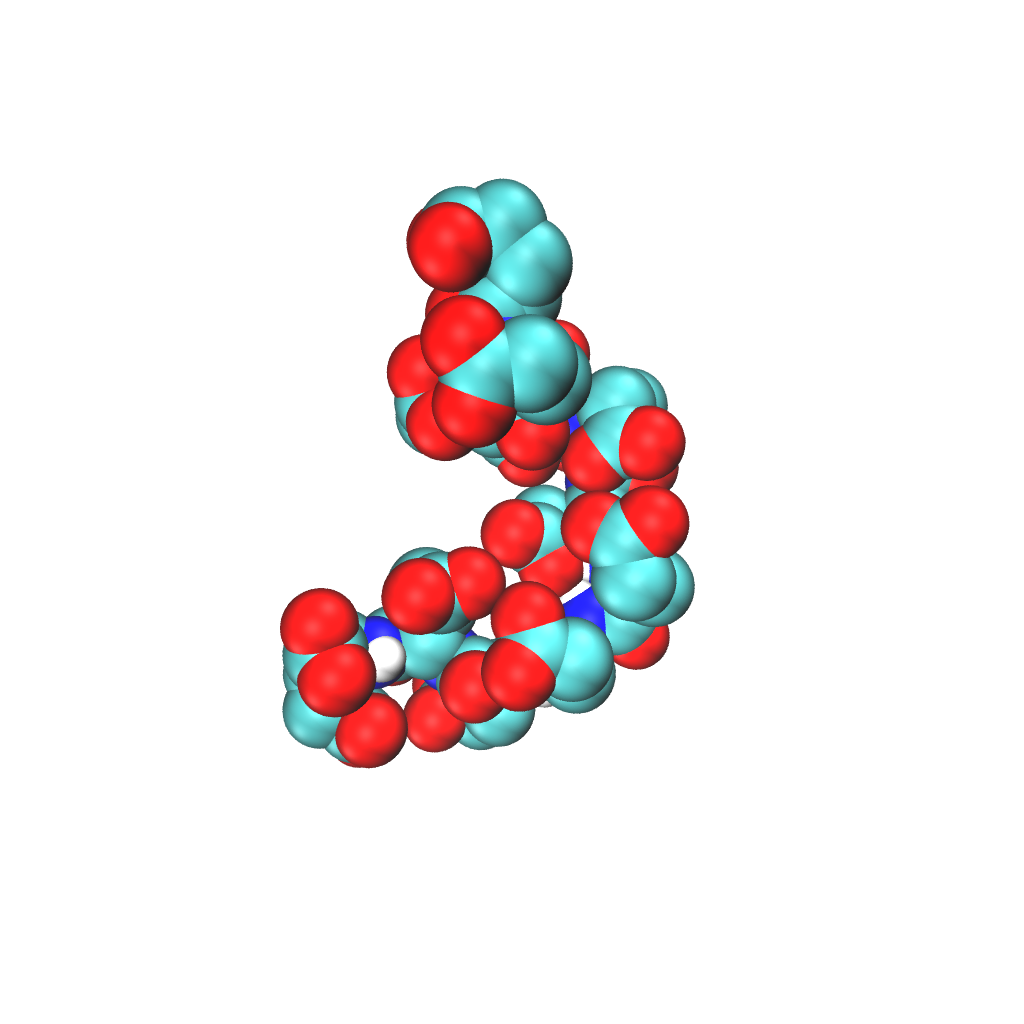}
    \caption{GLU11 in \ce{cC6H12}}
    \label{fig:fig3p}
\end{subfigure}
\hfill
\caption{Representative snapshots of the smallest $R_g$ conformers i.e. the most collapsed conformations. On the left the structures obtained in water \ce{H2O} and on the right those obtained in cyclohexane \ce{cC6H12}. From top to bottom the corresponding structures for GLY11, ALA11, ILE11, ASN11, LYS11, ARG11 and GLU11, respectively.}
\label{fig:fig3}
\end{figure*}

In summary, we observe a general tendency for those undeca-polypeptides folding in water not folding in cyclohexane, and conversely those folding in cyclohexane not folding in water. However,  LYS11 fails to follow this general rule as it remains essentially extended in both water \ce{H2O} and cyclohexane \ce{cC6H12}, albeit with side chains more parallel to the backbone in the latter case, see Fig. \ref{fig:fig3}. This behaviour might be ascribed to a steric hindrance of the long arms side chain densely parked around the relatively short undeca-homopeptide backbone, thus significantly reducing its conformational space, not allowing the proper collapse of the polymer within the simulated time considered here.     

In principle, the relative stability of each polypeptide with respect to a specific solvent can be also quantified by a direct calculation of the solvation free energy in both water \ce{H2O} and cyclohexane \ce{cC6H12}. This will be carried out in the next Section. However, in interpreting a comparison with the data reported here, the differences in the flexibility conditions (fully flexible here, fully constrained in the solvation free energy calculation reported below), plays an important role as noted earlier \cite{Kokubo2013}.

Figure \ref{fig:fig3} reports snapshots of the most representative conformers in all considered cases, and Table \ref{tab:tab1}  summarizes these results in a synoptic form.

\begin{table}[h]
  \begin{center}
  \caption{Summary of the solvent property in relation to the polymers (undeca-mer) considered here in water \ce{H2O} and cyclohexane \ce{cC6H12}. Good and poor are used to point out whether the solvent tends to promote the extension or the collapse of the solute, respectively. Furthermore, the sign $+$ is an indication of either a fully extended or fully compact conformation, without any significant structural fluctuations that characterize those cases without the $+$ sign.}
  \label{tab:tab1}
    \begin{tabular}{l| r r r r r r r }
    \hline
    \text{polymer}  & \text{GLY11}   & \text{ALA11} & \text{ILE11} & \text{ASN11}  & \text{LYS11}  & \text{ARG11}  & \text{GLU11} \\
    \ce{H2O}        & \text{poor+}   & \text{poor}  & \text{poor}  & \text{good}   & \text{good}   & \text{good}  & \text{poor} \\
     \ce{cC6H12}    & \text{good+}   & \text{good+}  & \text{good+}  & \text{poor+}   & \text{good}   & \text{poor}  & \text{good} \\   
    \hline
    \end{tabular}
  \end{center}
\end{table}

\begin{figure}[h!]
\centering
\captionsetup{justification=raggedright,width=\linewidth}
\begin{subfigure}{9cm}
\includegraphics[width=0.85\linewidth]{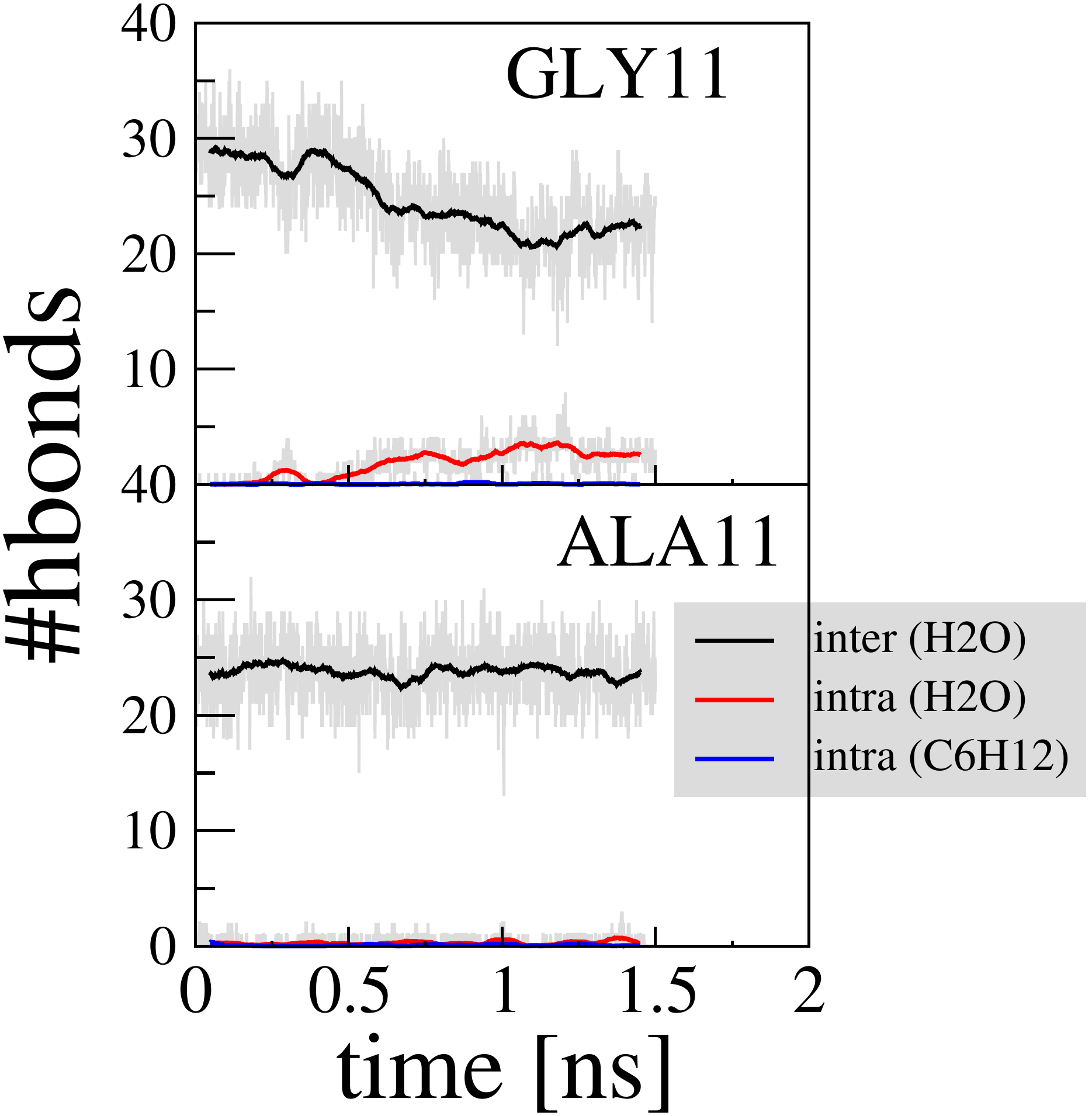}
\caption{}\label{fig:fig4a}
\end{subfigure}
\begin{subfigure}{9cm}
\includegraphics[width=0.85\linewidth]{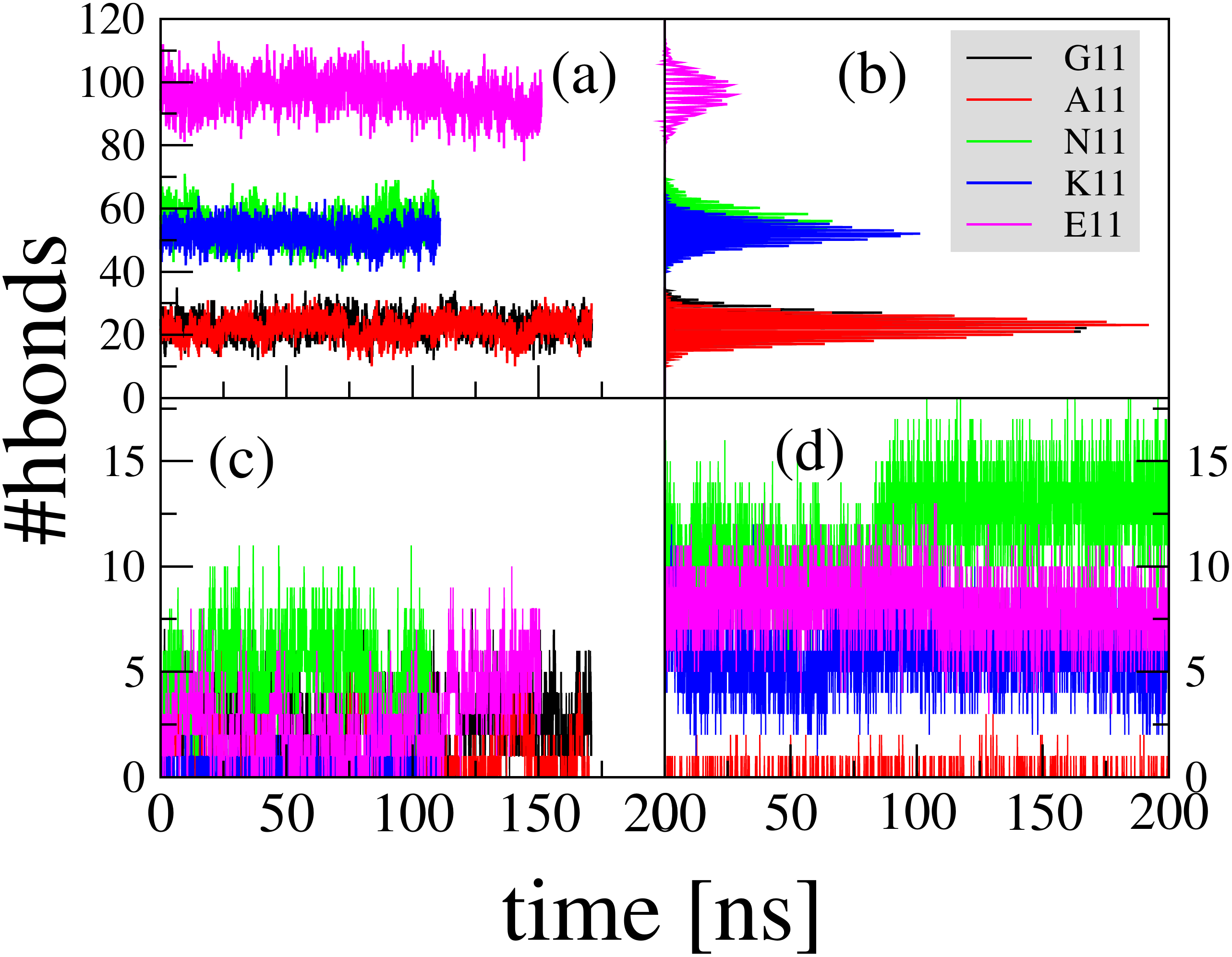}
\caption{}\label{fig:fig4b}
\end{subfigure}
\caption{{\color{black} Top panel: Initial stage evolution of the number of hydrogen bonds for GLY11 (top) and ALA11 (bottom). Both inter- (\ce{H2O}-solute black line) and intra- (solute-solute in \ce{H2O} red line and \ce{cC6H12} blue line) molecular hydrogen bonds are plotted. Bottom panel: Long time evolution of the number of hydrogen bonds of GLY11 (black line) and ALA11 (red line). (a) Solute-\ce{H2O} hydrogen bonds (GLY11 black line, ALA11 red line); (b) Histogram distribution of (a) (GLY11 black line, ALA11 red line); (c) Solute-Solute hydrogen bonds in \ce{H2O} (GLY11 black line, ALA11 red line); (d) Solute-Solute hydrogen bonds in \ce{cC6H12} (GLY11 black line, ALA11 red line) }}
\label{fig:fig4}
\end{figure}

Additional insights can be obtained by monitoring the evolution in the fractions of peptide-solvent and intra-peptide hydrogen bonds. Confining our attention to the initial equilibration stage of few nanoseconds first, we report the total number of hydrogen bonds with water \ce{H2O} for both GLY11  in Fig.\ref{fig:fig4} (black line of the top panel) and ALA11 (Fig.\ref{fig:fig4a} black line of the top panel). Correspondingly, the total number of intra-chain hydrogen bonds are also reported in for GLY11 (Fig.\ref{fig:fig4a} red line in top panel) and for ALA11 (Fig.\ref{fig:fig4a} red line in bottom panel).  For GLY11, the number of hydrogen bonds with water shows a fast drop (Fig.\ref{fig:fig4a} black line top panel) consistent with a folding of GLY11 being further stabilized by an increase of the number of intra-chain hydrogen bonds (Fig.(\ref{fig:fig4a} red line top panel). This does not seem the case for ALA11 where the number of hydrogen bonds with water does not show any drop with time ( Fig.(\ref{fig:fig4a}) black line bottom panel) and the number of intra-chain hydrogen bonds remains essentially unchanged (Fig.(\ref{fig:fig4a}  red line bottom panel).
It is worth noticing that on assuming an approximate average value of (\SI{20}{\kilo\joule\per\mole})  for each hydrogen bond, a typical total energy involved for approximately 30 bonds (see Figure \ref{fig:fig4a}) is of the order of \SI{600}{\kilo\joule\per\mole}  which is comparable with the solvation free energy discussed in the next Section. This confirms the fundamental role played by the hydrogen bonds in stabilizing the protein fold as discussed in detail in Ref. \cite{Rose06}. 
At equilibrium, the above findings are confirmed. Fig.\ref{fig:fig4b}-(a) and (c) report the fluctuations of the number solute-water \ce{H2O} and solute-solute hydrogen bonds, respectively. Black lines refer to GLY11 and red lines to ALA11. Note that the total number of hydrogen bonds with water \ce{H2O} is of the order of 25 for both GLY11 and ALA11 whereas the total number of internal hydrogen bonds is stably of the order of 2.5 and for GLY11 and it is highly fluctuating between 0 and 2.5 in the case of ALA11, coherently with the lack of a stable fold for ALA11 in water. Fig.\ref{fig:fig4b}(b) display the distributions of the total number of hydrogen bonds of both GLY11 (black) and ALA11 (red) with water that turns out to be nearly identical, as visible.

In cyclohexane \ce{cC6H12} the behaviour is clearly different. Fig.\ref{fig:fig4b}(d) shows the fluctuations of the number of solute-solute hydrogen bonds in cyclohexane \ce{cC6H12} for GLY11 (black line) and ALA11 (red line). Here the total number of intra-chain hydrogen bonds is significantly higher for GLY11 (black line) than for ALA11 (red line), indicating a much more stable fold in the GLY11 case. When compared to water \ce{H2O}, the total number of intra-chain hydrogen bond for ALA11 is smaller in cyclohexane \ce{cC6H12} than in water \ce{H2O} (compare the red lines in Figure \ref{fig:fig4b}(c) and Figure \ref{fig:fig4b})(d), so ALA11 is still relatively less stable in \ce{cC6H12} than in \ce{H2O}.

SI report the same quantities for ILE11, ASN11, LYS11, ARG11, and GLU11.
\sfref{supp-fig:hbond_rest}(a) display the total number of hydrogen bonds of ILE11 (black line), ASN11 (red line), LYS11 (green line), ARG11 (blue line), and GLU11 (magenta line) with water \ce{H2O}. While a nearly constant trend is observed in all cases, the actual total number decreases from GLU11 (the largest) to ILE11 (the smallest), with \sfref{supp-fig:hbond_rest}(b) displaying the corresponding equilibrium distribution. The total number of solute-solute intrachain hydrogen bonds, depicted in  \sfref{supp-fig:hbond_rest}(c), also shows a constant trend with slightly variable absolute number. This number increases in the case of cyclohexane \ce{cC6H12}, again due to the absence of an alternative provided by the solvent, and again decreases from GLU11 (the largest) to ILE11 (the smallest), thus confirming a stabilization effect of cyclohexane \ce{cC6H12} decreasing from the charged GLU11 to the hydrophobic ILE11.

\subsection{Solvation free energy}
\label{subsec:solvation}
\begin{figure}[h!]
\centering
 \captionsetup{justification=raggedright,width=\linewidth}
  \begin{subfigure}{9cm}
     \includegraphics[width=\linewidth]{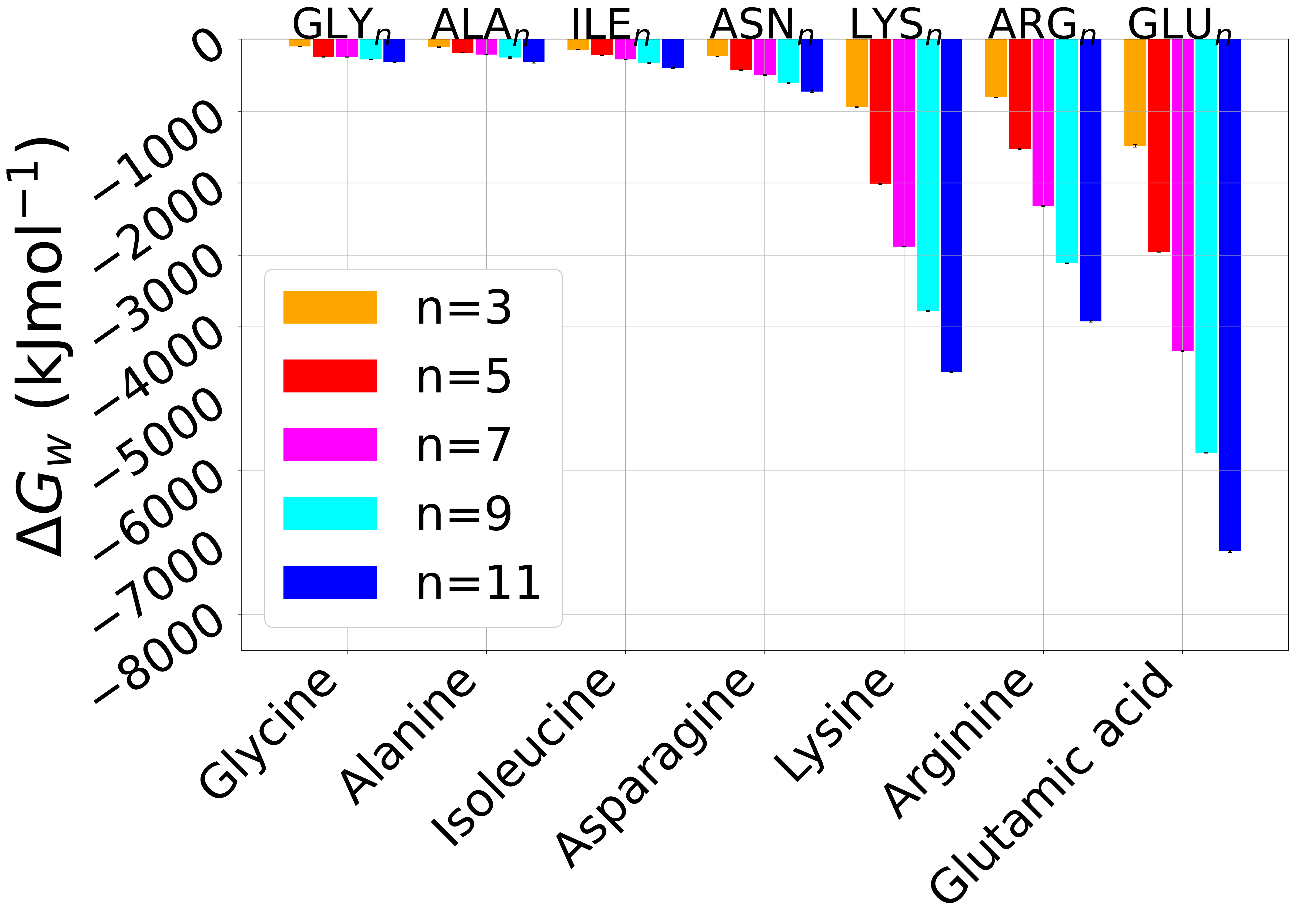}
    \caption{}\label{fig:fig5a}
  \end{subfigure}
  \begin{subfigure}{9cm}
     \includegraphics[width=\linewidth]{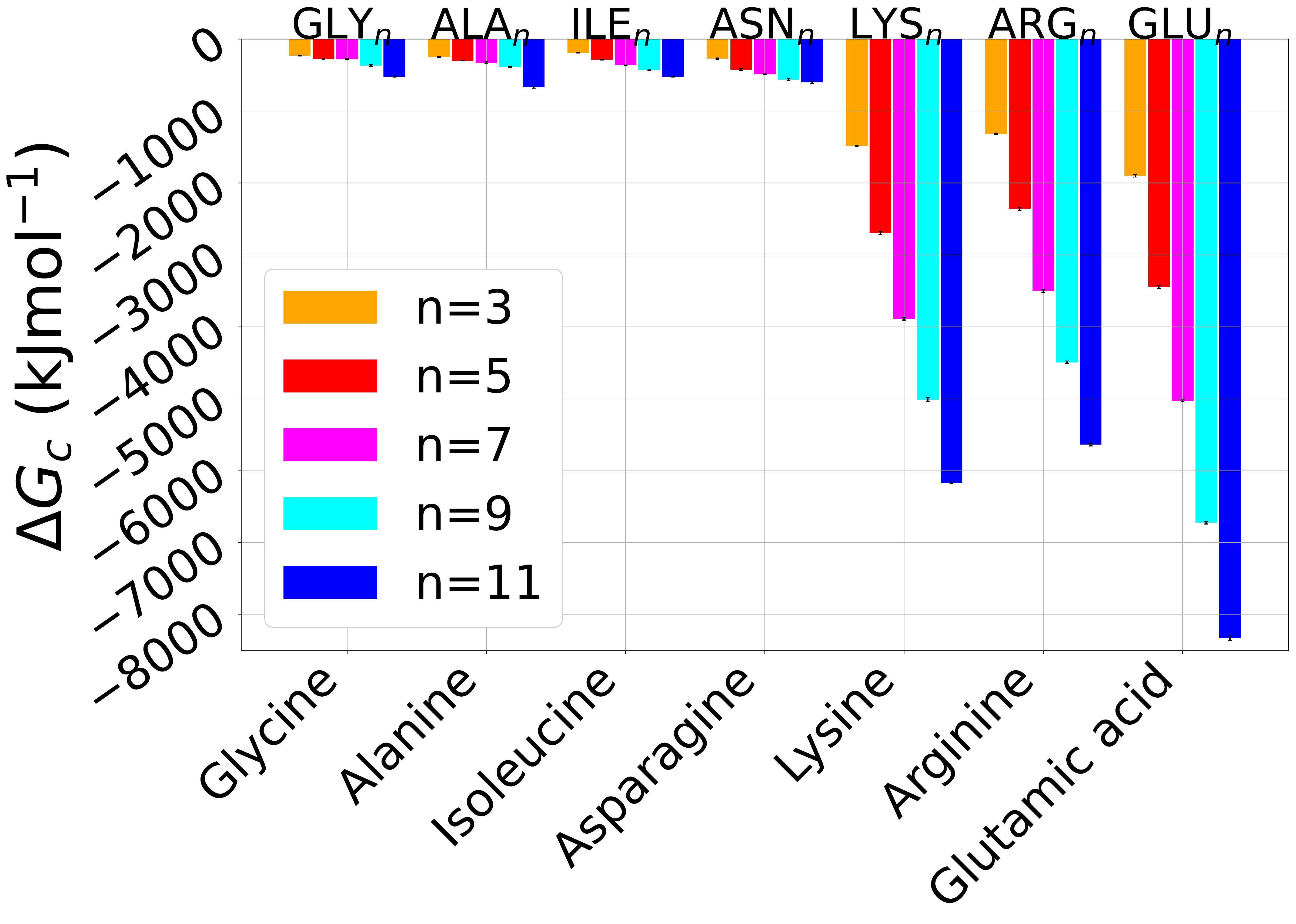}
    \caption{}\label{fig:fig5b}
  \end{subfigure}
   \caption{Solvation free energy $\Delta G_{solv} $ : (a) $\Delta G_{w} $ from vacuum to water \ce{H2O} at $25^{\circ}$C and (b)  $\Delta G_{c}$ from vacuum to cyclohexane \ce{cC6H12}. The polypeptides shown in the $x$-axis are {\color{black} representative of the full hydrophobic scale} following previous work \cite{Dongmo2020}. Their lengths vary from tri- ($n=3$) to undeca- ($n=11$) polypeptides. Note that all plots are in the same scale.
  \label{fig:fig5}}
\end{figure}

In Section \ref{subsec:good} we have seen how different polypeptides behave in solvents with different polarities. This analysis highlights that the definition of 'good' and 'poor' solvent is not an absolute property but has to be related to the specificities of the polypeptides. For example, water \ce{H2O} is a poor solvent for polyglycine, polyanaline, polyisoleucine and polyglutamic acid, but it is a good solvent for polyasparagine, polylysine and polyarginine. Conversely, cyclohexane \ce{cC6H12} is a poor solvent for polyasparagine and polyarginine, and it is a good solvent for polyglycine, polyalanine, polyisoleucine, and polylysine. In most cases these findings agree with our intuition and with the common view that " like dissolves like" but this is not always the case. For instance, polyglutamic acid collapses in water \ce{H2O} and remains extended in cyclohexane \ce{cC6H12}, whereas a reversed behavior could be expected on the basis of the charged nature of  the glutamic acid GLU residue. An even more notable exception is provided by polylysine which shows no collapse in either cyclohexane \ce{cC6H12} or water \ce{H2O}, in spite of the charged nature of the lysine residue.

{\color{black}In drafting these conclusions two additional points must be born in mind. First, none of the investigated homo polypeptides are really hydrophobic irrespective of the polarities of their residues. Indeed, we have shown that each of the considered polypeptides form a number of hydrogen bonds with the solvent ranging from $2-3$ bonds/residue for ILE11 to more than 10 hydrogen bonds/residue for GLU11 (see \sfref{supp-fig:hbond_rest}(a)). This is also evident from the snapshots of the initial conformation that shows in all cases a significant hydrogen bonding with the solvent, as explicitly displayed in \sfref{supp-fig:solvated_snapshots}. Accordingly, none of them with the exception of GLY11 is shown to have a stable fold in water \ce{H2O} (see representative snapshots in Fig.\ref{fig:fig3}), although clearly ILE11 has a stronger tendency to fold compared to GLU11. The second point that is worth stressing is that the difference between extended/swollen and compact/globule is well defined only for sufficiently longer polypeptides compared to those analyzed in the present work.}

Next, we turn our attention to the corresponding solvation free energies that can be computed via thermodynamic integration.
As anticipated, the aims here are twofold. First, we would like to extend our previous calculation \cite{Dongmo2020} for a single amino acid side chain equivalent  -- a single amino acid where the backbone part of the amino acid has been replaced with a single hydrogen atom, to include the effect of the backbone as well as the dependence of the number $n$ of included residues.  Relevant questions here are possible non-linear effects of the solvation free energy as a function of the number of repeated units, and whether there is a mirror symmetry in by changing a highly polar solvent such as water \ce{H2O} to an apolar organic solvent such as cyclohexane \ce{cC6H12}. For instance, is the solvation free energy of $(G)_n$ equal to $n$ times the solvation free energy of a single amino acid $(G)_1$? And is this depending on the polarity properties of the amino acids and/or the polarity of the solvent? Both questions will be addressed in the present section.

A second issue of paramount importance is what is the main driving force to solvation. A conventional simple accepted picture is that solvation includes two different and competing processes: the entropically unfavourable creation of a cavity, and the enthalpically favourable attractive dispersion contributions arising by the introduction of the solute. Note that in  water this picture is known to be affected above a critical solute size of \SI{1}{nm} in view of the fact that sufficiently small solutes (smaller that \SI{1}{nm}) do not affect the water hydrogen bond network \cite{Chandler2005}. 

While we will not consider all the 18 side chains studied in \citeauthor{Dongmo2020} \cite{Dongmo2020}, our representative results will be sufficient to understand the emerging pattern.

Figure \ref{fig:fig5} displays the solvation free energy for water \ce{H2O}
(Figure \ref{fig:fig5a}) and cyclohexane  \ce{cC6H12} (Figure \ref{fig:fig5b}), at room temperature ($25^{\circ}$C) in both cases.
All corresponding values can be found in  \stref{supp-tab:water_solv}, and \stref{supp-tab:chex_solv}. The ordering is according the nominal character of the amino acid from hydrophobic (left) to polar (right). Glycine GLY is listed first as the simplest case. 

{\color{black} As visible in Fig.\ref{fig:fig5a} all considered polypeptides display negative solvation free energy, indicating that in water \ce{H2O} the onset of attractive energies originating upon the insertion of the polypeptides overwhelms the entropic cost of creating a cavity.  The effect is more pronounced for polar and charged amino acids, with the $\Delta G$ decreasing with the increase of the number of identical amino acids $n$ from 3 to 11 (i.e. the length of the peptide).}

The same trend is observed for the solvation free energy in cyclohexane \ce{cC6H12}, as reported in Fig. \ref{fig:fig5b}. 
While in water \ce{H2O} this behaviour is in marked contrast with that of single amino acids equivalents \cite{Dongmo2020} where the solvation free energy $\Delta G_{w}$ is found to be large and positive for hydrophobic amino acids side chains equivalent, and large and negative for polar ones \cite{Dongmo2020}, it is in accord with a similar computational study of tripeptides in water \cite{Hajari15}.

In cyclohexane \ce{cC6H12}, however, this behaviour is more intriguing. We note that \textit{both} hydrophobic \textit{and} polar peptides have negative solvation free energy $\Delta G_{c}$ in cyclohexane \ce{cC6H12}, more negative for polar than for hydrophobic ones \cite{Dongmo2020}. A calculation of the transfer free energy $\Delta \Delta G_{w>c}$ from  water \ce{H2O} to cyclohexane \ce{cC6H12}, however restores our intuitive picture in terms of the relative stability.

\begin{figure}[htpb]
\centering
 \captionsetup{justification=raggedright,width=\linewidth}
     \includegraphics[width=\linewidth]{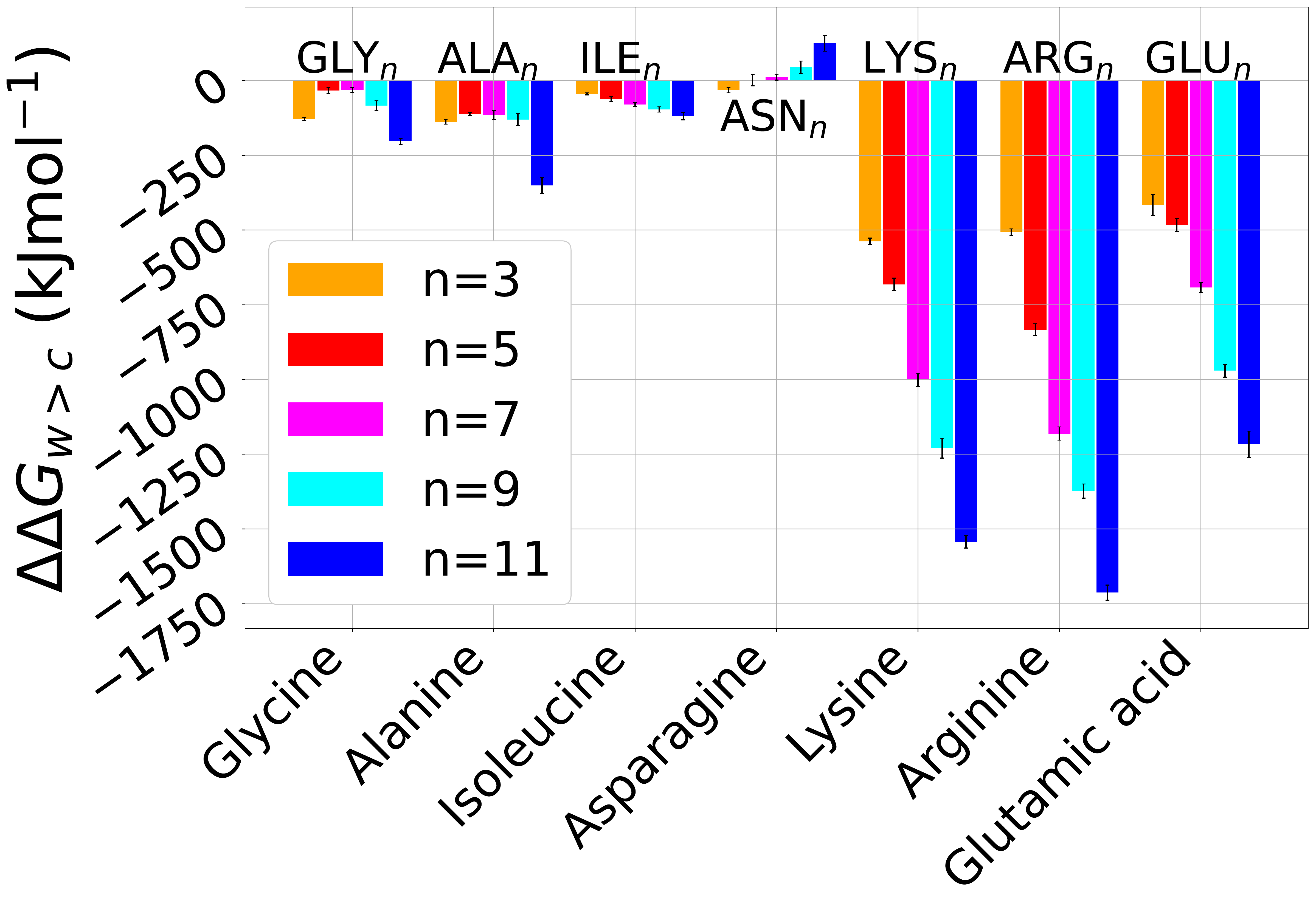}
  \caption{$\Delta \Delta G_{w>c}$ from water \ce{H2O} to cyclohexane \ce{cC6H12} at $25^{\circ}$C. Ordering is the same as in Fig. \ref{fig:fig7}.}
  \label{fig:fig6}
\end{figure}

Fig.\ref{fig:fig6} reports $\Delta \Delta G_{w>c}$ for polypeptides from water \ce{H2O} to cyclohexane \ce{cC6H12} with the same arrangement and ordering of Figs. \ref{fig:fig5}, hydrophobic (left) and polar (right), at different peptide length $n$. With the exclusion of asparagine (ASN), all $\Delta \Delta G_{w>c}$ are negative, significantly larger for polar than for hydrophobic polypeptides although $n=3$ is clearly an outlier for hydrophobic polypeptides likely due to its small size. As anticipated, previously alluded to in Fig. \ref{fig:fig2} and reported in \stref{supp-tab:poly_lenght}, all tripeptides ($n=3$) have sizes smaller that \SI{1}{\nm} that is known to be a critical value for solvation in water \cite{Chandler2005}, whereas all peptides with $n>3$ have sizes larger than this value.
In this respect, present results are complementary to those on tripeptides reported in Ref. \cite{Hajari15}.


Consider GLY$_{n}$ first (the outermost left in Fig.\ref{fig:fig6}). Here $\Delta \Delta G_{w>c}$ is small and negative, indicating a stabilizing effect of cyclohexane \ce{cC6H12} compared to water \ce{H2O}. This agrees with the calculations of Section \ref{subsec:good} and confirms findings from previous studies \cite{Tran2008,Kokubo2013}. However, the trend is not linear: $\Delta \Delta G_{w>c}$ increases from $n=3$ to $n=7$ and then decreases again for higher $n=9,11$. Polyalanine ALA$_{n}$ and polyisoleucine ILE$_{n}$ show a more regular increasing trend, whereas polyasparagine ASN$_{n}$ switches from negative to positive $ \Delta \Delta G_{w>c}$ as $n$ increases. Polar and charged polypeptides, on the other hand, display a much more significantly negative $ \Delta \Delta G_{w>c}$ with a monotonic increase with $n$, a result that defies with our physical intuition, but it is again in agreement with results on tripeptides \cite{Hajari15}.
 
 {\color{black} The emerging scenario is then that the stability of a (homo) polypeptide is mainly dictated by the polarity of the solute, with the polarity of the solvent playing a minor role  }
\subsection{Entropy-enthalpy compensation}
\label{subsec:compensation}
{\color{black}Two remaining issues are left from the results of previous sessions.
The first issue is whether any observed process is predominantly enthalpically or entropically driven, and it will be discussed in the present Session. This can be conveniently obtained by the analysis of the solvation free energy at different temperatures that allows to separate out the entropy and the enthalpy contributions, as anticipated in Sect.\ref{sec:theory}.}

{\color{black} As anticipated, the solvation free energy $\Delta G$ can be factorized in two terms.}  First, the creation of a cavity in the solvent to accomodate the solute. This process is clearly entropically unfavourable so $T \Delta S <0$ ($-T \Delta S> 0$). However, attractive interactions may form upon inserting the solute in the cavity, thus leading to a favourable process with $\Delta H <0$. If the two processes happen to balance each other, then $\Delta G \approx 0$ and $-T\Delta S = -\Delta H$, thus leading to a perfect anticorrelation in the $-T\Delta S$ versus $\Delta H$ plane, known as "entropy-enthalpy compensation" with a slope $=-1$ (see \sfref{supp-fig:cartoon_entropy_enthalpy}). If the slope is $>-1$, then the system is entropically driven, conversely is enthalpically driven.

\sfref{supp-fig:water} and \sfref{supp-fig:chex} display the temperature dependence of $\Delta G_{w}$ in water \ce{H2O} and $\Delta G_{c}$ in cyclohexane \ce{cC6H12} respectively. Both are increasing function of the temperature as expected since both $T \Delta S_{w}$ and  $T \Delta S_{c}$ are entropic positive costs irrespective of the solvent polarity, in agreement with the results from the single amino acid side chain equivalents as well as past experimental results \cite{Dongmo2020}. Curvatures are however different depending of the specific solvent and also on the length $n$ of the polypeptide, indicating a very complex patchwork of interactions that in water may also depend on the size of the polypeptide \cite{Chandler2005}.   

In Ref.\cite{Dongmo2020}, we reported this calculation for each single amino acid side chain equivalent. In water \ce{H2O}, hydrophobic amino acid side chain equivalents were found to comply the entropy-enthalpy compensation rule reasonably well, with a wide distribution of values along the line with slope $\approx -1$ in the $-T \Delta S$ vs $\Delta H$ plane, depending on the specificity of each single residue. Polar amino acid side chain equivalents showed instead a tendency to lump together around a specific region of this line, with the exception of arginine ARG. In cyclohexane \ce{cC6H12} the tendency to lump around similar state points was found to be even more pronounced for both polar and hydrophobic amino acids \cite{Dongmo2020}. 

The values of the slopes along with the intercepts to origin and the  corresponding correlation coefficients are reported in \stref{supp-tab:corr_coeffs} for all considered polypeptides and for both \ce{H2O} and \ce{cC6H12}. Interestingly, all slopes are found $<1$ indicating that all these solvation processes are largely enthalpically dominated.

\begin{figure*}[h!]
\centering
 \captionsetup{justification=raggedright,width=\linewidth}
  \begin{subfigure}{0.4\linewidth}
     \includegraphics[width=\linewidth]{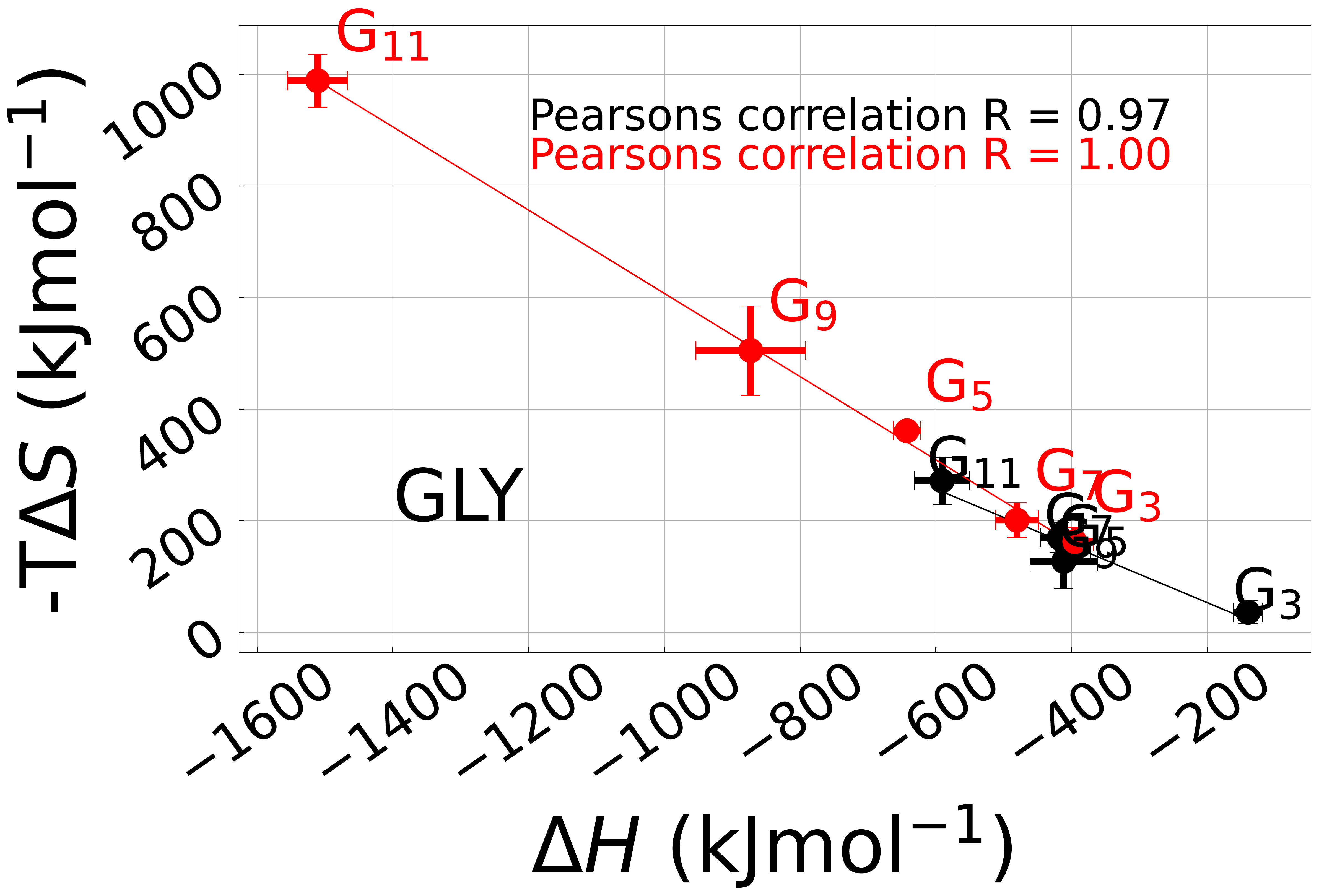}
    \caption{}\label{fig:fig7a}
  \end{subfigure}
  \hfill
  \begin{subfigure}{0.4\linewidth}
     \includegraphics[width=\linewidth]{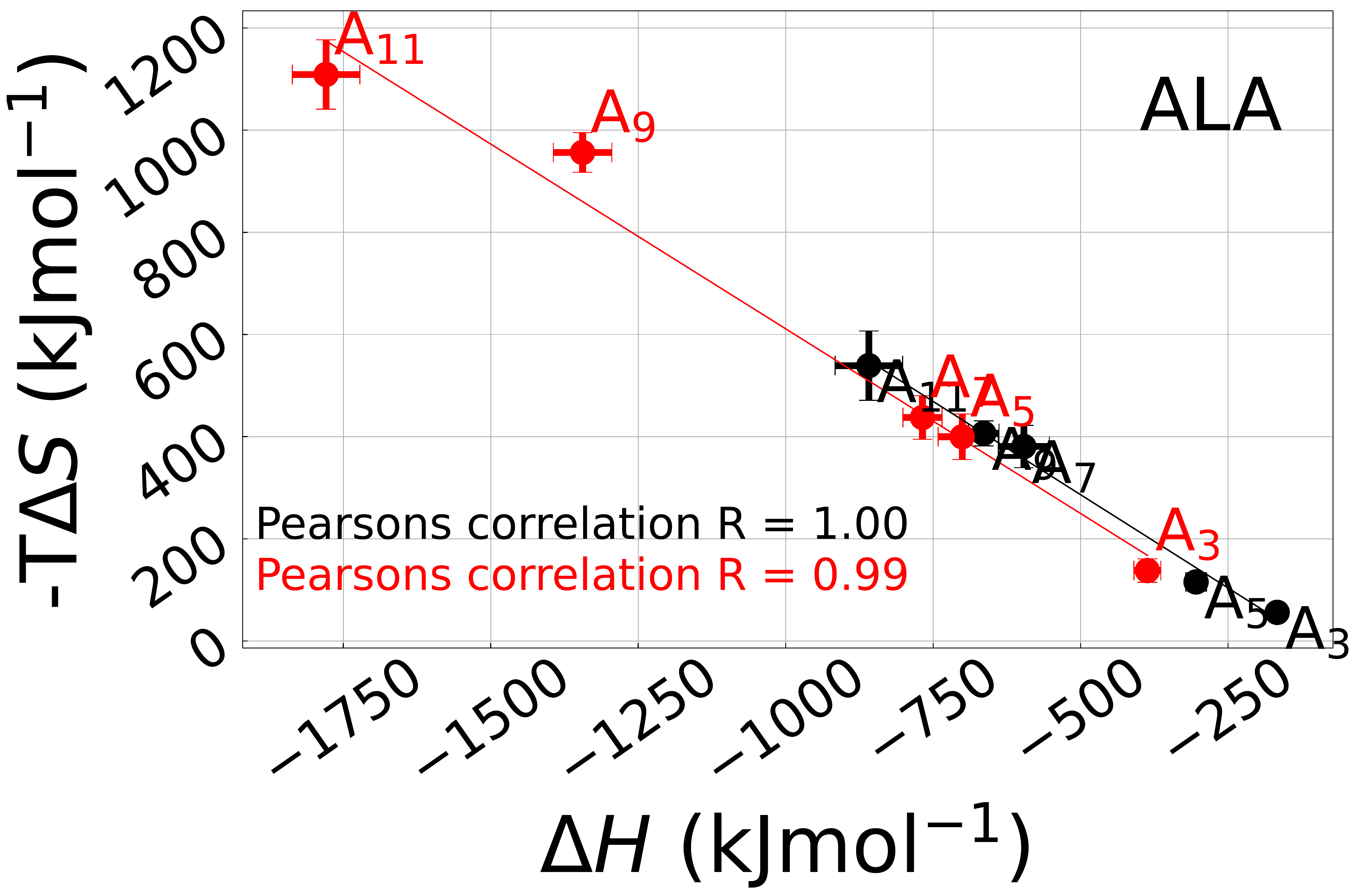}
    \caption{}\label{fig:fig7b}
  \end{subfigure}
  \hfill
  \begin{subfigure}{0.5\linewidth}
     \includegraphics[width=\linewidth]{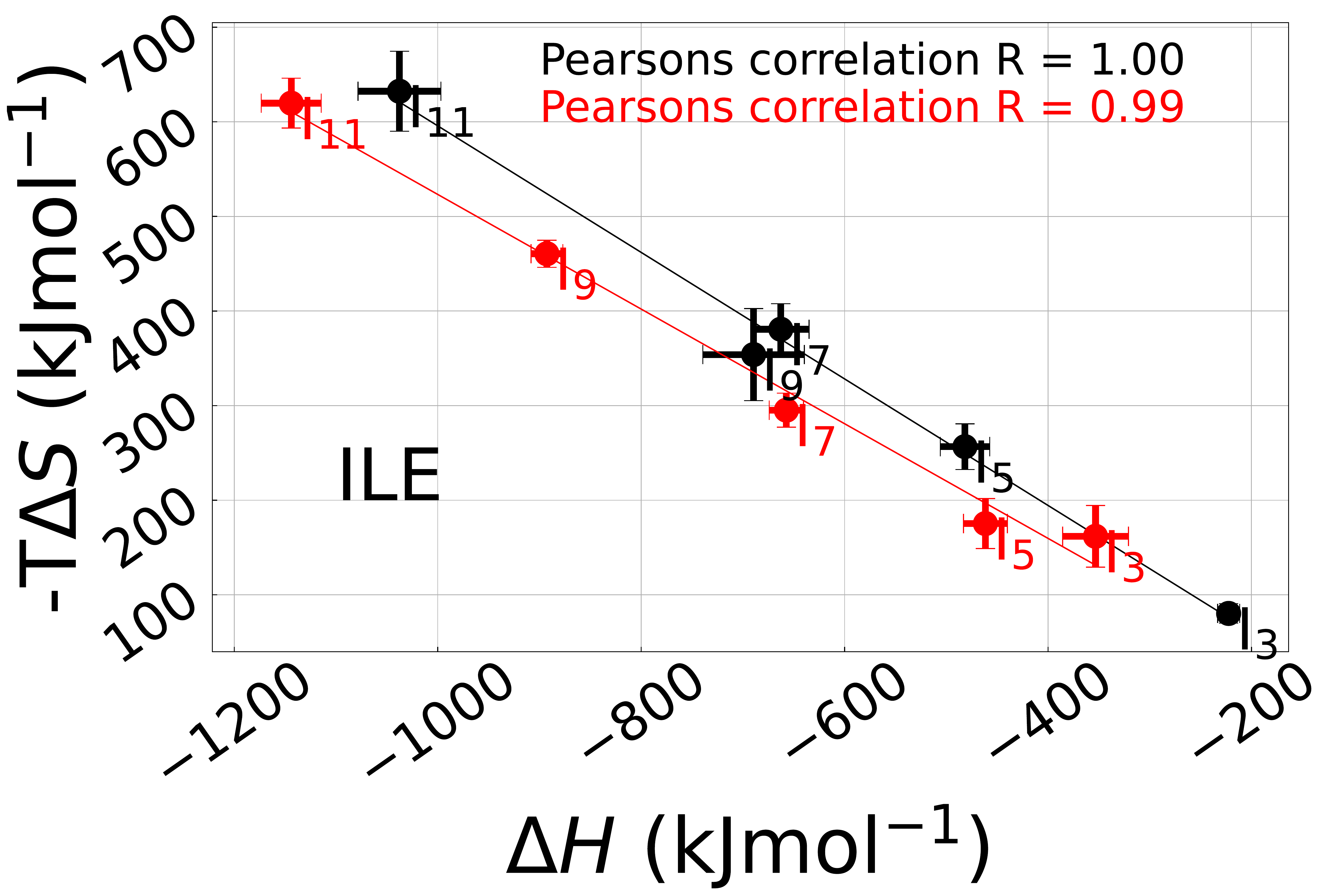}
    \caption{}\label{fig:fig7c}
  \end{subfigure}
  \hfill
    \begin{subfigure}{0.4\linewidth}
     \includegraphics[width=\linewidth]{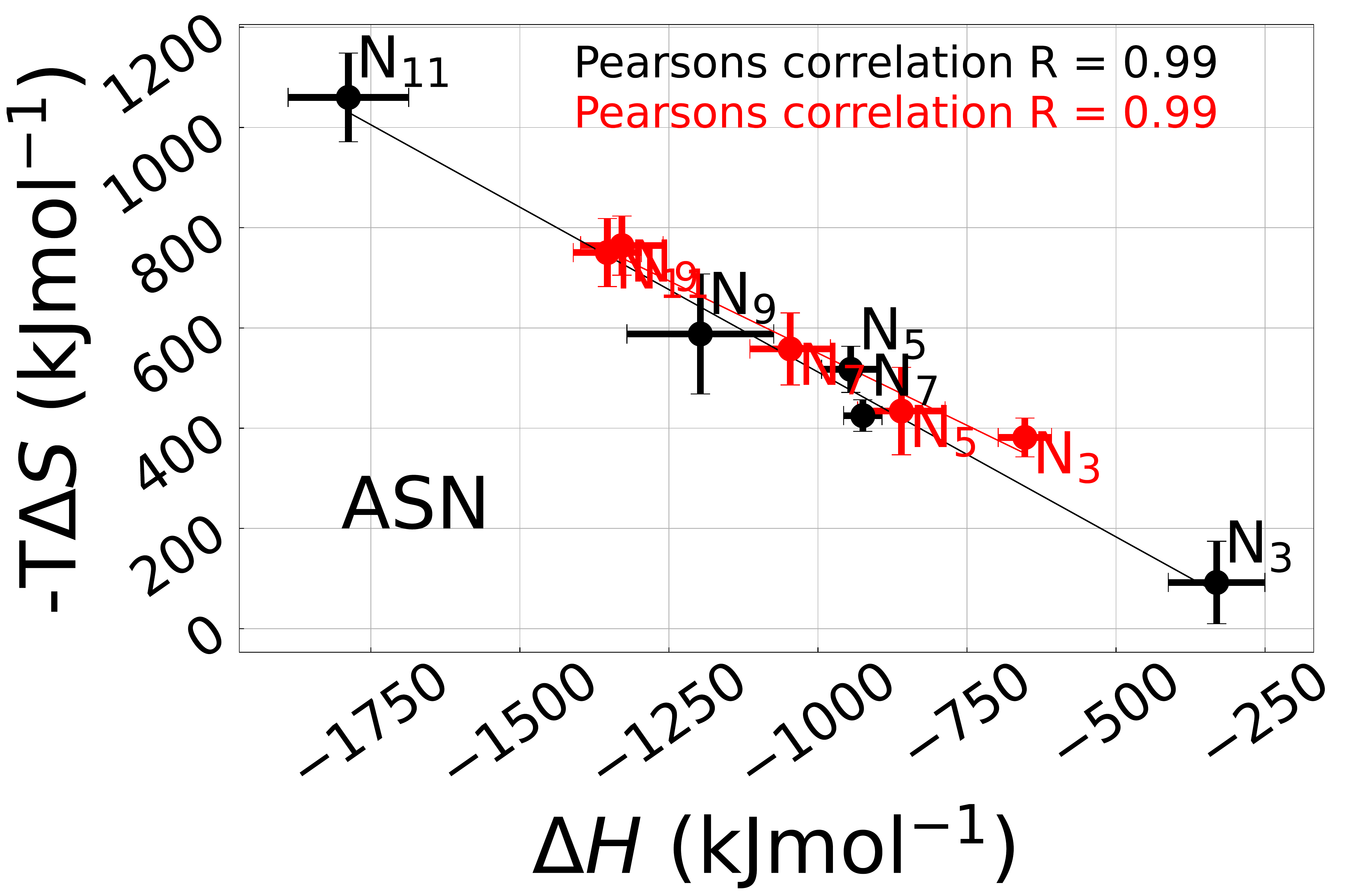}
    \caption{}\label{fig:fig7d}
    \end{subfigure}
    \hfill
    \begin{subfigure}{0.4\linewidth}
     \includegraphics[width=\linewidth]{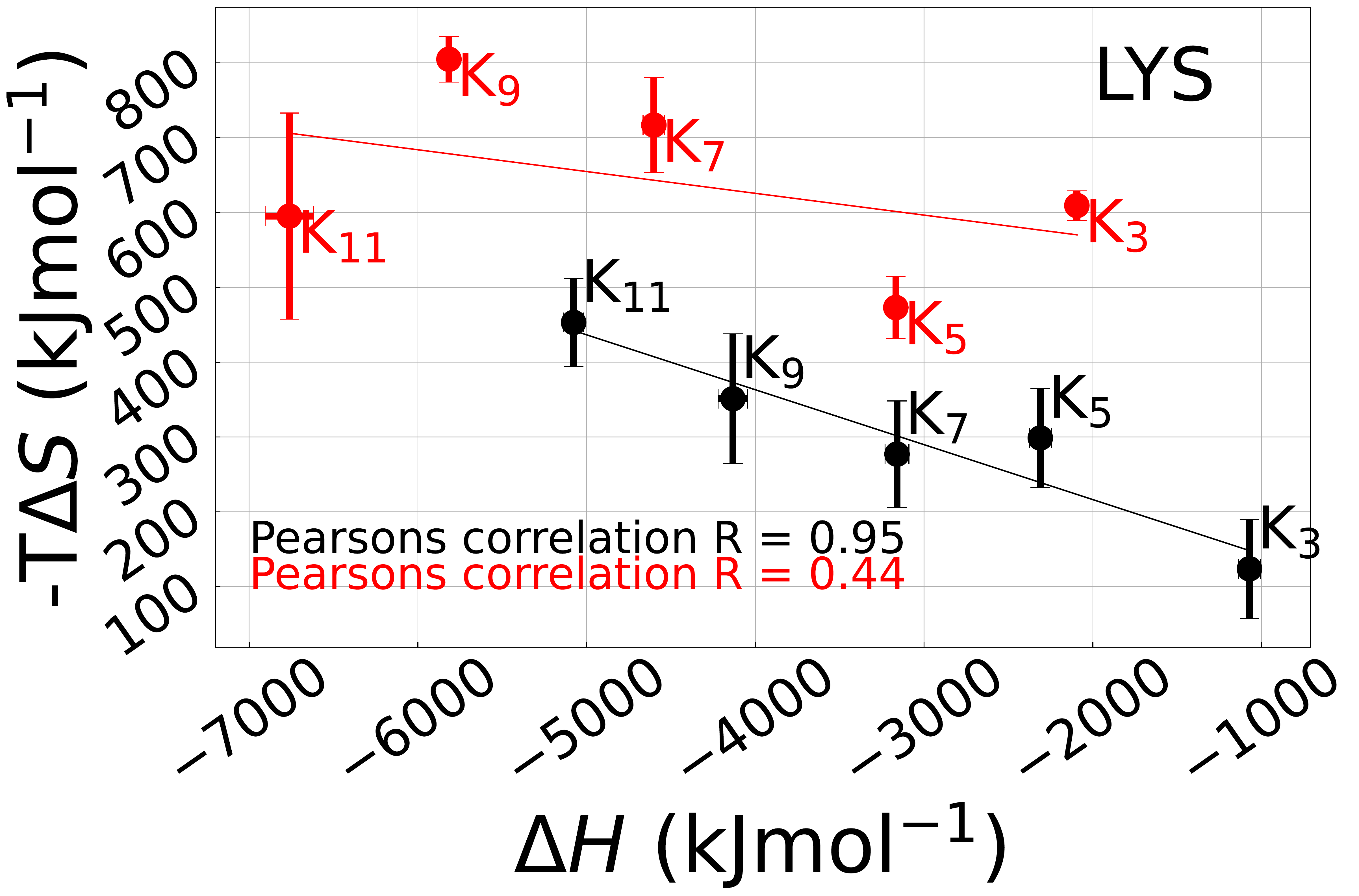}
   \caption{}\label{fig:fig7e}
    \end{subfigure}
    \hfill
   \begin{subfigure}{0.4\linewidth}
     \includegraphics[width=\linewidth]{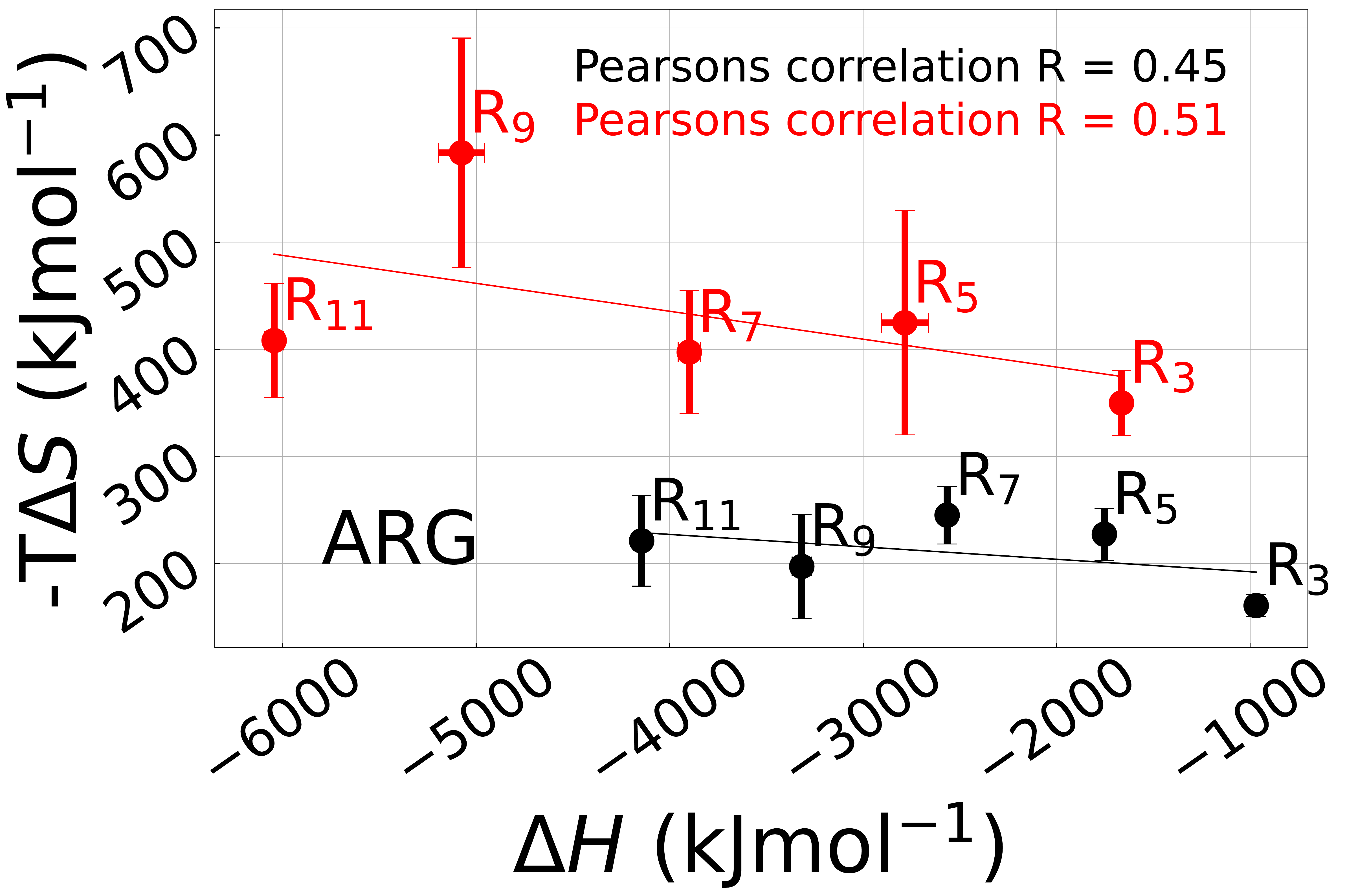}
    \caption{}\label{fig:fig7f}
   \end{subfigure}
   \hfill
   \begin{subfigure}{0.4\linewidth}
     \includegraphics[width=\linewidth]{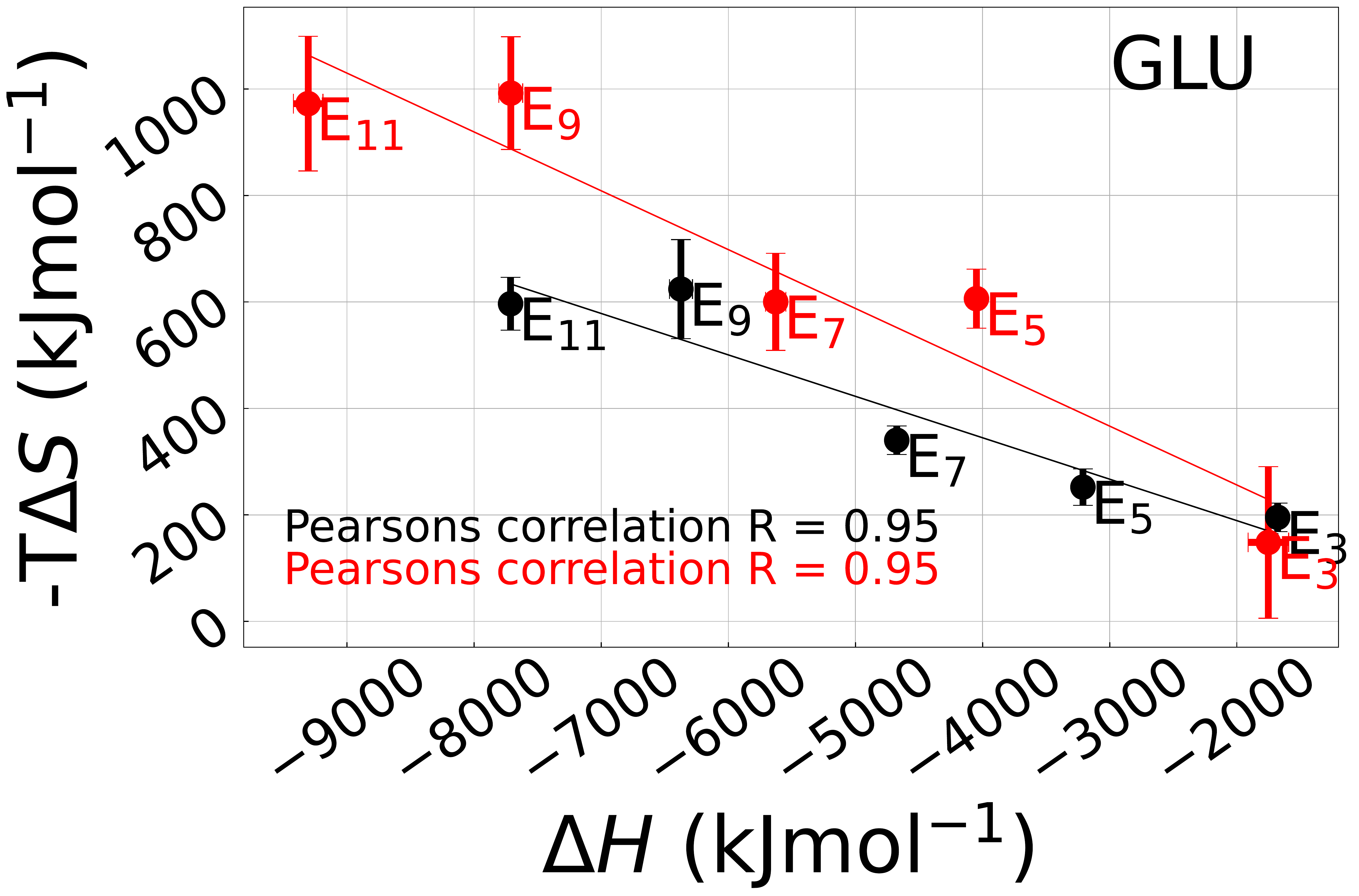}
   \caption{}\label{fig:fig7g}
   \end{subfigure}
   \hfill
   \caption{Entropic contribution $-T \Delta S$ of the solvation free energy $\Delta G$ as a function of the enthalpic counterpart $\Delta H$ in the case of  water \ce{H2O} and cyclohexane \ce{cC6H12} for different polymers length. The solvation data in water \ce{H2O} are displayed in black and those in cyclohexane \ce{cC6H12} are plotted in red while the error bars represent the standard deviations. The subplots annotated from (a) to (g) correspond to each of the polypeptides used here. These are GLY, ALA, ILE, ASN, LYS, ARG, and GLU. Furthermore, the continuous lines represent the linear fits of the simulation data. Please note that different scales have been used in different cases.  
  \label{fig:fig7}}
\end{figure*}

Fig.\ref{fig:fig7} reports the results of this analysis, where the entropic part of the free energy $-T \Delta S$ is plotted as a function of the enthalpic part $\Delta H$. Each panel \ref{fig:fig7a}-\ref{fig:fig7g} includes points computed at different lengths from $n=3$ to $n=11$ for all the considered polypeptides. In all cases, data for water \ce{H2O} are in black, those for cyclohexane \ce{cC6H12} are in red.

Consider the glycine GLY case first, see Fig.\ref{fig:fig7a}. In water \ce{H2O} (black), nearly all different points G$_{3}-$G$_{11}$ lump very closely one another along a line with slope approximately $-1$. By contrast, in cyclohexane \ce{cC6H12} (Fig.\ref{fig:fig7a} red)  there is a very clear anti-correlation in the sense that $\Delta H$ is decreasing with increasing length $n$, with a corresponding increase of $-T \Delta S$. That is, a gain in enthalpy translates into a corresponding loss of entropy. This corresponds exactly to the entropy-enthalphy compensation usually found in water \ce{H2O} (see e.g. \cite{Hayashi17,Hayashi18}, this time in cyclohexane \ce{cC6H12} rather than in water, and it reflects the fact that cyclohexane \ce{cC6H12} is a good solvent for polyglycine whereas water \ce{H2O} is poor one, in agreement with the results of Section \ref{subsec:good}. The cases of  alanine ALA (Fig. \ref{fig:fig7b}) and isoleucine ILE (Fig. \ref{fig:fig7c}) are expected to follow a similar pattern on the basis of their hydrophobic character (Table \ref{tab:amino_acid}), but they appear to present a more complex behaviour. In the case of polyalanine ALA (Fig. \ref{fig:fig7b}) a rather similar behaviour in water \ce{H2O} (black) and cyclohexane \ce{cC6H12} (red) is found (note the two scales of Figs. \ref{fig:fig7a} and \ref{fig:fig7b} are nearly equivalent), suggesting a similar behaviour for polyglycine and GLY and polyalanine ALA. An additional notable feature of polyalanine ALA in water \ce{H2O} is the irregular dependence as a function of $n$, with $n=11$ very different from all others, in line with the same trend displayed for $\Delta  \Delta G_{w>c}$ (Fig.\ref{fig:fig8}). Polyisoleucine ILE (Fig. \ref{fig:fig7c}) also shows an entropy-enthalpy compensation for both  water \ce{H2O} and cyclohexane \ce{cC6H12}, but with a much more linear dependence on $n$. Interestingly, polyasparagine ASN also displays a similar pattern (Fig.\ref{fig:fig7d}) where for polylysine LYS (Fig.\ref{fig:fig7e}), polyarginine  ARG (Fig.\ref{fig:fig7f}), and polyglutamic acid GLU (Fig.\ref{fig:fig7f}) a rather different trend is observed for water \ce{H2O} and cyclohexane \ce{cC6H12}, in all cases with a slope significantly smaller than $-1$, indicating a predominant enthalpic role.
Here, we emphasize again that the assumed temperature dependence reported in Eq. \ref{sec2:eq4} is phenomenological and it might break down for some of the cases reported here, although it has been found to work rather well in past similar studies on single amino acid side chain equivalents both in water \ce{H2O} \cite{Hajari15,Dongmo2020} and in cyclohexane \ce{cC6H12}. More robust direct calculations are possible \cite{Lai2012} albeit much more computational demanding.

\subsection{Chain length dependence of solvation free energy, $\Delta G$ : implication on additivity}
{\color{black} The second point is related to the $n$ dependence of $\Delta G$ in \ce{H2O} and in cyclohexane \ce{cC6H12} that was anticipated in Fig.\ref{fig:fig7}. Here the relevant question is whether $\Delta G_{n} \propto n$ (linear dependence on the length) or there exist non-linear effects due to the backbone, as it was observed in the case of tripeptides \cite{Hajari15}. Note that in water marked change is expected when a hydrophobic solute size increases from below to above \SI{1}{nm} because below  \SI{1}{nm} a cavity able to accommodate the solute can be created without affecting the hydrogen bond network \cite{Chandler2005} and the tripeptides considered in Ref. \cite{Hajari15} were all smaller that \SI{1}{nm}.}

\begin{figure*}[h!]
\centering
 \captionsetup{justification=raggedright,width=\linewidth}
  \begin{subfigure}{0.20\textwidth}
     \includegraphics[width=\linewidth]{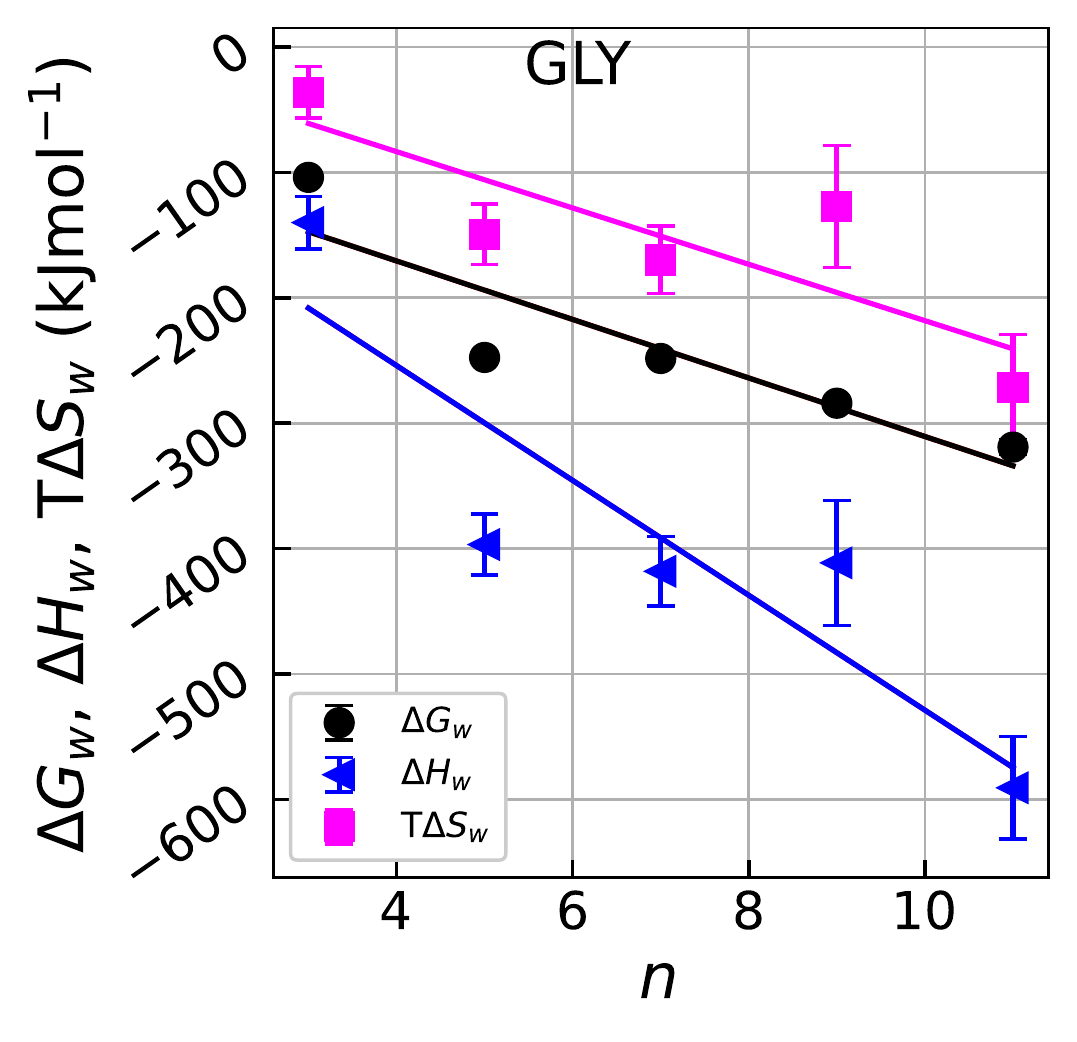}
    \caption{}\label{fig:fig8a}
  \end{subfigure}
  \begin{subfigure}{0.20\textwidth}
     \includegraphics[width=\linewidth]{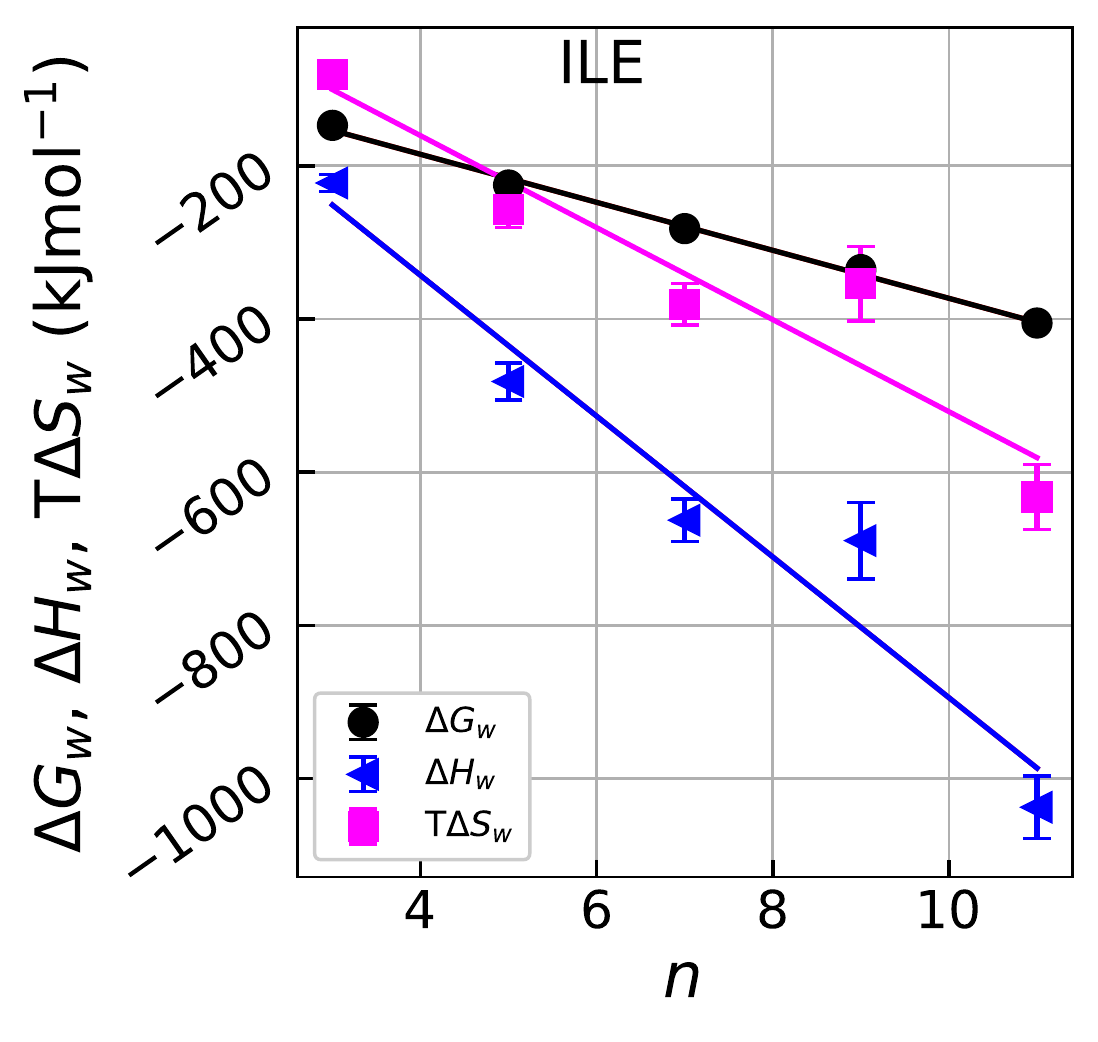}
    \caption{}\label{fig:fig8bb}
  \end{subfigure} 
  \begin{subfigure}{0.20\textwidth}
     \includegraphics[width=\linewidth]{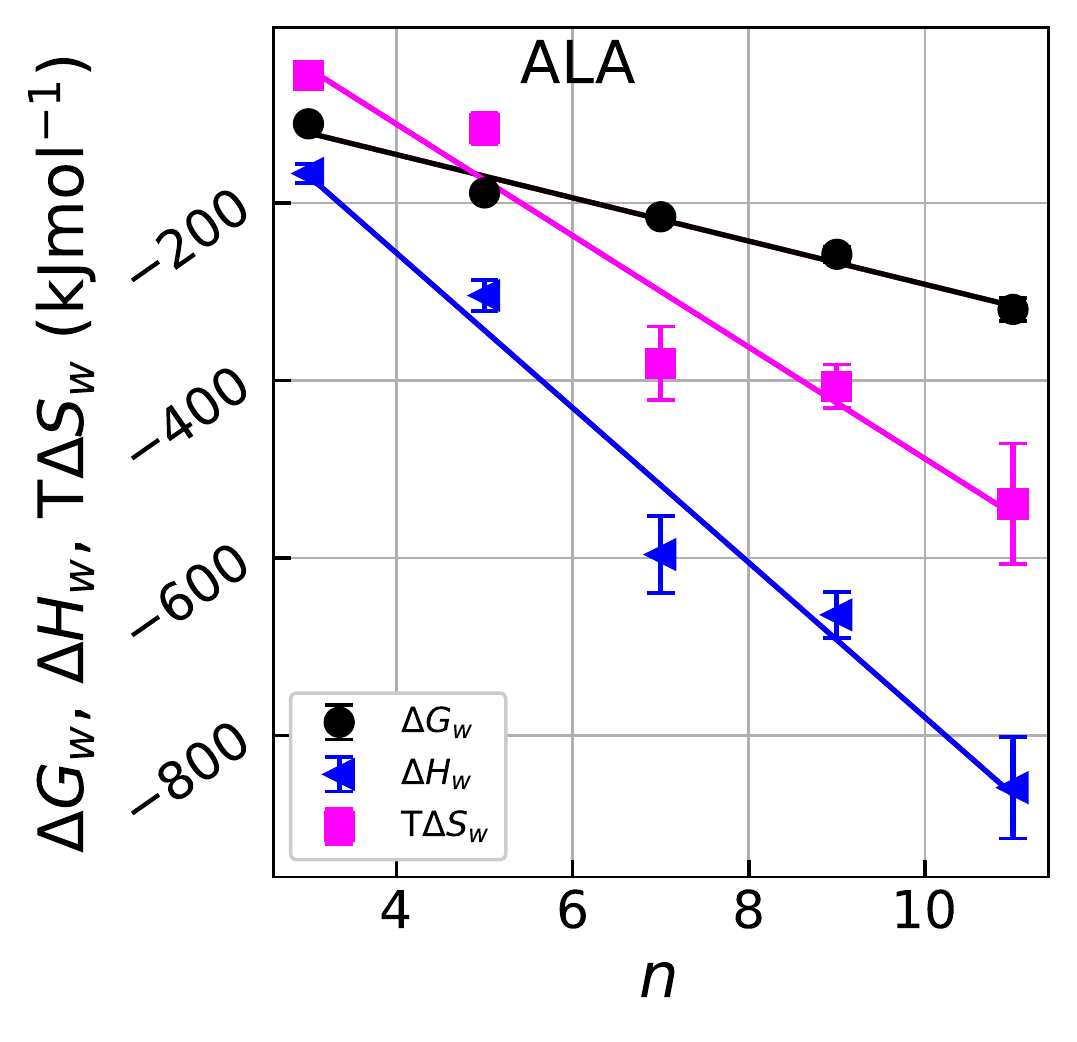}
    \caption{}\label{fig:fig8c}
    \end{subfigure}
    \begin{subfigure}{0.20\textwidth}
     \includegraphics[width=\linewidth]{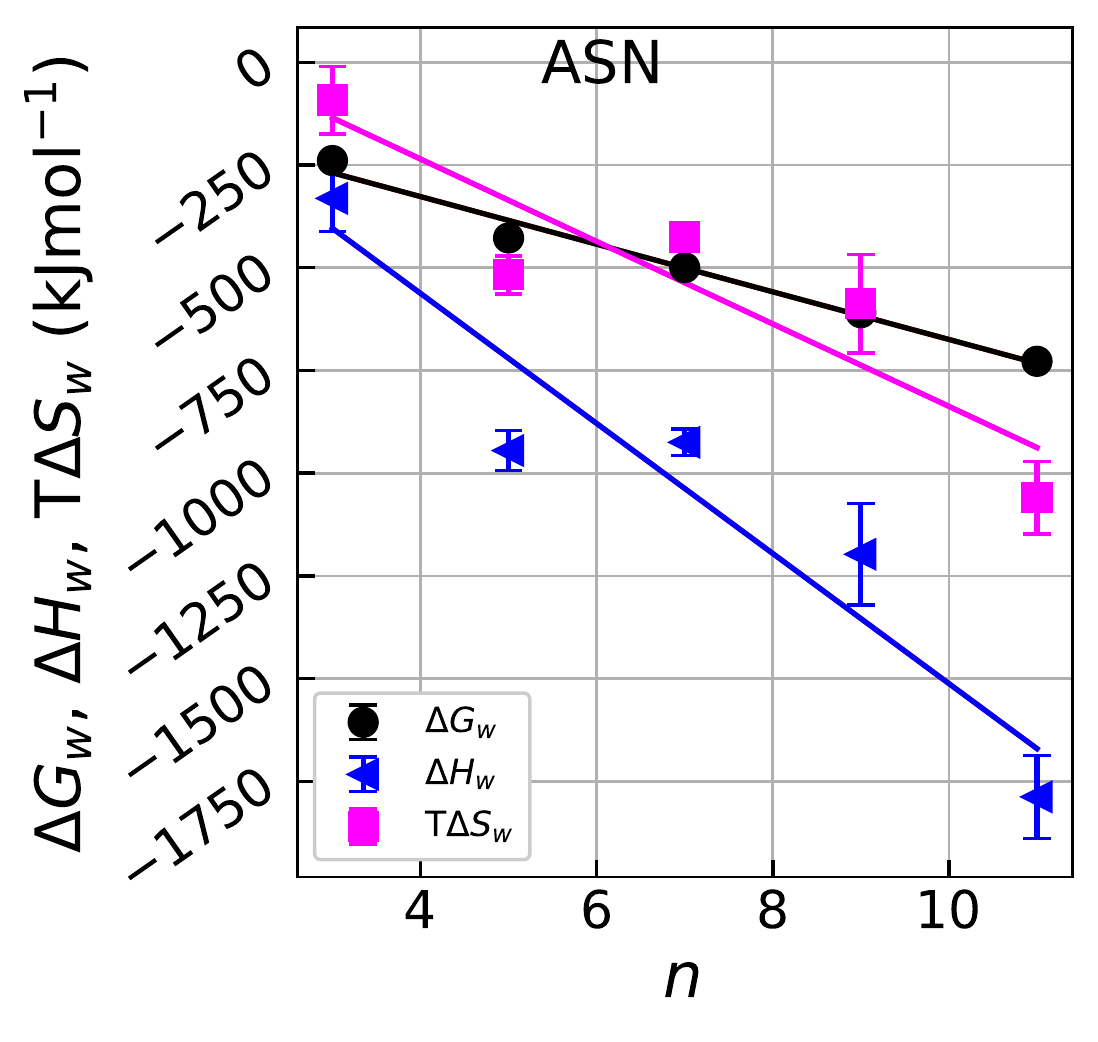}
    \caption{}\label{fig:fig8d}
    \end{subfigure} 
    \begin{subfigure}{0.20\textwidth}
     \includegraphics[width=\linewidth]{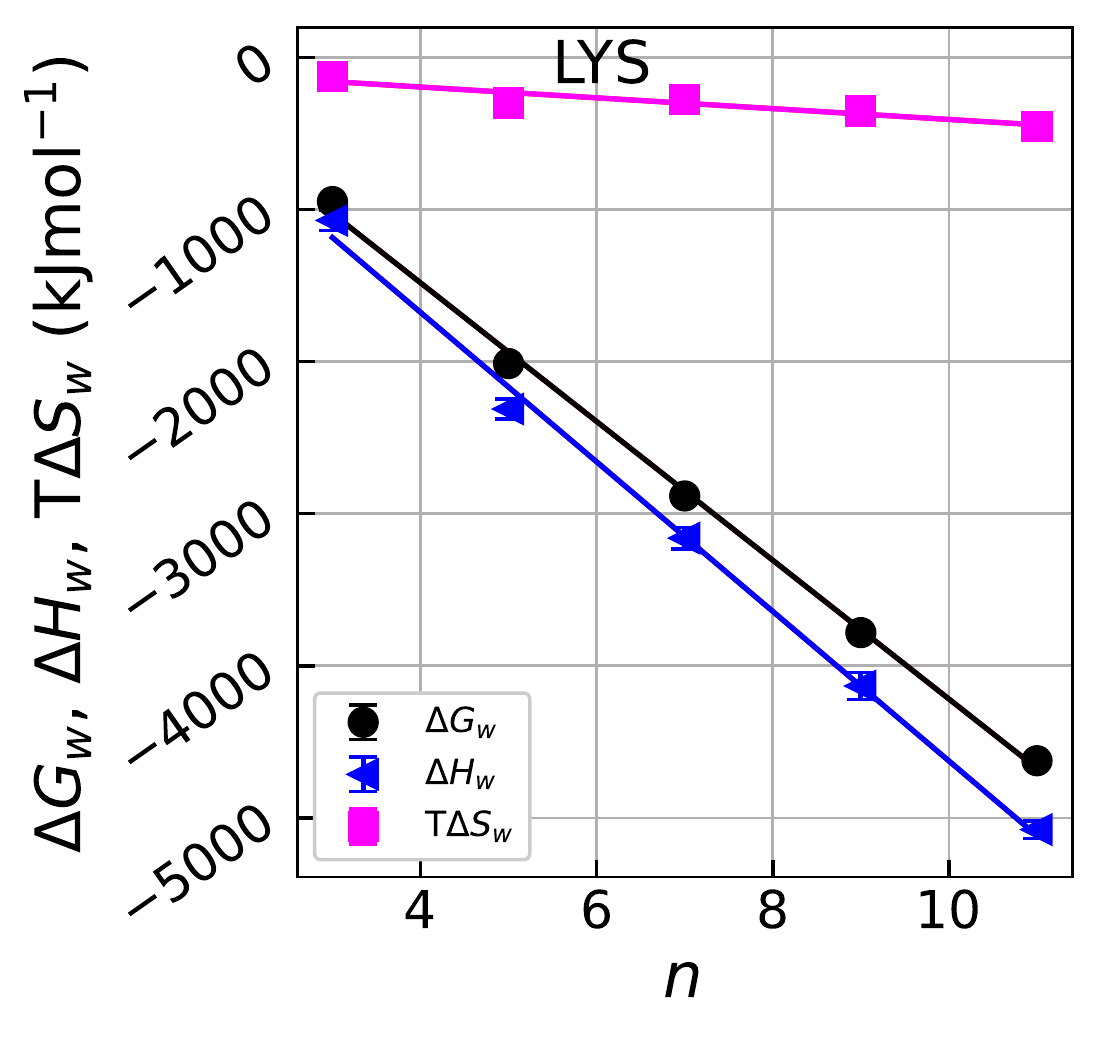}
   \caption{}\label{fig:fig8e}
   \end{subfigure}
   \begin{subfigure}{0.20\textwidth}
     \includegraphics[width=\linewidth]{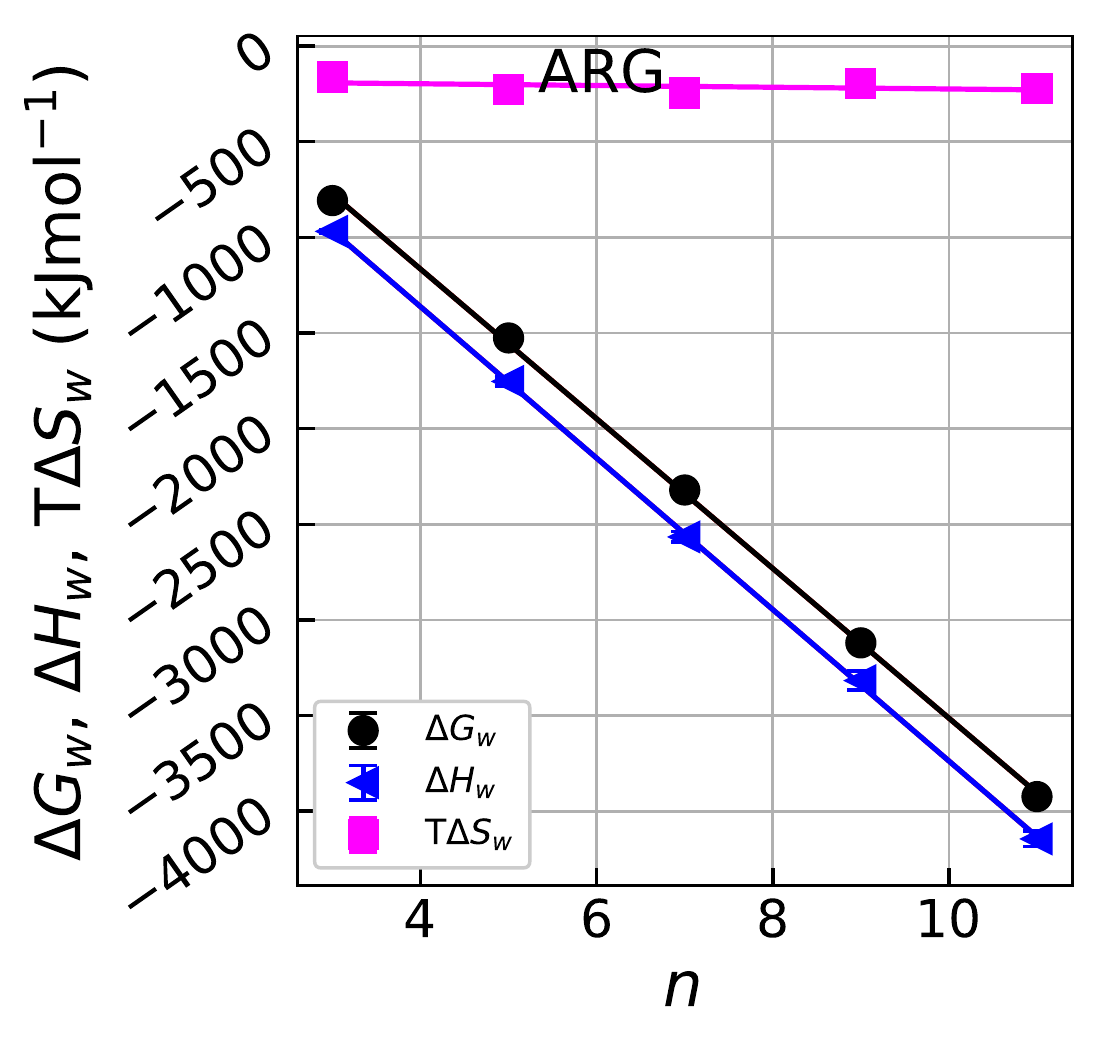}
   \caption{}\label{fig:fig8f}
   \end{subfigure}
   \begin{subfigure}{0.20\textwidth}
     \includegraphics[width=\linewidth]{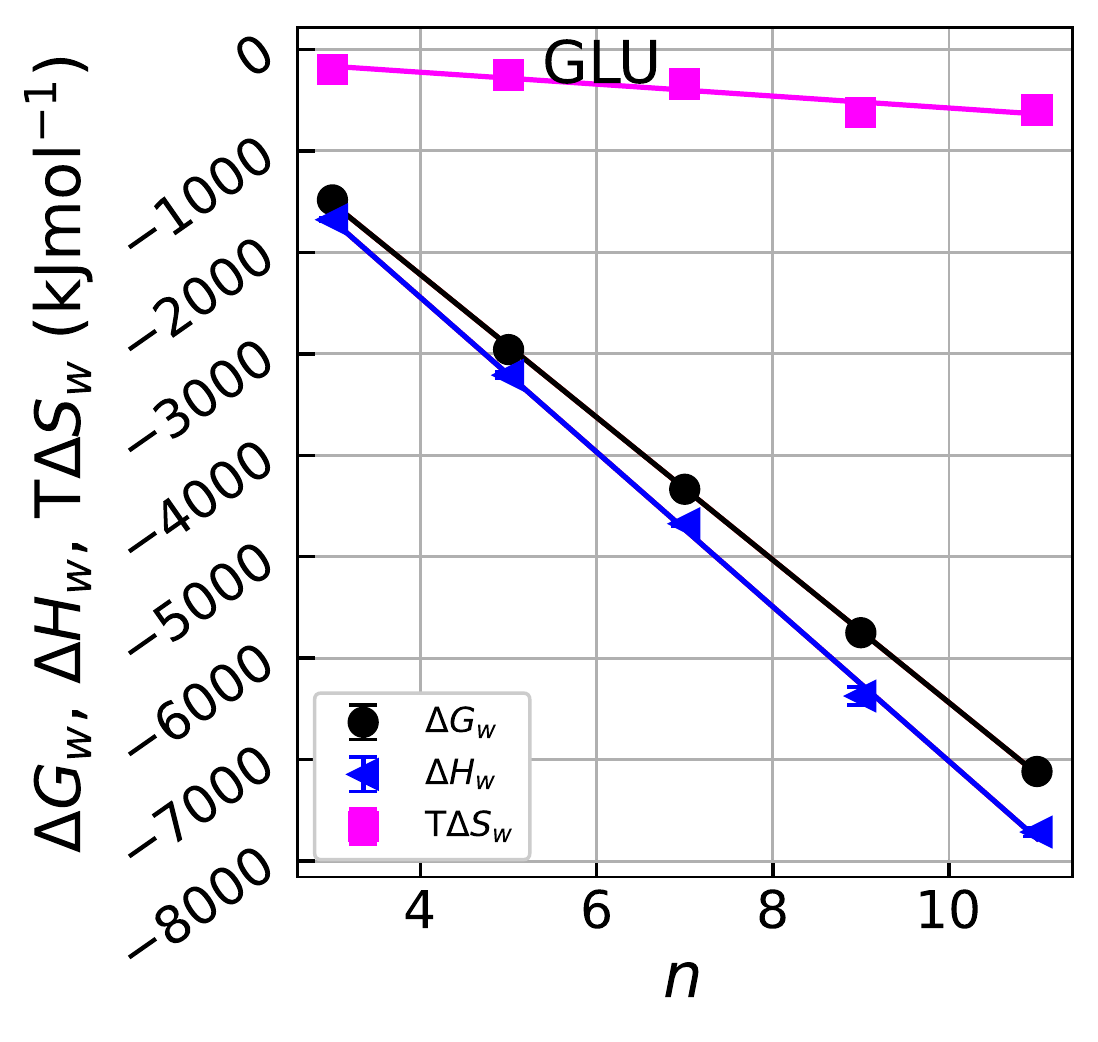}
   \caption{}\label{fig:fig8g}
   \end{subfigure}
  \begin{subfigure}{0.20\textwidth}
     \includegraphics[width=\linewidth]{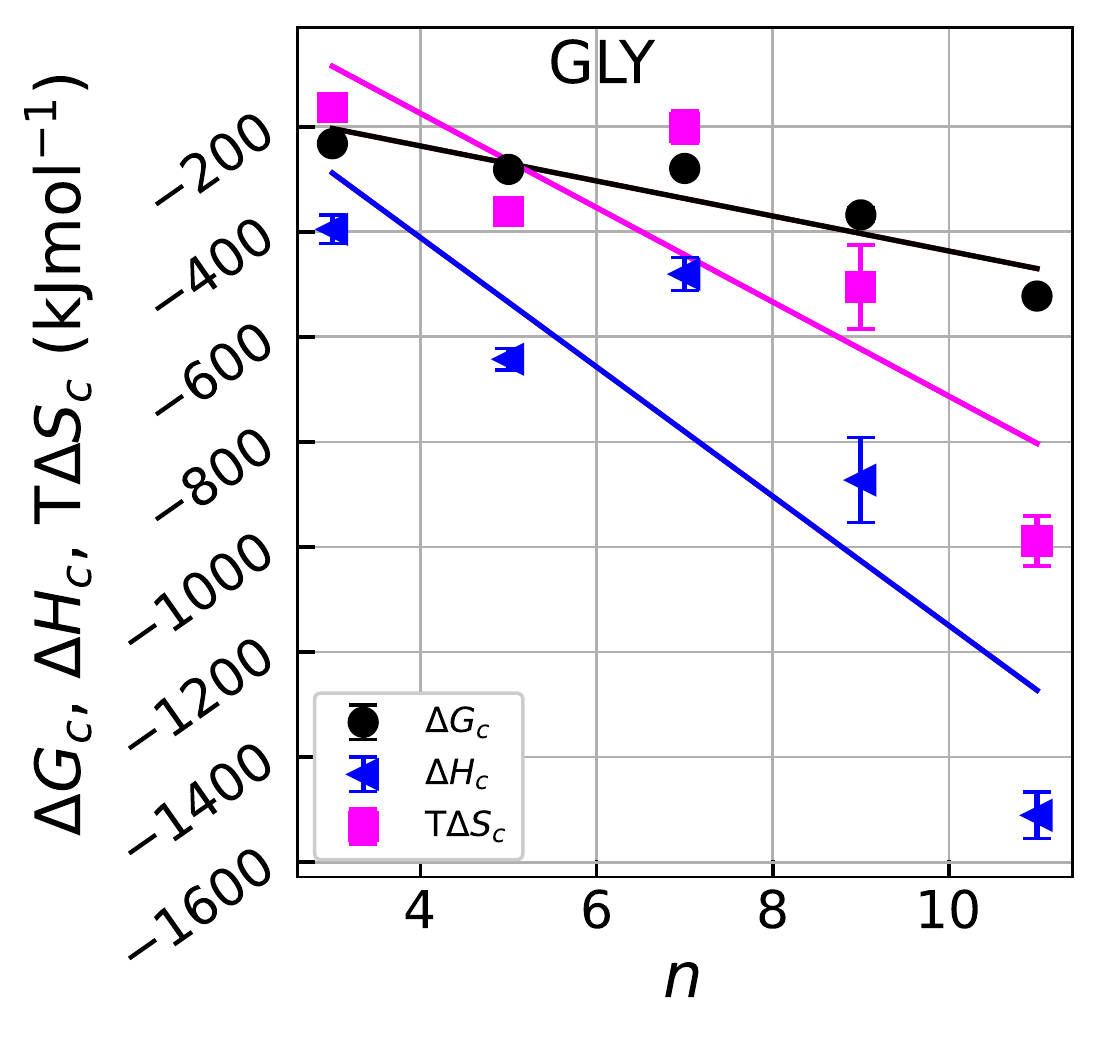}
    \caption{}\label{fig:fig8h}
  \end{subfigure}
  \begin{subfigure}{0.20\textwidth}
     \includegraphics[width=\linewidth]{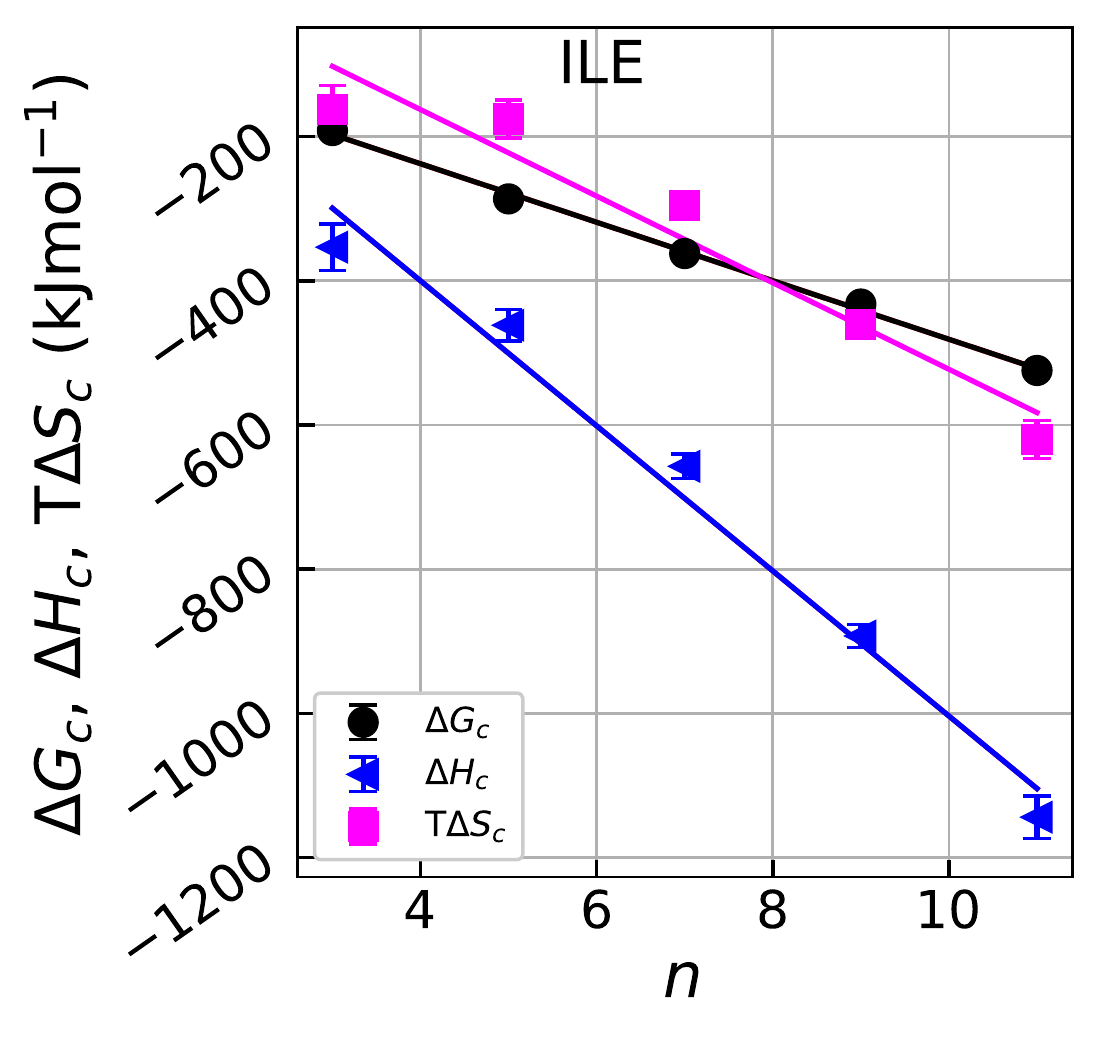}
    \caption{}\label{fig:fig8i}
  \end{subfigure}
  \begin{subfigure}{0.20\textwidth}
     \includegraphics[width=\linewidth]{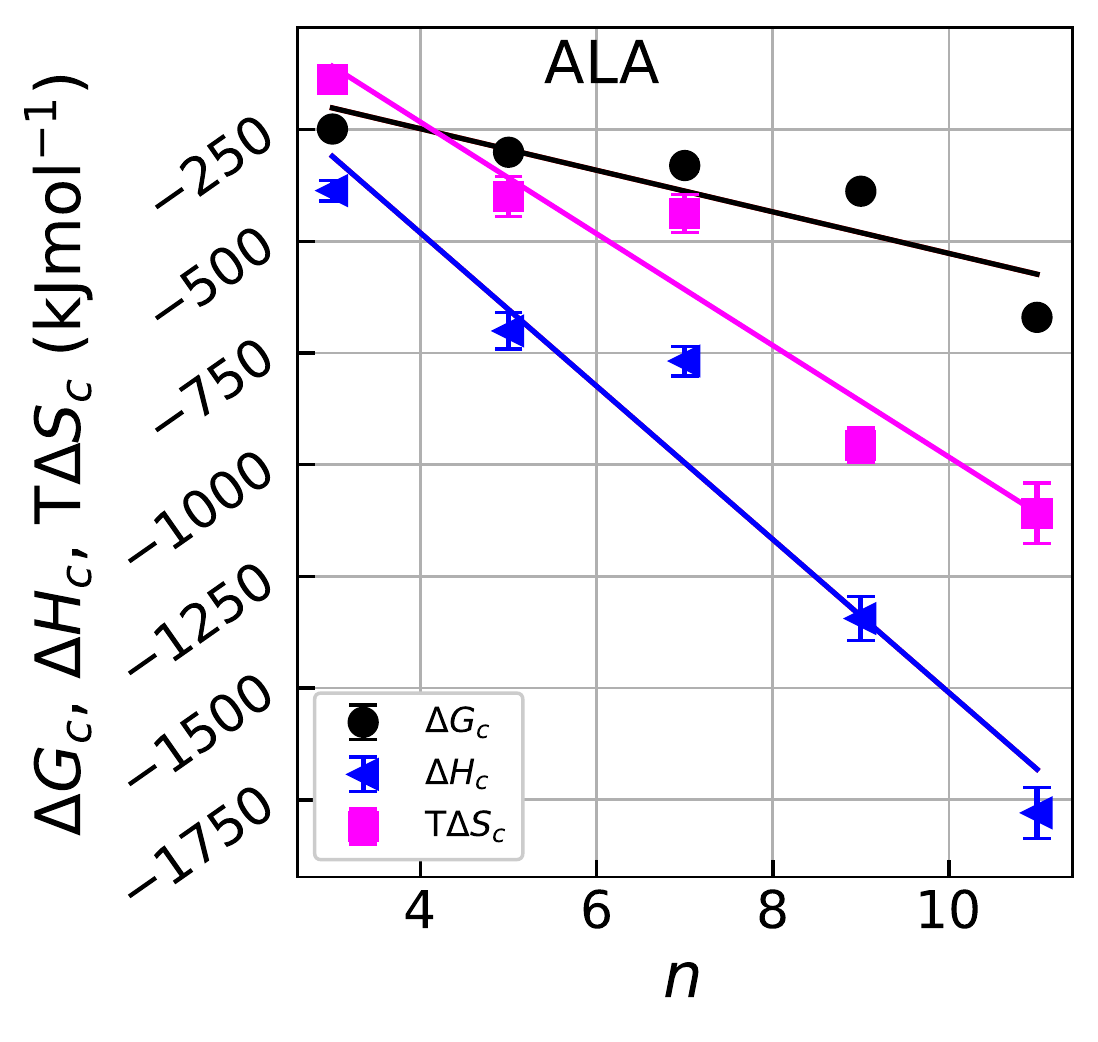}
    \caption{}\label{fig:fig8j}
    \end{subfigure}
    \begin{subfigure}{0.20\textwidth}
     \includegraphics[width=\linewidth]{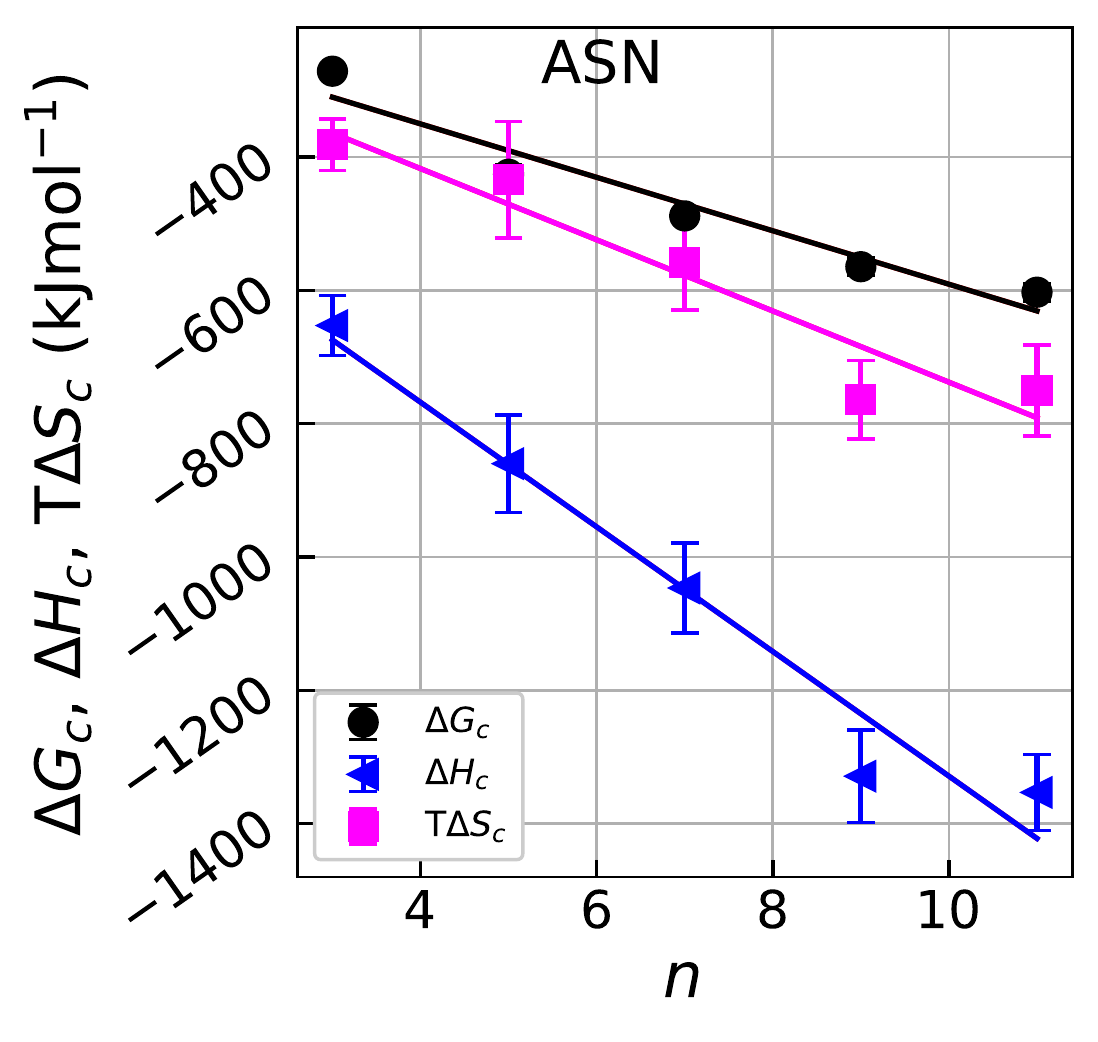}
     \caption{}\label{fig:fig8k}
    \end{subfigure}
    \begin{subfigure}{0.20\textwidth}
     \includegraphics[width=\linewidth]{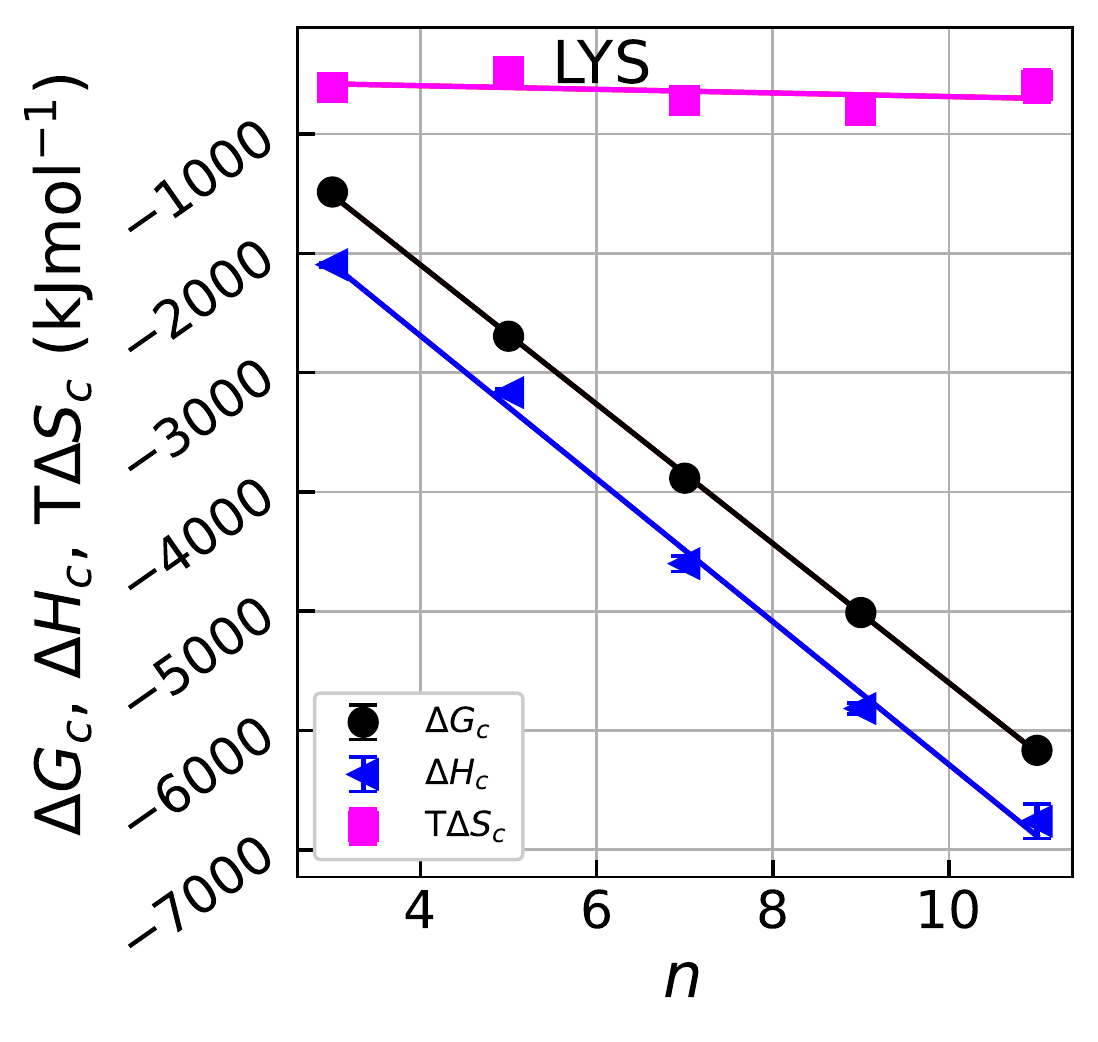}
   \caption{}\label{fig:fig8l}
   \end{subfigure} 
   \begin{subfigure}{0.20\linewidth}
     \includegraphics[width=\linewidth]{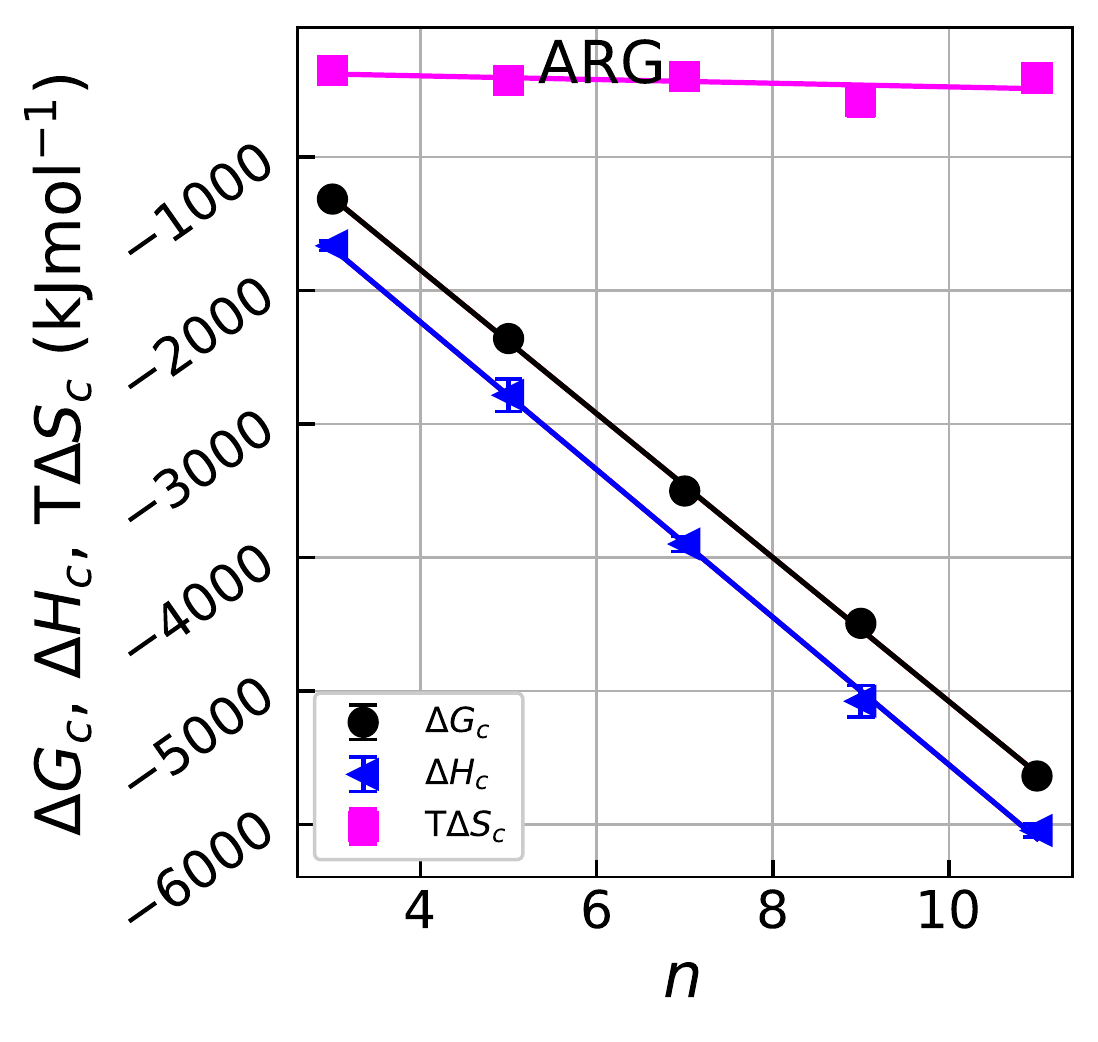}
   \caption{}\label{fig:fig8m}
   \end{subfigure}
   \begin{subfigure}{0.20\linewidth}
     \includegraphics[width=\linewidth]{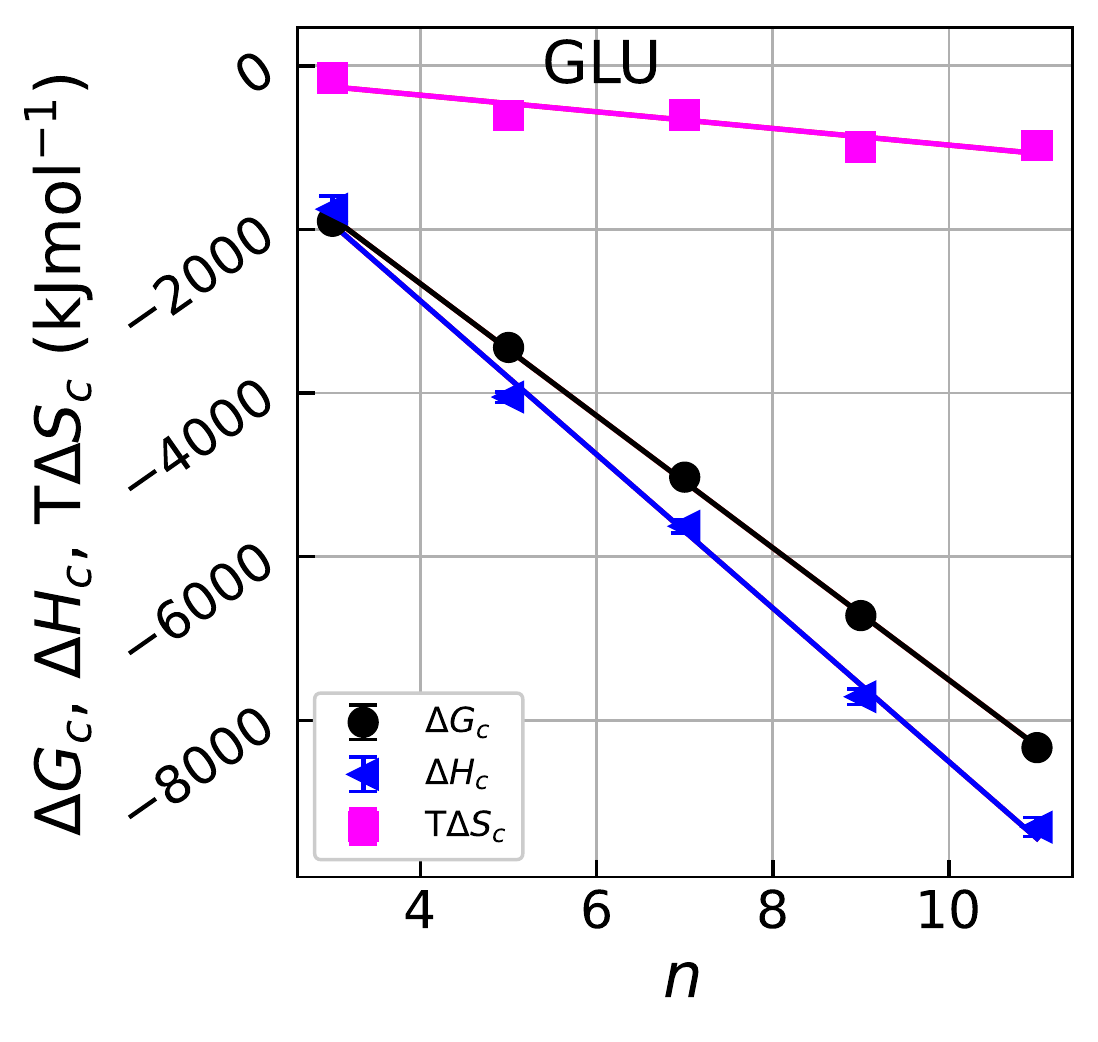}
   \caption{}\label{fig:fig8n}
   \end{subfigure}
   \caption{Solvation free energy, enthalpy, and entropy ($\Delta G $, $\Delta H $, $T\Delta$) changes with the polymer chain length \textit{n} in \ce{H2O} and in cyclohexane \ce{cC6H12} at $25^{\circ}$C for each of the polypeptides considered in this work (GLY, ALA, ILE,  ASN, LYS, ARG, GLU). The continuous lines connecting the points are the representative linear fitting. The data representing the hydration thermodynamics in water \ce{H2O} are shown from (a) to (g), whilst those corresponding to cyclohexane \ce{cC6H12} are plotted from (h) to (n) for each of the polypeptides, respectively. Negative $\Delta G$ and $\Delta H$ represent an energetic gain upon solvation, whereas a negative $T \Delta S$ represent an entropic loss, upon solvation.}
  \label{fig:fig8}
\end{figure*}

Fig.\ref{fig:fig8} reports our results for $\Delta G$ (black circles), and it includes also the corresponding dependence of $\Delta H$ (blue triangles) and $T\Delta S$ (magenta squares), in water \ce{H2O} panels (a)-(g) and in cyclohexane \ce{cC6H12} panels (h)-(n). In all cases solid lines represent a linear fit. Note that $T \Delta S$ and $\Delta H$ both decrease as a function of $n$ indicating an enthalpic gain and an entropic loss. Fig.\ref{fig:fig5} already suggested a linear dependence on $n$ of both $\Delta G_{w}$ and $\Delta G_{c}$ for all consider polypeptides. This is indeed confirmed by Fig.\ref{fig:fig8} (black lines) but with different slopes, smaller for the hydrophobic polypeptides (GLY, ALA, ILE, top three row panels (a) to (c) for water \ce{H2O} and (h) to (j) for cyclohexane \ce{cC6H12}), as well as for ASN ((d) for water \ce{H2O} and (k) for cyclohexane \ce{cC6H12}), larger in all cases for the polar polypeptides (LYS, ARG, GLU) (lower four panels (e) to (g) in water \ce{H2O} and (l) to (n) in cyclohexane. Upon splitting in the enthalpic and entropic terms, reveals however a rather different weight of the two contributions in the different cases. For GLY, ALA, ILE and ASN, the relative weights of $\Delta H$ and $T \Delta S$ appears to be comparable and results into a weak increase of $\Delta G_{w}$ and $\Delta G_{c}$ as a function of $n$ (Figs. \ref{fig:fig8a}-\ref{fig:fig8d}), in agreement with the findings of Fig.\ref{fig:fig5}. By contrast, LYS, ARG, GLU have a much stronger $n$ dependence stemming from $\Delta H$ as is clearly visible in Figs. \ref{fig:fig8e}-\ref{fig:fig8g}, so its additivity is purely enthalpically driven. While this is clearly consistent with the different trends observed in the enthalpy-entropy plots of Fig.\ref{fig:fig7e}-\ref{fig:fig7g}, the very similar behaviour in water \ce{H2O} and cyclohexane \ce{cC6H12} is rather surprising and it will require further analysis that are planned in the future. 

{\color{black}Another related relevant issues concerns the relation with past results referring to the solvation free energy $\Delta G_1$ for a single amino acid side chain equivalent \cite{Dongmo2020}, that is a single amino acid with the backbone part replaced with a single hydrogen atom. We show this analysis in \sfref{supp-fig:solvG_n-nsolvG} where $\Delta G_n$ is plotted versus $n \times \Delta G_1$ both in water \ce{H2O} and in cyclohexane \ce{cC6H12} for all considered polypeptides, with the exception of GLY for which there is clearly no amino acid side chain equivalent since it does not have a proper side chain. The results highlights rather clearly the importance of the backbone in particular for ALA and ILE for which a significant deviation from the naive expectation $\Delta G_n \propto n (\Delta G_1)$ is observed. Again, this is consistent with the relevant role of the backbone in the case of nominally hydrophobic polypeptides. }

\section{Conclusions}
\label{sec:conclusions}
In this paper, we have addressed the issue of "good" and "poor" solvents in the framework of polypeptides of different polarities, both hydrophobic and polar, and including polyglycines as a reference point. Here the definition of good and poor solvent refers to the common view of " like dissolves like": polar solutes dissolve in polar solvents and hydrophobic solutes dissolve in hydrophobic (apolar) solvents. Polar solvents have typically large dipole moments and high dielectric constants, a feature that can be easily rationalized by the fact that high dielectric constants favour the tendency to dissociate and hence forming dipoles. A paradigmatic example of polar solvent is water (dielectric constant $\approx 80$), and hence polar solvents typically mix with water. As a representative example of hydrophobic solvent, we have considered cyclohexane that has dielectric constant $\approx 2$ and hence can be considered at the opposite end of water.
A similar reasoning applies to solutes that can be classified in polar and hydrophobic based on the same rationale as the solvent. Hence good and poor solvents are to be defined with respect to a specific solute. A fully hydrophobic  polypeptide is expected to collapse in water (water is a poor solvent), but it remains extended in cyclohexane (cyclohexane is a then good solvent).
Conversely, a fully polar polypeptide usually folds in its own poor solvent such as cyclohexane, while remaining extended in its own good solvent such as water.
Hence, the solvent quality always requires the solvent polarity to be unambiguously defined. This is especially important for polypeptides as they are formed by an identical backbone part, plus a sum of single side chains that provide their hydrophobic/polar character. While the polarity character is usually attributed to the side chains on the basis of their chemical characters, a much more robust indication is given by their solvation free energies in solvents with different polarities, and we have studied their properties in the present paper. 

Our main findings can be summarized as follows.
\begin{itemize}
    \item[1)] {\color{black} There is no general mirror symmetry between the behaviour of hydrophobic/polar polypeptides in water/cyclohexane, due to the presence of the backbone, as well as of the different energy scales involved. Hence hydrophobic polypeptides in water do not behave as polar peptides in cyclohexane, nor the other way around. Polyglycine is formed by $n$ different residues having a single hydrogen atom as a side chain, and it is usually regarded as a rough model for the peptide backbone of a protein. We find that it collapses in water and it remains extended in cyclohexane, so water is a poor solvent for polyglycine (in line with past studies), and cyclohexane a good one.  Accordingly, the solvation free energy in cyclohexane $\Delta G_{c}$ is negative and decreases approximately linearly with the number $n$ of residues. Interestingly, a similar trend is also found for the solvation free energy in water $\Delta G_{w}$ with the transfer free energy $\Delta \Delta G_{w>c}$ negative and decreasing with the length of the polypeptide. Additional hydrophobic polypeptides, such as alanine ALA and isoleucine ILE, behave similarly to GLY, with some small differences. These results can be rationalized as follows. Firstly, none of these polypeptides are really hydrophobic irrespective of the polarities of the single side chain. This is evident since they all form at least 2-3 hydrogen bonds/residue. Secondly, the solvation free energy is composed by an entropically unfavourable term associated with the creation of a cavity, and an enthalpic favourable term originating upon the insertion of the polypeptides in the solvent. Our results indicate that the latter always dominate the former leading to negative solvation free energies. }
    \item[2)]  Polar polypeptides such as ASN$_n$, LYS$_n$, ARG$_n$, and GLU$_n$ markedly deviate from the mirror symmetry. GLU$_n$ collapses in water but not in cyclohexane, whereas both water and cyclohexane are good solvents for LYS$_n$. Accordingly, the transfer free energy $\Delta \Delta G_{w>c}$ from water to cyclohexane is found negative, linearly decreasing with the number $n$ of residues, and significantly more negative than the hydrophobic counterparts.  LYS$_n$, ARG$_n$, and GLU$_n$ are mostly enthalpically driven, whereas in ASN$_n$, as well as all the hydrophobic polypeptides, the driving force is a mixture of enthalpic and entropic contributions. {\color{black} These results suggest that the solvation process is mainly dominated by the polarity of the solute, with the solvent playing a minor role. } 
    \item[3)] For all hydrophobic polypeptides as well as for ANS$_n$, there is nearly a similar entropy-enthalpy compensation in both water and cyclohexane, whereas for the other polar polypeptides LYS$_n$, ARG$_n$, and GLU$_n$ there is a marked difference. Combined with previous point, this shows that ANS$_n$ hardly belongs to the same class as LYS$_n$, ARG$_n$, and GLU$_n$, and more generally that the rough polar/hydrophobic division of the amino acids scale is not representing well the complexity of the interactions, and additional features (e.g. charge, size, etc.) should be taken into account. The peculiar properties of ASN$_n$ reported throughout this study might also be related to its marked propensity together with aspartic acid ASP to populate loop regions in protein structures thus most often with no defined secondary structure \cite{Skrbic2021}.
\end{itemize}

While the present work is focused specifically on the solvation process of polypeptides and its dependence on both the solvent and peptide polarities, a similar study has been tackled by the present authors also for a specific synthetic polymer displaying a coil-helix transition and it will be presented elsewhere. Coupled with the present findings, the general scenario presents still some missing points requiring further studies. One promising route that has been already addressed in past studies \cite{Lai2012}, is the quantification of the individual entropic and enthalpic solute-solvent and solvent-solvent contributions, thus allowing a quantitative assessment on the exact putative cancellation of the solvent-solvent enthalpy and solvent-solvent entropy in water and not in cyclohexane.  We are planning to explore this possibility in a future dedicated study. 
All together it is hoped that a systematic analysis as those outlined above will provide new insights on the nuances solvation mechanism in different solvents, a process which is ubiquitous in biological systems.  
\section*{Conflicts of interest}
There are no conflicts to declare.

\section*{Acknowledgements}
The authors would like to acknowledge useful discussions and correspondence with Giuseppe Graziano, George Rose and Chris Oostenbrink, as well as networking support by the COST Action CA17139.
The use of the SCSCF multiprocessor cluster at  the Universit\`{a} Ca' Foscari Venezia is gratefully acknowledged. We also acknowledge the CINECA
projects HP10CGFUDT and HP10C1XOOJ for the availability of high performance computing resources through the ISCRA initiative. The work was supported by MIUR PRIN-COFIN2017 \textit{Soft Adaptive Networks} grant 2017Z55KCW (A.G). 

\clearpage

\begin{mcitethebibliography}{44}
\providecommand*{\natexlab}[1]{#1}
\providecommand*{\mciteSetBstSublistMode}[1]{}
\providecommand*{\mciteSetBstMaxWidthForm}[2]{}
\providecommand*{\mciteBstWouldAddEndPuncttrue}
  {\def\EndOfBibitem{\unskip.}}
\providecommand*{\mciteBstWouldAddEndPunctfalse}
  {\let\EndOfBibitem\relax}
\providecommand*{\mciteSetBstMidEndSepPunct}[3]{}
\providecommand*{\mciteSetBstSublistLabelBeginEnd}[3]{}
\providecommand*{\EndOfBibitem}{}
\mciteSetBstSublistMode{f}
\mciteSetBstMaxWidthForm{subitem}
{(\emph{\alph{mcitesubitemcount}})}
\mciteSetBstSublistLabelBeginEnd{\mcitemaxwidthsubitemform\space}
{\relax}{\relax}

\bibitem[Flory(1969)]{Flory69}
P.~Flory, \emph{Statistical mechanics of chain molecules}, Interscience
  Publishers, 1969\relax
\mciteBstWouldAddEndPuncttrue
\mciteSetBstMidEndSepPunct{\mcitedefaultmidpunct}
{\mcitedefaultendpunct}{\mcitedefaultseppunct}\relax
\EndOfBibitem
\bibitem[Doi and Edwards(1988)]{Doi88}
M.~Doi and S.~F. Edwards, \emph{The Theory of Polymer Dynamics (International
  Series of Monographs on Physics)}, Clarendon Press, 1988\relax
\mciteBstWouldAddEndPuncttrue
\mciteSetBstMidEndSepPunct{\mcitedefaultmidpunct}
{\mcitedefaultendpunct}{\mcitedefaultseppunct}\relax
\EndOfBibitem
\bibitem[Khokhlov \emph{et~al.}(2002)Khokhlov, Grosberg, and Pande]{Khokhlov02}
A.~R. Khokhlov, A.~Y. Grosberg and V.~S. Pande, \emph{Statistical Physics of
  Macromolecules (Polymers and Complex Materials)}, American Institute of
  Physics, 1994th edn, 2002\relax
\mciteBstWouldAddEndPuncttrue
\mciteSetBstMidEndSepPunct{\mcitedefaultmidpunct}
{\mcitedefaultendpunct}{\mcitedefaultseppunct}\relax
\EndOfBibitem
\bibitem[Rubinstein and Colby(2003)]{Rubinstein03}
M.~Rubinstein and R.~H. Colby, \emph{Polymer Physics (Chemistry)}, Oxford
  University Press, 1st edn, 2003\relax
\mciteBstWouldAddEndPuncttrue
\mciteSetBstMidEndSepPunct{\mcitedefaultmidpunct}
{\mcitedefaultendpunct}{\mcitedefaultseppunct}\relax
\EndOfBibitem
\bibitem[de~Gennes(1979)]{deGennes79}
P.~de~Gennes, \emph{Scaling Concepts in Polymer Physics}, Cornell University
  Press, 1979\relax
\mciteBstWouldAddEndPuncttrue
\mciteSetBstMidEndSepPunct{\mcitedefaultmidpunct}
{\mcitedefaultendpunct}{\mcitedefaultseppunct}\relax
\EndOfBibitem
\bibitem[Bhattacharjee \emph{et~al.}({2013})Bhattacharjee, Giacometti, and
  Maritan]{Bhattacharjee13}
S.~M. Bhattacharjee, A.~Giacometti and A.~Maritan, \emph{{JOURNAL OF
  PHYSICS-CONDENSED MATTER}}, {2013}, \textbf{{25}}, {503101/1--15}\relax
\mciteBstWouldAddEndPuncttrue
\mciteSetBstMidEndSepPunct{\mcitedefaultmidpunct}
{\mcitedefaultendpunct}{\mcitedefaultseppunct}\relax
\EndOfBibitem
\bibitem[Bolen and Rose(2008)]{Bolen2008}
D.~W. Bolen and G.~D. Rose, \emph{Annu. Rev. Biochem.}, 2008, \textbf{77},
  339--362\relax
\mciteBstWouldAddEndPuncttrue
\mciteSetBstMidEndSepPunct{\mcitedefaultmidpunct}
{\mcitedefaultendpunct}{\mcitedefaultseppunct}\relax
\EndOfBibitem
\bibitem[Wolynes(1995)]{Wolynes1995}
P.~G. Wolynes, \emph{Proceedings of the National Academy of Sciences}, 1995,
  \textbf{92}, 2426--2427\relax
\mciteBstWouldAddEndPuncttrue
\mciteSetBstMidEndSepPunct{\mcitedefaultmidpunct}
{\mcitedefaultendpunct}{\mcitedefaultseppunct}\relax
\EndOfBibitem
\bibitem[Meyer \emph{et~al.}(2013)Meyer, Gabelica, Grubmüller, and
  Orozco]{Meyer2013}
T.~Meyer, V.~Gabelica, H.~Grubmüller and M.~Orozco, \emph{WIREs Computational
  Molecular Science}, 2013, \textbf{3}, 408--425\relax
\mciteBstWouldAddEndPuncttrue
\mciteSetBstMidEndSepPunct{\mcitedefaultmidpunct}
{\mcitedefaultendpunct}{\mcitedefaultseppunct}\relax
\EndOfBibitem
\bibitem[Carrer \emph{et~al.}(2020)Carrer, Skrbic, Bore, Milano, Cascella, and
  Giacometti]{Carrer20}
M.~Carrer, T.~Skrbic, S.~L. Bore, G.~Milano, M.~Cascella and A.~Giacometti,
  \emph{The Journal of Physical Chemistry B}, 2020, \textbf{124},
  6448--6458\relax
\mciteBstWouldAddEndPuncttrue
\mciteSetBstMidEndSepPunct{\mcitedefaultmidpunct}
{\mcitedefaultendpunct}{\mcitedefaultseppunct}\relax
\EndOfBibitem
\bibitem[Hayashi \emph{et~al.}(2017)Hayashi, Yasuda, {\v{S}}krbi{\'c},
  Giacometti, and Kinoshita]{Hayashi17}
T.~Hayashi, S.~Yasuda, T.~{\v{S}}krbi{\'c}, A.~Giacometti and M.~Kinoshita,
  \emph{The Journal of Chemical Physics}, 2017, \textbf{147}, 125102\relax
\mciteBstWouldAddEndPuncttrue
\mciteSetBstMidEndSepPunct{\mcitedefaultmidpunct}
{\mcitedefaultendpunct}{\mcitedefaultseppunct}\relax
\EndOfBibitem
\bibitem[Hayashi \emph{et~al.}(2018)Hayashi, Inoue, Yasuda, Petretto,
  {\v{S}}krbi{\'c}, Giacometti, and Kinoshita]{Hayashi18}
T.~Hayashi, M.~Inoue, S.~Yasuda, E.~Petretto, T.~{\v{S}}krbi{\'c},
  A.~Giacometti and M.~Kinoshita, \emph{The Journal of Chemical Physics}, 2018,
  \textbf{149}, 045105\relax
\mciteBstWouldAddEndPuncttrue
\mciteSetBstMidEndSepPunct{\mcitedefaultmidpunct}
{\mcitedefaultendpunct}{\mcitedefaultseppunct}\relax
\EndOfBibitem
\bibitem[Karandur \emph{et~al.}(2014)Karandur, Wong, and Pettitt]{Karandur2014}
D.~Karandur, K.-Y. Wong and B.~M. Pettitt, \emph{The Journal of Physical
  Chemistry B}, 2014, \textbf{118}, 9565--9572\relax
\mciteBstWouldAddEndPuncttrue
\mciteSetBstMidEndSepPunct{\mcitedefaultmidpunct}
{\mcitedefaultendpunct}{\mcitedefaultseppunct}\relax
\EndOfBibitem
\bibitem[Dongmo~Foumthuim \emph{et~al.}(2020)Dongmo~Foumthuim, Carrer, Houvet,
  Škrbić, Graziano, and Giacometti]{Dongmo2020}
C.~J. Dongmo~Foumthuim, M.~Carrer, M.~Houvet, T.~Škrbić, G.~Graziano and
  A.~Giacometti, \emph{Phys. Chem. Chem. Phys.}, 2020, \textbf{22},
  25848--25858\relax
\mciteBstWouldAddEndPuncttrue
\mciteSetBstMidEndSepPunct{\mcitedefaultmidpunct}
{\mcitedefaultendpunct}{\mcitedefaultseppunct}\relax
\EndOfBibitem
\bibitem[Wolfenden \emph{et~al.}({2015})Wolfenden, Lewis, Yuan, and
  Carter]{Wolfenden15}
R.~Wolfenden, C.~A. Lewis, Y.~Yuan and C.~W. Carter, \emph{{Proceedings of the
  National Academy of Sciences of the United States of America}}, {2015},
  \textbf{{112}}, {7484--7488}\relax
\mciteBstWouldAddEndPuncttrue
\mciteSetBstMidEndSepPunct{\mcitedefaultmidpunct}
{\mcitedefaultendpunct}{\mcitedefaultseppunct}\relax
\EndOfBibitem
\bibitem[Hajari and van~der Vegt({2015})]{Hajari15}
T.~Hajari and N.~F.~A. van~der Vegt, \emph{{Journal of Chemical Physics}},
  {2015}, \textbf{{142}}, {144502}\relax
\mciteBstWouldAddEndPuncttrue
\mciteSetBstMidEndSepPunct{\mcitedefaultmidpunct}
{\mcitedefaultendpunct}{\mcitedefaultseppunct}\relax
\EndOfBibitem
\bibitem[Chandler(2005)]{Chandler2005}
D.~Chandler, \emph{Nature}, 2005, \textbf{437}, 640--647\relax
\mciteBstWouldAddEndPuncttrue
\mciteSetBstMidEndSepPunct{\mcitedefaultmidpunct}
{\mcitedefaultendpunct}{\mcitedefaultseppunct}\relax
\EndOfBibitem
\bibitem[Voet and Voet(2010)]{Voet2010}
D.~Voet and J.~G. Voet, \emph{Biochemistry}, John Wiley \& Sons, 2010\relax
\mciteBstWouldAddEndPuncttrue
\mciteSetBstMidEndSepPunct{\mcitedefaultmidpunct}
{\mcitedefaultendpunct}{\mcitedefaultseppunct}\relax
\EndOfBibitem
\bibitem[Tomar \emph{et~al.}(2013)Tomar, Asthagiri, and Weber]{Tomar2013}
D.~S. Tomar, D.~Asthagiri and V.~Weber, \emph{Biophysical Journal}, 2013,
  \textbf{105}, 1482--1490\relax
\mciteBstWouldAddEndPuncttrue
\mciteSetBstMidEndSepPunct{\mcitedefaultmidpunct}
{\mcitedefaultendpunct}{\mcitedefaultseppunct}\relax
\EndOfBibitem
\bibitem[Avbelj and Baldwin(2009)]{Avbelj2009}
F.~Avbelj and R.~L. Baldwin, \emph{Proceedings of the National Academy of
  Sciences}, 2009, \textbf{106}, 3137--3141\relax
\mciteBstWouldAddEndPuncttrue
\mciteSetBstMidEndSepPunct{\mcitedefaultmidpunct}
{\mcitedefaultendpunct}{\mcitedefaultseppunct}\relax
\EndOfBibitem
\bibitem[Kokubo \emph{et~al.}(2013)Kokubo, Harris, Asthagiri, and
  Pettitt]{Kokubo2013}
H.~Kokubo, R.~C. Harris, D.~Asthagiri and B.~M. Pettitt, \emph{The Journal of
  Physical Chemistry B}, 2013, \textbf{117}, 16428--16435\relax
\mciteBstWouldAddEndPuncttrue
\mciteSetBstMidEndSepPunct{\mcitedefaultmidpunct}
{\mcitedefaultendpunct}{\mcitedefaultseppunct}\relax
\EndOfBibitem
\bibitem[Staritzbichler \emph{et~al.}(2005)Staritzbichler, Gu, and
  Helms]{Staritzbichler2005}
R.~Staritzbichler, W.~Gu and V.~Helms, \emph{The Journal of Physical Chemistry
  B}, 2005, \textbf{109}, 19000--19007\relax
\mciteBstWouldAddEndPuncttrue
\mciteSetBstMidEndSepPunct{\mcitedefaultmidpunct}
{\mcitedefaultendpunct}{\mcitedefaultseppunct}\relax
\EndOfBibitem
\bibitem[König \emph{et~al.}(2010)König, Bruckner, and Boresch]{Konig2013}
G.~König, S.~Bruckner and S.~Boresch, \emph{Biophysical Journal}, 2010,
  \textbf{104}, 453--463\relax
\mciteBstWouldAddEndPuncttrue
\mciteSetBstMidEndSepPunct{\mcitedefaultmidpunct}
{\mcitedefaultendpunct}{\mcitedefaultseppunct}\relax
\EndOfBibitem
\bibitem[Hu \emph{et~al.}(2010)Hu, Kokubo, Lynch, Bolen, and Pettitt]{Char2010}
C.~Y. Hu, H.~Kokubo, G.~C. Lynch, D.~W. Bolen and B.~M. Pettitt, \emph{Protein
  Science}, 2010, \textbf{19}, 1011--1022\relax
\mciteBstWouldAddEndPuncttrue
\mciteSetBstMidEndSepPunct{\mcitedefaultmidpunct}
{\mcitedefaultendpunct}{\mcitedefaultseppunct}\relax
\EndOfBibitem
\bibitem[Frenkel and Smit(2001)]{Frenkel01}
D.~Frenkel and B.~Smit, \emph{Understanding Molecular Simulation, Second
  Edition: From Algorithms to Applications (Computational Science Series, Vol
  1)}, Academic Press, 2nd edn, 2001\relax
\mciteBstWouldAddEndPuncttrue
\mciteSetBstMidEndSepPunct{\mcitedefaultmidpunct}
{\mcitedefaultendpunct}{\mcitedefaultseppunct}\relax
\EndOfBibitem
\bibitem[Fogolari \emph{et~al.}(2016)Fogolari, Dongmo~Foumthuim, Fortuna,
  Soler, Corazza, and Esposito]{Fogolari2016}
F.~Fogolari, C.~J. Dongmo~Foumthuim, S.~Fortuna, M.~A. Soler, A.~Corazza and
  G.~Esposito, \emph{Journal of Chemical Theory and Computation}, 2016,
  \textbf{12}, 1--8\relax
\mciteBstWouldAddEndPuncttrue
\mciteSetBstMidEndSepPunct{\mcitedefaultmidpunct}
{\mcitedefaultendpunct}{\mcitedefaultseppunct}\relax
\EndOfBibitem
\bibitem[Fogolari \emph{et~al.}(2018)Fogolari, Maloku, Dongmo~Foumthuim,
  Corazza, and Esposito]{Fogolari2018}
F.~Fogolari, O.~Maloku, C.~J. Dongmo~Foumthuim, A.~Corazza and G.~Esposito,
  \emph{Journal of Chemical Information and Modeling}, 2018, \textbf{58},
  1319--1324\relax
\mciteBstWouldAddEndPuncttrue
\mciteSetBstMidEndSepPunct{\mcitedefaultmidpunct}
{\mcitedefaultendpunct}{\mcitedefaultseppunct}\relax
\EndOfBibitem
\bibitem[Lai and Oostenbrink(2012)]{Lai2012}
B.~Lai and C.~Oostenbrink, \emph{Theoretical Chemistry Accounts}, 2012,
  \textbf{131}, 1--13\relax
\mciteBstWouldAddEndPuncttrue
\mciteSetBstMidEndSepPunct{\mcitedefaultmidpunct}
{\mcitedefaultendpunct}{\mcitedefaultseppunct}\relax
\EndOfBibitem
\bibitem[Škrbić \emph{et~al.}(2021)Škrbić, Maritan, Giacometti, and
  Banavar]{Skrbic2021}
T.~Škrbić, A.~Maritan, A.~Giacometti and J.~R. Banavar, \emph{Protein
  Science}, 2021, \textbf{30}, 818--829\relax
\mciteBstWouldAddEndPuncttrue
\mciteSetBstMidEndSepPunct{\mcitedefaultmidpunct}
{\mcitedefaultendpunct}{\mcitedefaultseppunct}\relax
\EndOfBibitem
\bibitem[Hanwell \emph{et~al.}(2012)Hanwell, Curtis, Lonie, Vandermeersch,
  Zurek, and Hutchison]{Hanwell2012}
M.~D. Hanwell, D.~E. Curtis, D.~C. Lonie, T.~Vandermeersch, E.~Zurek and G.~R.
  Hutchison, \emph{Journal of Cheminformatics}, 2012, \textbf{4}, 1--17\relax
\mciteBstWouldAddEndPuncttrue
\mciteSetBstMidEndSepPunct{\mcitedefaultmidpunct}
{\mcitedefaultendpunct}{\mcitedefaultseppunct}\relax
\EndOfBibitem
\bibitem[Schmid \emph{et~al.}(2011)Schmid, Eichenberger, Choutko, Riniker,
  Winger, Mark, and van Gunsteren]{Schmid11}
N.~Schmid, A.~P. Eichenberger, A.~Choutko, S.~Riniker, M.~Winger, A.~E. Mark
  and W.~F. van Gunsteren, \emph{European Biophysics Journal}, 2011,
  \textbf{40}, 843\relax
\mciteBstWouldAddEndPuncttrue
\mciteSetBstMidEndSepPunct{\mcitedefaultmidpunct}
{\mcitedefaultendpunct}{\mcitedefaultseppunct}\relax
\EndOfBibitem
\bibitem[Oostenbrink \emph{et~al.}(2004)Oostenbrink, Villa, Mark, and
  Van~Gunsteren]{Oostenbrink2004}
C.~Oostenbrink, A.~Villa, A.~E. Mark and W.~F. Van~Gunsteren, \emph{Journal of
  computational chemistry}, 2004, \textbf{25}, 1656--1676\relax
\mciteBstWouldAddEndPuncttrue
\mciteSetBstMidEndSepPunct{\mcitedefaultmidpunct}
{\mcitedefaultendpunct}{\mcitedefaultseppunct}\relax
\EndOfBibitem
\bibitem[Villa and Mark({2002})]{Villa02}
A.~Villa and A.~Mark, \emph{{Journal of Computational Chemistry}}, {2002},
  \textbf{{23}}, {548--553}\relax
\mciteBstWouldAddEndPuncttrue
\mciteSetBstMidEndSepPunct{\mcitedefaultmidpunct}
{\mcitedefaultendpunct}{\mcitedefaultseppunct}\relax
\EndOfBibitem
\bibitem[Reif \emph{et~al.}(2012)Reif, Hünenberger, and Oostenbrink]{Reif2012}
M.~M. Reif, P.~H. Hünenberger and C.~Oostenbrink, \emph{Journal of Chemical
  Theory and Computation}, 2012, \textbf{8}, 3705--3723\relax
\mciteBstWouldAddEndPuncttrue
\mciteSetBstMidEndSepPunct{\mcitedefaultmidpunct}
{\mcitedefaultendpunct}{\mcitedefaultseppunct}\relax
\EndOfBibitem
\bibitem[Shirts \emph{et~al.}(2003)Shirts, Pitera, Swope, and
  Pande]{Shirts2003}
M.~R. Shirts, J.~W. Pitera, W.~C. Swope and V.~S. Pande, \emph{The Journal of
  Chemical Physics}, 2003, \textbf{119}, 5740--5761\relax
\mciteBstWouldAddEndPuncttrue
\mciteSetBstMidEndSepPunct{\mcitedefaultmidpunct}
{\mcitedefaultendpunct}{\mcitedefaultseppunct}\relax
\EndOfBibitem
\bibitem[Abraham \emph{et~al.}(2015)Abraham, Murtola, Schulz, Páll, Smith,
  Hess, and Lindahl]{Abraham15}
M.~J. Abraham, T.~Murtola, R.~Schulz, S.~Páll, J.~C. Smith, B.~Hess and
  E.~Lindahl, \emph{SoftwareX}, 2015, \textbf{1-2}, 19 -- 25\relax
\mciteBstWouldAddEndPuncttrue
\mciteSetBstMidEndSepPunct{\mcitedefaultmidpunct}
{\mcitedefaultendpunct}{\mcitedefaultseppunct}\relax
\EndOfBibitem
\bibitem[Foumthuim \emph{et~al.}(2018)Foumthuim, J., Corazza, Berni, Esposito,
  and Fogolari]{Dongmo2018}
D.~Foumthuim, C.~J., A.~Corazza, R.~Berni, G.~Esposito and F.~Fogolari,
  \emph{BioMed Research International}, 2018, \textbf{2018}, 1--14\relax
\mciteBstWouldAddEndPuncttrue
\mciteSetBstMidEndSepPunct{\mcitedefaultmidpunct}
{\mcitedefaultendpunct}{\mcitedefaultseppunct}\relax
\EndOfBibitem
\bibitem[Eisenhaber \emph{et~al.}(1995)Eisenhaber, Lijnzaad, Argos, Sander, and
  Scharf]{Eisenhaber1995}
F.~Eisenhaber, P.~Lijnzaad, P.~Argos, C.~Sander and M.~Scharf, \emph{Journal of
  computational chemistry}, 1995, \textbf{16}, 273--284\relax
\mciteBstWouldAddEndPuncttrue
\mciteSetBstMidEndSepPunct{\mcitedefaultmidpunct}
{\mcitedefaultendpunct}{\mcitedefaultseppunct}\relax
\EndOfBibitem
\bibitem[Gong and Rose(2008)]{Gong2008}
H.~Gong and G.~D. Rose, \emph{Proceedings of the National Academy of Sciences},
  2008, \textbf{105}, 3321--3326\relax
\mciteBstWouldAddEndPuncttrue
\mciteSetBstMidEndSepPunct{\mcitedefaultmidpunct}
{\mcitedefaultendpunct}{\mcitedefaultseppunct}\relax
\EndOfBibitem
\bibitem[Tran \emph{et~al.}(2008)Tran, Mao, and Pappu]{Tran2008}
H.~T. Tran, A.~Mao and R.~V. Pappu, \emph{Journal of the American Chemical
  Society}, 2008, \textbf{130}, 7380--7392\relax
\mciteBstWouldAddEndPuncttrue
\mciteSetBstMidEndSepPunct{\mcitedefaultmidpunct}
{\mcitedefaultendpunct}{\mcitedefaultseppunct}\relax
\EndOfBibitem
\bibitem[Karandur \emph{et~al.}(2016)Karandur, Harris, and
  Pettitt]{Karandur2016}
D.~Karandur, R.~C. Harris and B.~M. Pettitt, \emph{Protein Science}, 2016,
  \textbf{25}, 103--110\relax
\mciteBstWouldAddEndPuncttrue
\mciteSetBstMidEndSepPunct{\mcitedefaultmidpunct}
{\mcitedefaultendpunct}{\mcitedefaultseppunct}\relax
\EndOfBibitem
\bibitem[Merlino \emph{et~al.}(2017)Merlino, Pontillo, and
  Graziano]{Merlino2017}
A.~Merlino, N.~Pontillo and G.~Graziano, \emph{Physical Chemistry Chemical
  Physics}, 2017, \textbf{19}, 751--756\relax
\mciteBstWouldAddEndPuncttrue
\mciteSetBstMidEndSepPunct{\mcitedefaultmidpunct}
{\mcitedefaultendpunct}{\mcitedefaultseppunct}\relax
\EndOfBibitem
\bibitem[Pace \emph{et~al.}(2004)Pace, Trevino, Prabhakaran, and
  Scholtz]{Pace04}
C.~N. Pace, S.~Trevino, E.~Prabhakaran and J.~M. Scholtz, \emph{Philosophical
  Transactions of the Royal Society of London B: Biological Sciences}, 2004,
  \textbf{359}, 1225--1235\relax
\mciteBstWouldAddEndPuncttrue
\mciteSetBstMidEndSepPunct{\mcitedefaultmidpunct}
{\mcitedefaultendpunct}{\mcitedefaultseppunct}\relax
\EndOfBibitem
\bibitem[Rose \emph{et~al.}({2006})Rose, Fleming, Banavar, and Maritan]{Rose06}
G.~D. Rose, P.~J. Fleming, J.~R. Banavar and A.~Maritan, \emph{{PROCEEDINGS OF
  THE NATIONAL ACADEMY OF SCIENCES OF THE UNITED STATES OF AMERICA}}, {2006},
  \textbf{{103}}, {16623--16633}\relax
\mciteBstWouldAddEndPuncttrue
\mciteSetBstMidEndSepPunct{\mcitedefaultmidpunct}
{\mcitedefaultendpunct}{\mcitedefaultseppunct}\relax
\EndOfBibitem
\end{mcitethebibliography}

\providecommand*{\mcitethebibliography}{\thebibliography}
\csname @ifundefined\endcsname{endmcitethebibliography}
{\let\endmcitethebibliography\endthebibliography}{}

\end{document}